\documentclass[12pt]{article}

\usepackage{amsmath,graphicx}
\usepackage{multirow}
\usepackage{amsfonts}
\usepackage{amssymb}
\usepackage{amscd}
\usepackage{cite}
\usepackage{amsmath}
\usepackage{bbold}
\usepackage{bbm}
\usepackage{nicefrac}
\usepackage[vcentermath]{youngtab}
\usepackage{hyperref}

\newcommand{\ft}[2]{{\textstyle\frac{#1}{#2}}}

\def\Re{\text{Re}}
\def\Im{\text{Im}}

\def\cV{{\cal V}}

\def\Ma{M}
\def\Mb{N}
\def\Mc{P}

\def\ja{i}
\def\jb{j}
\def\jc{k}
\def\jd{l}

\def\jja{i}
\def\jjb{j}

\def\xa{a}

\def\aa{\alpha}
\def\ab{\beta}

\def\ta{\tilde{a}}
\def\tb{\tilde{b}}
\def\tc{\tilde{c}}
\def\td{\tilde{d}}
\def\te{\tilde{e}}
\newcommand{\y}[1]{\tilde{#1}}

\def\hybrid{\topmargin -20pt    \oddsidemargin 0pt
        \headheight 0pt \headsep 0pt
        \textwidth 6.25in       
        \textheight 9.5in       
        \marginparwidth .875in
        \parskip 5pt plus 1pt   \jot = 1.5ex}

\hybrid

\numberwithin{equation}{section}
\numberwithin{table}{section}

\newcommand{\be}{\begin{equation}}
\newcommand{\ee}{\end{equation}}

\newcommand{\bea}{\begin{eqnarray}}
\newcommand{\eea}{\end{eqnarray}}

\def\cN{{\cal N}}

\renewcommand{\Re}{\operatorname{Re}}
\renewcommand{\Im}{\operatorname{Im}}

\newenvironment{matr}[1]
{\left[ \begin{array}{{#1}}}{\end{array} \right]}

\def\fp{f} 
\def\rep#1{[{\bf #1}]}
\def\ba{\begin{aligned}}
\def\ea{\end{aligned}}

\begin{document}
\begin{titlepage}
\begin{center}
\rightline{\small ZMP-HH/12-27}
\vskip 1cm

{\Large \bf
Electrically gauged $\cN=4$ supergravities\\in $D=4$ with $\cN=2$ vacua}
\vskip 1.2cm
{\bf Christoph Horst$^{a,b}$, Jan Louis$^{a,b}$ and Paul Smyth$^{c}$}

\vskip 0.8cm

$^{a}${\em II. Institut f\"ur Theoretische Physik der Universit\"at Hamburg, Luruper Chaussee 149, D-22761 Hamburg, Germany}
\vskip 0.4cm

$^{b}${\em Zentrum f\"ur Mathematische Physik,
Universit\"at Hamburg,\\
Bundesstrasse 55, D-20146 Hamburg}
\vskip 0.4cm

$^{c}${\em Institut de Th\'eorie des Ph\'enom\`enes Physiques, EPFL, \\
\mbox{CH-1015 Lausanne, Switzerland}}
\vskip 0.8cm

{\tt christoph.horst@desy.de,jan.louis@desy.de\\paul.smyth@epfl.ch}

\end{center}

\vskip 1cm

\begin{center} {\bf ABSTRACT } \end{center}

\noindent

We study $\cN=2$ vacua in spontaneously
broken $\cN=4$ electrically gauged supergravities in four
space-time dimensions. We argue that the classification
of all such solutions amounts to solving a system of 
purely algebraic equations.
We then explicitly construct a special class of
consistent $\cN=2$ solutions and study their properties.
In particular we find that the spectrum assembles
in $\cN=2$ massless or BPS supermultiplets.
We show that
(modulo $U(1)$ factors) arbitrary unbroken gauge groups can be
realized provided that the number of $\cN=4$ vector multiplets is
large enough. 
Below the scale of partial
supersymmetry breaking we calculate the relevant terms of
the low-energy effective action and argue that the special K\"ahler
manifold for vector multiplets 
is completely determined, up to its dimension, and lies in the
unique series of special K\"ahler product manifolds.

\vfill
\noindent December 2012

\end{titlepage}

\section{Introduction}
The issue of spontaneous partial breaking in  theories  with extended supersymmetry has  long been studied \cite{Witten:1981nf,Cecotti:1984rk,Cecotti:1984wn}. The case of spontaneous $\cN=2\to \cN=1$ breaking in Minkowski vacua is of particular interest due to its phenomenological relevance and the early no-go theorems 
of \cite{Witten:1981nf,Cecotti:1984rk}.  
In $\cN=2$ globally  supersymmetric theories the no-go theorems could be
evaded in the presence of electric and magnetic Fayet-Iliopoulos terms that are not aligned \cite{Bagger:1994vj,Antoniadis:1995vb}. 
In supergravity the no-go theorem was circumvented in simple examples
by formulating the problem in a symplectic frame in which no
prepotential exists for the special geometry of the vector multiplets
\cite{Ferrara:1995gu,Ferrara:1995xi,Fre:1996js}. Recently, a
systematic analysis in $\cN=2$ supergravity with general matter
content was carried out
\cite{Louis:2009xd,Louis:2010ui,Cortes:2011ut} using the embedding
tensor formalism \cite{deWit:2005ub}. 

Spontaneous partial supersymmetry breaking in $\cN=4$ gauged
supergravity has been much less studied since the original examples
were found \cite{deRoo:1986yw,Wagemans:1987zy}. 
Motivated by the fact that the
original $\cN=4$ supergravities did not have vacua with
non-zero cosmological constant $\Lambda$, more general deformations
were introduced via a set of $SU(1,1)$ phases associated to the angles between the semi-simple factors of the gauge group \cite{Gates:1983ha,deRoo:1985jh}, now known as de Roo-Wagemans angles. In the embedding tensor language, non-trivial de Roo-Wagemans angles correspond to particular non-vanishing embedding tensor components which imply the simultaneous appearance of electric and magnetic gaugings \cite{Schon:2006kz,Weidner:2006rp}. These additional deformations were seen to allow for vacua with non-zero $\Lambda$ which can spontaneously break supersymmetry to all $\cN<4$ \cite{deRoo:1986yw}. The problem of partially breaking $\cN=4$ supersymmetry in Minkowski vacua was then studied \cite{Wagemans:1987zy}, where it was found that one could break to $\cN=1$ and $\cN=2$ supersymmetry, but not $\cN=3$.

More recently, examples of vacua with supersymmetry spontaneously
broken to $\cN < 4$ have been found and their relation to string
theory compactifications have been studied in some detail (see, for
example,
\cite{Tsokur:1994gr,Andrianopoli:2002rm,Dall'Agata:2009gv,Dibitetto:2011gm,Cassani:2011fu}
and references therein), but a systematic analysis of the problem has
yet to be carried out. The purpose of this paper is to initiate such
an analysis in $\cN=4$ gauged supergravity 
by solving the supersymmetry conditions for the charges and gaugings
that allow for a specified amount of preserved supersymmetry. 
As a first step, we shall focus on the specific case of spontaneous $\cN=4 \rightarrow \cN=2$ breaking with only electric gaugings.
 
In ungauged  $\cN=4$ supergravity with $n$ Abelian vector multiplets
the scalar field space is fixed to be the homogeneous space\cite{Bergshoeff:1985ms,deRoo:1985np,deRoo:1985jh}
\begin{equation} \label{s2:product_mf}
M = \nicefrac{SL(2)}{SO(2)}\times \nicefrac{SO(6,n)}{SO(6)\times
  SO(n)}\ , 
\end{equation}
where the first factor is spanned by the two scalars in the $\cN=4$
gravitational multiplet (the dilaton and axion), while the second factor
is spanned by the $6n$ scalars of the vector multiplets.
No scalar potential is allowed and thus all values of the scalar fields 
correspond to degenerate $\cN=4$ backgrounds.
 
This situation changes if one considers 
gauged $\cN=4$ supergravities \cite{deRoo:1985jh,Schon:2006kz}.
For simplicity, we confine our interest  to $\cN=4$ supergravity
coupled to $n$ vector multiplets transforming in the adjoint representation of an electric gauge group $G_{\cN=4}$.\footnote{That is, we do not consider the situation where (some of) 
the vector fields carry charges under dual magnetic gauge fields.}
This induces additional couplings and, in particular, a scalar potential $V$
which  is characterized by the structure constants of $G_{\cN=4}$.
In this case the analysis of possible backgrounds and the amount of supersymmetry they preserve becomes non-trivial.
The order parameters of supersymmetry breaking
are the scalar parts of the fermionic supersymmetry transformations,
which generically depend on the scalar fields and
the structure constants $f_{MNP}$.

Spontaneous $\cN=4\to \cN=2$ supersymmetry breaking occurs at points 
in the $\cN=4$ field space where the supersymmetry transformations of
two supercharges vanish (or are proportional to the square root of the
cosmological constant) while the remaining two are non-zero. This will
impose a set of conditions on the structure constants $f_{MNP}$, which
must also satisfy a complicated set of constraints (termed quadratic
constraints in the following) such that the theory itself is gauge
invariant and supersymmetric. The supersymmetry conditions are
significantly simplified by using the symmetries of the theory and the
fact that $M$ is a homogeneous space and therefore we can always
choose to perform our analysis at the origin of field space
\cite{Borghese:2010ei}. We shall see that this allows us to find a
purely algebraic reformulation of the problem, part of which can be
discussed in terms of the representation theory of a solvable Lie
algebra. We find that all maximally symmetric vacua of the
electrically gauged theory with $\cN =1$ or $\cN=2$ supersymmetry
preserved are necessarily Minkowski and that $\cN=3$ vacua do not
exist, as was already observed in \cite{Wagemans:1987zy}. We then turn
to solving the quadratic constraints, which prove too complicated to
solve in complete generality. In order to progress, we impose an
additional condition on the $f_{MNP}$, which holds automatically when
the number of vector multiplet $n$ is less or equal than six. It
corresponds to a particular choice of gauging which minimizes the
mixing between the gaugini and the gravitini in the
Lagrangian. Indeed, we shall see that in this case one can arrange for
only one $\cN = 4$ vector multiplet to contribute to the
gravity/Goldstini sector. For this class of gaugings we give the
explicit solutions of the quadratic constraints and the unbroken gauge
groups when $n\leq 6$. Moreover, for arbitrary $n$  we give solutions
with an additional set of gaugings (and couplings) turned off. In the
appendix we show that if any other solution were to exist, then it
would necessarily require the number of vector multiplet to be $n >
6$.  

Well below the scale  of the partial supersymmetry breaking $m_{3/2}$
one can derive a low-energy effective  theory
 by integrating out the two
heavy gravitini together with all other  fields
which gain a mass of order  $m_{3/2}$.
This effective theory is an $\cN=2$ supergravity 
which only contains light 
(with respect to $m_{3/2}$) $\cN=2$ multiplets.
We observe that all fields come in complete  $\cN=2$ supermultiplets
with appropriate mass degeneracies.
Furthermore, the two heavy gravitini which gain a mass  $m_{3/2}$ via
the super-Higgs mechanism have to be in a single $\cN=2$, spin-$\tfrac32$
BPS multiplet.

In the scalar sector we find that one of the $\cN=2$ vector multiplets contains the dilaton/axion of the original $\cN=4$ gravitational multiplet and that its field space $SL(2)/SO(2)$ descends
unchanged to the effective $\cN=2$ theory.
In particular  no mixing  with the other
scalar fields occurs in the kinetic terms. As the scalars in 
$\cN=2$ vector multiplets must span
a special K\"ahler manifold, we can use this observation to conclude that the $\cN=2$ scalar field space lies in the unique series of special K\"ahler product manifolds. This follows by a theorem of  \cite{Ferrara:1989py}, where it was shown that the only special K\"ahler manifolds which split into a direct product are the manifolds
\begin{equation}\label{skp}
M_{\rm SK} = 
\nicefrac{SL(2)}{SO(2)}\times \nicefrac{SO(2,k)}{SO(2)\times
  SO(k)}\ . 
\end{equation}
Since for the case at hand the first factor of $M$ coincides with the first factor of $M_{\rm SK} $, we can conclude that the 
$\cN=2$ vector multiplet field space is given by \eqref{skp}.

The rest of this paper is organized as follows. In Section 2 we review the main properties of electrically gauged 
$\cN=4$ supergravities. In Section 3 we formulate the conditions for
supersymmetry preserving vacua, focussing on the case of spontaneous
$\cN=4\to \cN=2$ breaking. We then present the solution of the $\cN=2$
vacuum conditions for a particular subclass of possible gaugings,
leaving the derivations and the discussion of the general case to the
appendices. In Section 4 we investigate the structure of the mass
terms and their consistency with the unbroken $\cN=2$
supersymmetry. We then discuss the possible unbroken gauge groups and 
comment on the geometry of the scalar manifold of the low energy effective $\cN=2$ theory. Our conventions and further technical details are gathered in the appendices.

\section{Electrically gauged $\cN=4$ supergravities in $D=4$}

Let us briefly recall some properties of $\cN=4$ gauged supergravity in four
dimensions. The generic spectrum consists of the gravity multiplet together with $n$ vector
multiplets.  The graviton multiplet contains the graviton $g_{\mu\nu}$, four
gravitini $\psi^i_{\mu},\, (i=1,\ldots,4)$, six vectors $A_{\mu}^m,\,
(m=1,\ldots,6)$, four spin-1/2 fermions $\chi^i$ and two scalars. We label the vector multiplets with the index $a=1,\ldots,n$ and each contains a vector $A_{\mu}^a$, $4$ spin-1/2 fermions $\lambda^{ai}$
and $6$ scalars.   
In this paper we only consider theories where the above fields carry
charges with respect to the electric gauge bosons.\footnote{More generally,
  one could also allow for charges with respect to dual magnetic
  gauge bosons. Such magnetically gauged theories can be described
  by means of the embedding tensor formalism
  \cite{deWit:2005ub,Weidner:2006rp}. Here, we choose a symplectic frame such that  
  the $A_{\mu}^m,A_{\mu}^a$ are the electric gauge bosons and restrict
  ourselves to electric gaugings only. 
}  
The bosonic Lagrangian for this class of theories is given by
\cite{Weidner:2006rp} 
\begin{equation}\begin{aligned}\label{Ldef}
  e^{-1}\mathcal{L}_{\text{bos.}}  
\ =\ & \tfrac{1}{2}R 
- \tfrac{1}{4} {\text{Im}} (\tau) M_{MN} {H_{\mu\nu}}^M H^{\mu\nu N} +
 \tfrac{1}{8} {\text{Re}} (\tau)\eta_{MN} \epsilon^{\mu\nu\rho\lambda}
  {H_{\mu\nu}}^M {H_{\rho\lambda}}^N 
  \\
 & + \tfrac{1}{16}(D_{\mu}M_{MN})(D^\mu M^{MN}) -\tfrac{1}{4 \text{Im}(\tau)^2}
  (\partial_{\mu} \tau)(\partial^{\mu}\tau^*)- V \ ,
\end{aligned}\end{equation}
where $R$ is the Ricci-scalar of the spacetime metric $g_{\mu\nu}$
and $e= \sqrt{|{\rm det}\, g|}$. The field strengths of
the vectors are defined by  
\begin{equation}
  {H_{\mu\nu}}^M = 2\,\partial_{[\mu} {A_{\nu]}}^M -  {\fp_{NP}}^M
  {A_{[\mu}}^N {A_{\nu]}}^P \ ,
\end{equation}
where the index $M=(m,a)=1,\ldots,6+n$ labels all the vector fields 
$A_{\mu}^M = (A_{\mu}^m,A_{\mu}^a)$. In \eqref{Ldef}, the matrix
$M=(M_{MN})=\cV \cV^T$ with $\cV\in SO(6,n)$ describes a (left) coset 
of $\nicefrac{SO(6,n)}{SO(6)\times SO(n)}$ 
which is the target
manifold of the scalars of the vector multiplets.
Similarly, $\tau\in\mathbb{C}$ with ${\rm Im}\,\tau > 0$ 
parametrizes
$\nicefrac{SL(2)}{SO(2)}$ which is the target manifold for the two
scalars of the gravity multiplet (see Appendix \ref{a:scalars} for further
details). 

The gauge covariant derivative acting on the vector multiplet
scalars is defined as
\begin{equation}
  \label{s2:DMMN}
  D_{\mu} M_{MN} = \partial_{\mu} M_{MN} + 2 {A_{\mu}}^P {\fp_{P(M}}^Q
  M_{N)Q}\ ,
\end{equation}
where $f_{MNP}$ are the real deformation parameters of the theory
(with $f_{MNP}=0$ in the ungauged theory). 
Supersymmetry and closure of the gauge
Lie algebra require the $f_{MNP}$  to satisfy the following linear and quadratic 
constraints
\cite{Bergshoeff:1985ms,Weidner:2006rp}     
\begin{equation}
  \label{s2:quadconstr}
 \fp_{MNP}=\fp_{[MNP]}\ ,\qquad \fp_{R[MN}\, {\fp_{PQ]}}^R=0 \ ,
\end{equation}
where the indices are raised and lowered with the $SO(6,n)$ invariant
metric  
\begin{equation}
  \eta = (\eta_{MN}) =  (\eta^{MN}) =
  \text{diag}(\underbrace{-1,\ldots,-1}_{6 \text{ 
      times}}, \underbrace{1,\ldots,1}_{n \text{ times}})\ .
\end{equation}

In the formalism used in \cite{Weidner:2006rp} the
$f_{MNP}$ are specific components of the embedding tensor, which is a spurionic matrix of charges. 
For purely electric gaugings the ${f_{MN}}^P$ are the structure
constants of the gauge Lie algebra and the quadratic constraint in
\eqref{s2:quadconstr} is the Jacobi identity. Note, however, that not
all gauge algebras can occur since the $f_{MNP} = {f_{MN}}^L \eta_{LP}$
have to be completely antisymmetric. Here, the occurrence of the
$SO(6,n)$ invariant metric $\eta_{MN}$ puts constraints on the
possible Lie algebras that can be gauged
\cite{Bergshoeff:1985ms,Weidner:2006rp}.\footnote{In contrast, for a semisimple Lie algebras
  with structure constants ${f_{ab}}^c$ the Killing form $\kappa_{ab}$
  is non-degenerate and can therefore be used to raise/lower
  indices. Then $f_{abc} = {f_{ab}}^d \kappa_{cd}$ would be
  automatically completely antisymmetric.} In the following we will
not initially specify the gauge group, but rather carry out the analysis for arbitrary
$\fp_{MNP}$. Later, when we discuss a restricted class of solutions for vacua with $\cN=2$ supersymmetry, we shall also be able to
determine the possible gauge groups. Finally, the scalar potential is given by 
\begin{equation}
\label{s2:scalarpot}
V=\tfrac{1}{16} \ft {1} {\Im \tau}  \fp_{MNP}\, \fp_{QRS}\, 
   \left[\tfrac{1}{3}M^{MQ}M^{NR}M^{PS} + \left(\tfrac{2}{3}
        \eta^{MQ}-M^{MQ}\right )\eta^{NR}\eta^{PS} \right]\ .
\end{equation}

For our analysis we also need the fermionic bilinear couplings,
which for the gravitini are \cite{Weidner:2006rp}\footnote{In Appendix 
  \ref{a:weyl} we will give our spinor conventions and relate the Weyl
  spinors used here to Dirac spinors which are used frequently in the
  literature. Also note that in \eqref{32couplings} we removed factors
  of $i$ in the mixed terms of gravitini and spin-1/2 fermions given
  in \cite{Weidner:2006rp}.} 
\begin{eqnarray}\label{32couplings}
   e^{-1} {\cal L}_{3/2} & = &
                           \ft 2 3 \,  A_1^{\jja\jjb} \,
                           (\psi_\mu^{i})^* \bar{\sigma}^{\mu\nu}
                           \epsilon\, (\psi_\nu^{j})^* +
                           \text{h.c.}\nonumber\\ 
                          && + \ft 1 3 \, A_2^{\jja\jjb}
                           (\psi_\mu^i)^* \sigma^\mu \epsilon\, (\chi^j)^*
			       - {A_{2\,\xa\jja}}^\jjb \,
                               \psi_\mu^i\,\epsilon\, \bar{\sigma}^\mu
                               \epsilon\, (\lambda^{aj})^*  
			       + \text{h.c.} \ ,
\end{eqnarray}
while the bilinear couplings of the spin-1/2 fermions read
\begin{eqnarray}  \label{s3:massspin12}
  e^{-1} {\cal L}_{1/2} &=& - {A_{2ai}}^j
  \chi^i(\lambda^{aj})^* +\text{h.c.}\nonumber\\
  & & + \tfrac{1}{3}\, A_2^{ij}
  (\lambda^{ai})^*\epsilon\, ({\lambda_a}^j)^*  
  + {A_{ab}}^{ij} (\lambda^{ai})^* \epsilon\, (\lambda^{bj})^* 
  +\text{h.c.}\ .
\end{eqnarray}
The scalar shift matrices $A$ appearing in \eqref{32couplings} and
\eqref{s3:massspin12} depend on the vielbein $\cV$ for $SO(6,n)$ and
$(\cV_{\alpha})=(\cV_-,\cV_+)$ for $SL(2)$ which are defined in
Appendix \ref{a:scalars}. They are given by
\begin{equation}\ba
  \label{s2:fermionshiftmatrices}
   A_1^{ij} &= ({\cV}_-)^* 
   {\fp_{\Ma}}^{\Mb\Mc} \, {{\cV^\Ma}_{[\jc\jd]}} {{\cV}_\Mb}^{[\ja\jc]} {{\cV}_\Mc}^{[\jb\jd]}    \ ,
		  \\
   A_2^{\ja\jb} &= {\cV}_-\,
              {\fp_{\Ma}}^{\Mb\Mc}      {{\cV^\Ma}_{[\jc\jd]}} {{\cV}_\Mb}^{[\ja\jc]} {{\cV}_\Mc}^{[\jb\jd]}  ,    \\
 	  {A_{2\, \xa\ja}}^\jb &={\cV}_-\,
      {\fp_{\Ma\Mb}}^\Mc\, {{\cV^{\Ma}}_\xa} {{\cV}^\Mb}_{[\ja\jc]}
      {{\cV}_\Mc}^{[\jb\jc]}\ ,\\
	     {A_{ab}}^{ij} &= \cV_-\, 
 {\fp_{MN}}^P\, {\cV^M}_a\,
  {\cV^N}_b\, {\cV_P}^{[ij]} \; ,
\ea\end{equation}
where we again use a double index notation $M=(m,a)$ with
$m=1,\ldots,6,\, a= 1, \ldots, n$. Indices $i,j,k,\ldots$
run from 1 to 4 and will turn out to be $SU(4)$ indices. More
precisely, objects with upper/lower indices transform under the $\bf 4$ and
$\bf\bar{4}$ of $SU(4)$, respectively, and complex conjugation interchanges 
upper and lower indices, e.g.\ $(\psi^i_\mu)^*$ transforms as a
$\bf\bar{4}$. 
Note that supersymmetry relates the
$A$-matrices in \eqref{32couplings} to the scalar potential via the generalized Ward identity \cite{Weidner:2006rp} 
\begin{equation}
  \label{s2:ward}
   \ft 1 3 \, A_1^{ik} \, (A_{1}^{jk})^*  - \, \ft 1 9 \,
   A_2^{ik} \, (A_{2}^{jk})^* 
   - \, \ft 1 2 \, {A_{2aj}}^k \, ({A_{2ai}}^k)^* \,  = \, - \,
   \ft 1 4 \, \delta^i_j \, V 
   \ ,
\end{equation}
with $V$ given in \eqref{s2:scalarpot}.

The full Lagrangian $\mathcal{L}$ (\eqref{Ldef} + fermionic terms) is
gauge invariant under local gauge transformations of a gauge group
which satisfies \eqref{s2:quadconstr}. In addition, ${\cal L}$ has a
global $G=SO(6,n)$ symmetry under which the vectors and
matter scalars (i.e.\ the scalars in the vector multiplets)
transform in the fundamental representation of $SO(6,n)$ provided that
the $\fp_{MNP}$ transform as a completely antisymmetric rank 3 tensor
with respect to $SO(6,n)$. It is in this sense that capitalized
indices $M,N,\ldots$ are referred to as $SO(6,n)$
indices. Furthermore, ${\cal L}$ is also invariant under the local
(i.e.\ spacetime dependent) symmetry $H=SU(4)\times SO(n)$ acting
non-trivially on fermionic fields and matter scalars. Indices
$a,b,\ldots=1,\ldots,n$ and $m,n,\ldots$ are indices with respect to
$SO(n)\subset H$ and $SO(6)\sim SU(4)\subset H$,
respectively. In addition to $H$, there is another local $U(1)$ symmetry
 acting both on fermions and on the vielbein $\cV_{\alpha}$ of 
$\nicefrac{SL(2)}{SO(2)}$ by multiplication with phase factors. 
The representations of the fields with respect to the
two groups $G$ and $H$, as well as the additional $U(1)$ symmetry are
summarized in Table~\ref{table_GHreps}. 

\begin{table}[htb]
\begin{center}
{\small\begin{tabular}{|c|c|c|c|}
\hline
        field & $G=SO(6,n)$ & $H=SU(4)\times SO(n)$ & $U(1) \text{
          charges}$\\ 
        \hline
        $g_{\mu\nu}$ & $\mathbf{1}$ & $(\mathbf{1},\mathbf{1})$ &
        $0$\\ 
        $\psi^i_{\mu}$ & $\mathbf{1}$ & $(\mathbf{4},\mathbf{1})$ &
        $-1/2$ \\
        $A_{\mu}^M$ & {\tiny $\yng(1)$} & $(\mathbf{1},\mathbf{1})$ & $0$\\ 
        $\chi^i$ & $\mathbf{1}$ & $(\mathbf{4},\mathbf{1})$ &
        $3/2$ \\
        $\lambda^{ai}$ & $\mathbf{1}$ &
        $(\mathbf{4},\mathbf{n})$ & $1/2$ \\ 
        $\cV= \cV_{SO(6,n)}$ & $\cV\rightarrow g\, \cV$ & $\cV\rightarrow \cV\,
        h(x)$ & $0$\\
        $\cV_{\alpha}$ & $\mathbf{1}$ &
        $(\mathbf{1},\mathbf{1})$ & 1 \\ 
\hline
     \end{tabular}}
\caption{$G$ and $H$ representations of the fields.
Here $g\in SO(6,n)$ and $h(x)\in SO(6)\times SO(n)$, i.e.\ in
particular, matter scalar representatives $\cV$ are charged with respect to
$SO(6)\sim SU(4)\subset H$.} 
\label{table_GHreps}
\end{center}
\end{table}

Since we are interested in vacua with a reduced number of supercharges
we need to identify the order parameters of this spontaneous
supersymmetry breaking. In a maximally symmetric
background they are the scalar
parts of the fermionic supersymmetry transformations which depend on
the $A$-matrices and are given by  
\cite{Weidner:2006rp} \footnote{Here the fields are understood to be
  background configurations.}
\begin{equation}\begin{aligned}
  \label{sec2:var_ferm}
  \delta_{\epsilon} \psi_{\mu}^i & =  2D_{\mu}
  \epsilon^i+\ft{2}{3}\, A_1^{ij}\bar\sigma_{\mu}\epsilon\, (\epsilon^j)^*\,, \\
  \delta_{\epsilon} \chi^i & =
  \ft{4}{3}i\,A_2^{ji}\epsilon\,(\epsilon^j)^*\,,\\ 
  \delta_{\epsilon} {\lambda_a}^i & =  2i\, A_{2aj}^i \epsilon^j\,.
\end{aligned}\end{equation}
Here the supersymmetry parameter $\epsilon^i$ is a Weyl spinor that
forms the right-handed spinor part of a Dirac spinor. It can be
decomposed into a product of a spacetime independent (complex) $SU(4)$ vector
$q^i$ and a Killing spinor $\eta$ of the spacetime according to
$\epsilon^i=q^i \eta$. The covariant derivative is then given by   
\begin{equation}
  D_\mu \epsilon^i = D_\mu q^i \eta 
= - q^i \sqrt{-\ft{1}{12}V}\, \bar\sigma_\mu \epsilon\, \eta^*\ .
\end{equation}
For supersymmetric vacua the background value of the scalar potential $V$ is either
zero (Minkowski) or negative (anti-de Sitter).

\section{Supersymmetric vacua and partial supersymmetry breaking}
\subsection{Preliminaries}

Let us first recall that an $\cN=4$ supersymmetric background is
defined by the conditions 
\begin{equation}\label{susyb}
  \delta_\epsilon \psi^\ja  =  0\ , \qquad
  \delta_\epsilon \chi^\ja =  0\ , \qquad
  \delta_\epsilon \lambda^{\xa\ja}  = 0
\end{equation}
for all free indices $i, a$ and for all supersymmetry parameters
$\epsilon^i$. Using the decomposition $\epsilon^i=q^i \eta$ introduced
in the previous section this translates into
\begin{equation}
  \label{s:killing_spinor}
   A_1^{\ja\jb} q_\jb = \sqrt{- \ft 3 4 V} \, q^\ja\ , \qquad
   q_\jb A_2^{\jb\ja} = 0\ ,\qquad 
   {A_{2\xa\jb}}^\ja q^\jb = 0,\qquad \forall i,a\, .  
\end{equation}
Using that in electrically gauged theories
the symmetric matrices $(A_1^{\ja\jb})$ and $(A_2^{\ja\jb})$ differ
only by an overall phase factor, see \eqref{s2:fermionshiftmatrices},
we can immediately conclude that in this class of theories an $\cN=4$
background has 
\begin{equation}
 (A_1^{\ja\jb})= (A_2^{\ja\jb}) =({A_{2\xa\jb}}^\ja)= 0,\,\forall a
 \,\qquad \textrm{and}\qquad V=0\ . 
\end{equation}
We also observe that by the same token \eqref{s:killing_spinor}
implies that in electrically gauged theories any background with at
least one preserved supersymmetry is necessarily Minkowski, i.e.\
$V=0$ \cite{Wagemans:1987zy}.\footnote{This result is implicit in \cite{Wagemans:1987zy}, in
  that the $SU(1,1)$ phases are set equal there. This is equivalent to saying that there are only electric gaugings
  \cite{Weidner:2006rp}.}   
We will not investigate $\cN=4$ backgrounds any further here, but instead shift our 
attention to backgrounds with less supersymmetry. Examples of vacua of
$\cN=4$  gauged supergravity with various amounts of preserved
supersymmetry have been discussed, for example, in
\cite{Dall'Agata:2009gv,Dibitetto:2011gm} and references therein. 

Our goal here is to classify the solution of \eqref{s:killing_spinor}
which preserve only two out of the four supercharges in a
maximally-symmetric background. Ordinarily, one should first pick a particular
$\cN=4$ supergravity theory, i.e. a specific gauging, and then look
for solutions of the Killing spinor
equations. Rather than take that approach, we shall follow
\cite{Louis:2009xd} and first specify the vacuum and the amount of
preserved supersymmetry, and then use the Killing spinors equations to
solve for the embedding tensor components, i.e. the gaugings, that
give rise to this vacuum. In this way we are solving the Killing
spinor equations to find the theory, rather than the vacuum. To this
end, one requires that \eqref{s:killing_spinor} hold for any 
preserved supersymmetry associated to a supercharge $\epsilon^i = q^i
\eta$, while for spontaneously broken supersymmetries
\eqref{s:killing_spinor} shall not be satisfied. Such a system of
equations and inequalities at an arbitrary critical point of the
scalar manifold is best solved (for the $f_{MNP}$) by using the
symmetry to go, without loss of generality, to the origin of the matter
scalar manifold, cf.\ \cite{Borghese:2010ei}, and, secondly, (by using
the residual symmetry) to diagonalize the gravitini mass matrix at
such a critical point.

\subsubsection{Going to the origin of the matter scalar manifold}
\label{s3:section_origin}
We assume that a given consistent electrically gauged theory has a
stable scalar vacuum, i.e.\ a critical point of the scalar potential
at some point in the scalar manifold: 
\begin{equation}
  (\cV_{SL(2)}, \cV_{SO(6,n)}) \in SL(2) \times SO(6,n)\,.
\end{equation}
Using the $G=SO(6,n)$ symmetry that, in particular, acts on scalar
fields according to Table \ref{table_GHreps} we can transform $ \cV_{SO(6,n)}$ to
the unit matrix $\mathbb{1}_{6+n} \in \text{Mat}_{6+n,6+n}$
and, hence, obtain a theory given in terms of redefined fields, new
components $f_{MNP}$ and a critical point
\begin{equation}
  \label{s3:origin}
  ((\cV_{SL(2)}, \mathbb{1}_{6+n}) \in SL(2) \times SO(6,n)\,.
\end{equation}
This is of help because the shift-matrices in \eqref{s:killing_spinor}
evaluated at \eqref{s3:origin} end up being disentangled with respect
to certain components of $f_{MNP}$, as we will see below. The residual
symmetry in a theory with vacuum \eqref{s3:origin} is a combination of
$SO(6)\times SO(n) \subset G$ and global $H$ symmetries such that
their compositions leave $\mathbb{1}_{6+n}$ invariant. In contrast,
$SL(2)$ is not part of the global symmetry of the Lagrangian and,
thus, cannot be used to also transform $\cV_{SL(2)}$
to the origin $\mathbb{1}_2\in SL(2)$ without loss of generality . Being an on-shell symmetry that
maps the system of equations of motion and Bianchi identities onto
another such system, general $SL(2)$ transformations would lead to
non-electrically gauged theories which (for simplicity) we do not want
to consider here.\footnote{General magnetic gaugings are described in
  terms of embedding tensors $f_{\alpha MNP} = f_{\alpha [MNP]}$ and
  $\xi_{\alpha M}$ where the index $\alpha$ is a vector index with
  respect to $SL(2)$ \cite{Schon:2006kz}. In our symplectic frame purely electric gaugings
  are given by $f_{-MNP}=0$ and $\xi_{\alpha M}=0$; $f_{MNP} \equiv
  f_{+MNP}$.} Using the additional local symmetry $U(1)\sim SO(2)$ of
the Lagrangian which acts as in Table \ref{table_GHreps} both on gravity
scalar representatives $\cV_{\alpha}$ and on fermions, we can bring the gravity vielbein to a form such that
$\cV_- =\nicefrac{1}{\sqrt{\Im\tau}} > 0$ without loss of generality (see Appendix
\ref{a:scalars} for details). This comes at the cost of redefining the  
fermion fields but simplifies the $A$-matrices in
\eqref{s2:fermionshiftmatrices}, in that $\cV_->0$ becomes an overall 
scaling factor. As a result, the components of the $A$-matrices at the
critical point \eqref{s3:origin} can be expressed as  
\begin{equation}\ba\label{shiftO}
  A_1^{ij} = A_2^{ij} & =  \tfrac{1}{8} \cV_-\,
  ([G_m]_{ik})^*[G_n]_{kl}([G_p]_{lj})^*f_{mnp} \ ,\\
  {A_{2ai}}^j & =  -\tfrac{1}{4} \cV_-\, [G_m]_{ik} ([G_n]_{kj})^* f_{amn}\
  ,\\
  {A_{ab}}^{ij} & = - \ft 1 2 \cV_-\, [G_m]^{ij} f_{abm}\ ,
\ea\end{equation}
where $G$ are the 't Hooft matrices, which we review in Appendix
\ref{a:scalars}. Note that at a critical point $(\mathbb{1}_2, 
\mathbb{1}_{6+n})$ one has $\cV_-=1$.

From the generalized Ward identity \eqref{s2:ward} or the explicit
form of the scalar potential given in \eqref{s2:scalarpot} one finds
that the scalar potential scales with a factor $(\cV_-)^2$. As a
consequence, the Killing spinor equations \eqref{s:killing_spinor} do
not depend on $\cV_-$, i.e.\ the analysis of partial supersymmetry
breaking does not depend on the critical point $\cV_{SL(2)} \in SL(2)$ in the
gravity scalar manifold. However, we observe that a generic $\cV_{SL(2)}$ leads 
to an overall scaling of all mass terms. Note that it is only upon
canonically normalizing the gauge kinetic terms that the mass
terms for the vector bosons also  scale appropriately. Also recall that for
any value of $\cV_{SL(2)}$ (or $\tau$) the background potential vanishes and,
hence, $\tau$ is a flat complex direction of the potential.

\subsubsection{Gravitino masses}

For all supergravity theories unbroken supersymmetries are in one-to-one
correspondence with massless gravitini \cite{Cecotti:1984wn}. Therefore, 
it will be instructive to first consider mass terms for the gravitini, 
\begin{equation}\begin{aligned}
  \label{s2:gravitini_mass}
  e^{-1}\mathcal{L}_{m_{3/2}} & =  \ft 2 3 \,  A_1^{\jja\jjb} \,
                           (\psi_\mu^{i})^* \bar{\sigma}^{\mu\nu}
                           \epsilon\, (\psi_\nu^{j})^* +
                           \text{h.c.}\ . 
\end{aligned}\end{equation}
An arbitrary symmetric complex matrix $(A_1^{ij})$ can be
diagonalized by means of an $SU(4)$ transformation. This is a
consequence of the Autonne decomposition \cite{Horn:1985}:
One can always find an $S\in SU(4)$ such that   
\begin{equation}
  \label{s2:diagA1}
  S (A_1^{ij}) S^T = \text{diag}(|a_1|e^{i\phi},|a_2|,|a_3|,|a_4|)\, ,
\end{equation}
with $|a_1|\le \ldots \le |a_4|$. Note, however, that diagonalizing a
non-diagonal matrix $(A_1^{ij})$ at the origin transforms also the
matrices $(A_2^{ij})$ and $({A_{2aj}}^i)$, and affects the vacuum by
an $SO(6)\subset H$ rotation moving it away from the critical point
\eqref{s3:origin}. Of course, the scalar vacuum always
remains in the same coset of $G/H$. We now think of such an $H$
transformation as acting globally and apply its inverse as a $G$
transformation on the vacuum, the $\fp_{MNP}$, and the vector
bosons. In doing so, one returns to the origin of $SO(6,n)$ and at the 
same time has a diagonal gravitino mass matrix. Moreover, one now
knows the $A$-matrices in terms of the transformed $\fp_{MNP}$. We
therefore may assume that, without loss of generality, $(A_1^{ij})$ is
of the form \eqref{s2:diagA1} and the $A$-matrices are explicitly
given as in Appendix \ref{a:amatrices}. Inspecting
\eqref{s2:gravitini_mass} we see that the gravitini mass parameters
are given by $2/3\cdot |a_1|,\ldots,2/3\cdot |a_4|$.

According to the Killing spinor equations \eqref{s:killing_spinor}
(with $V=0$ for electric gaugings) one requires for any unbroken
supersymmetry labelled by $q_i$ a zero diagonal entry of
$(A_1^{ij}) = (A_2^{ij})$. In contrast, for a broken supersymmetry
direction $q_i$ it is necessary that the diagonal entries be
positive. Furthermore, for each unbroken $q_i$ one needs a zero row in
matrices $({A_{2ai}}^j)$ for all $a$. It is apparent from the explicit
form of the shift matrices given in \eqref{s2:A_1}, \eqref{s2:A2a} and  
\eqref{s2:A2atrans} that the $f_{MNP}$ can be chosen in such a way
that the Killing spinor equations (and their inequalities) are
fulfilled at the critical point \eqref{s3:origin} for any number of
preserved supersymmetries. Note that this system of equations and
inequalities is linear in the $f_{MNP}$ and, hence, can easily be
solved. On the other hand, consistency of the gauged supergravities
requires solving the quadratic equations given in
\eqref{s2:quadconstr} which we discuss in Section 
\ref{s3:solvingquadconstraints}. Note that it is by means of
\eqref{s2:ward} that the first equation of \eqref{s:killing_spinor}
already implies the other two\footnote{This is always true for
  extended supergravity theories, see \cite{Cecotti:1984wn}.} which
means that in principle we need not demand zero rows in
$({A_{2ai}}^j)$ since this will follow from a solution of the
quadratic constraints. However, solving the constraints is difficult 
and introducing zero rows in the $({A_{2ai}}^j)$ is a useful measure
to simplify computations.

\subsection{$\cN=2$ vacuum}
\label{s2:n2vacuumcountingdofs}

Let us now turn to our main problem, which is to study spontaneous
breaking of $\cN=4$ to $\cN=2$ supersymmetry. For unbroken $\cN=2$
supersymmetry, one generically has  
\begin{equation}
  \label{s2:N=2A1}
  (A_1^{ij}) =  (A_2^{ij}) = \text{diag}(0,0,\mu_1,\mu_2)  \ ,
\qquad  {A_{2a1}}^i = {A_{2a2}}^i =  0\ ,\quad \forall i,a\ .
\end{equation}
Recall that the vacuum is necessarily Minkowski which implies that
the first two eigenvalues of $(A_1^{ij})$ are zero. Before we solve
\eqref{s2:N=2A1} let us study the decomposition of $\cN=4$ multiplets
in terms of $\cN=2$ multiplets. This is of interest as partial
supersymmetry breaking requires massive gravitini to be embedded into 
massive supermultiplets of the preserved
supersymmetry~\cite{Ferrara:1983gn}.  



\subsubsection{Multiplets of $\cN=4,2$ and their relations}
In this section we review the decomposition of the two massless
$\cN=4$ multiplets into $\cN=2$ multiplets. Let us denote a multiplet 
of $\cN$-extended supersymmetry with mass $m$ and highest
spin/helicity $s$ in Minkowski space by $M_{{\cal N},s,m}$. Using this  
terminology the $\cN=4$ gravitational multiplet and the massless
vector multiplet together with their component spectrum read 
\begin{equation}\ba\label{masslessN4}
\cN=4\ \textrm{gravitational multiplet:}&\qquad
M_{4,2,0} = \Big(\rep 2, 4\rep{\tfrac32},
   6\rep1,4\rep{\tfrac12},2\rep0\Big)\ ,\\
\cN=4\ \textrm{vector multiplet:}&\qquad
M_{4,1,0} = \Big(\rep 1, 4\rep{\tfrac12},
   6\rep0\Big)\ , 
\ea\end{equation}
where $\rep s$ denotes the spin/helicity of the component and the
number in front is its multiplicity. The massless $\cN=2$ multiplets are 
\begin{equation}\ba\label{masslessN2}
\cN=2\ \textrm{gravitational multiplet:}&\qquad
M_{2,2,0} = \Big(\rep 2, 2\rep{\tfrac32},\rep1\Big)\ ,\\
\cN=2\ \textrm{gravitino multiplet:}&\qquad
M_{2,3/2,0} = \Big(\rep{\tfrac32},2\rep1,\rep{\tfrac12}\Big)\ ,\\ 
\cN=2\ \textrm{vector multiplet:}&\qquad
M_{2,1,0} = \Big(\rep 1, 2\rep{\tfrac12},2\rep0\Big)\ ,\\
 \cN=2\ \textrm{hypermultiplet:}&\qquad
M_{2,1/2,0} = \Big(2\rep{\tfrac12},4\rep0\Big)\ ,
\ea\end{equation}
while the massive $\cN=2$ multiplets read
\begin{equation}\ba\label{massiveN2}
\cN=2\ \textrm{massive gravitino multiplet:}&\qquad
M_{2,3/2,m\neq 0} = \Big(\rep{\tfrac32},4\rep1,6\rep{\tfrac12},4\rep0\Big)\
,\\ 
\cN=2\ \textrm{BPS gravitino multiplet:}&\qquad
M_{2,3/2,{\rm BPS}} =
\Big(2\rep{\tfrac32},4\rep1,2\rep{\tfrac12}\Big)\ ,\\ 
\cN=2\ \textrm{massive vector multiplet:}&\qquad
M_{2,1,m\neq 0} = \Big(\rep 1, 4\rep{\tfrac12},5\rep0\Big)\ ,\\
\cN=2\ \textrm{BPS vector multiplet:}&\qquad
M_{2,1,{\rm BPS}} = \Big(\rep 1, 2\rep{\tfrac12},1\rep0\Big)\ ,\\
 \cN=2\ \textrm{BPS hypermultiplet:}&\qquad
M_{2,1/2,{\rm BPS}} = \Big(2\rep{\tfrac12},4\rep0\Big)\ .
\ea\end{equation}
Note that there are two distinct $\cN=2$ massive gravitino multiplets, 
the BPS gravitino multiplet $M_{2,3/2,{\rm BPS}}$ and the long massive
gravitino multiplet $M_{2,3/2,m\neq 0}$. They differ in that only the
BPS gravitino multiplet transforms under a central charge of the  
supersymmetry algebra in precisely the way that leads to multiplet
shortening. BPS gravitini can only occur in pairs as each of them
carries a non-vanishing BPS charge which by itself would not be 
CPT-invariant. This implies that $\cN=4$ cannot be broken to
  $\cN=3$ with a BPS gravitino multiplet. 

The branching rules of the two $\cN=4$ multiplets in terms of massless 
$\cN=2$ multiplets are as follows
\begin{equation}
M_{4,2,0} = M_{2,2,0}+2 M_{2,3/2,0}+M_{2,1,0}\ ,\qquad
M_{4,1,0}= M_{2,1,0}+M_{2,1/2,0}\ ,
\end{equation}
from which we see that in breaking $\cN=4 \rightarrow \cN=2$ the
gravity multiplet gives rise to a vector multiplet containing the
dilaton and axion in the $\cN=2$ spectrum.

As all degrees of freedom must 
be embedded into complete $\cN=2$ multiplets, the two 
heavy gravitini must lie in  massive $\cN=2$ 
supermultiplets. We thus have two options regarding the type of the
gravitino multiplet(s). For the situation where the heavy $\cN=2$
gravitini are in non-BPS multiplets one has
\begin{equation}
M_{4,2,0} + n M_{4,1,0} \to
M_{2,2,0}+ 2 M_{2,3/2,m\neq 0}+ n'_{\rm v} M_{2,1,m\neq 0}+ n_{\rm v}
M_{2,1,\cdot} 
+ n_{\rm h}M_{2,1/2,\cdot}\ ,
\end{equation}
where $n'_{\rm v}$ counts long massive vector multiplets, $n_{\rm v}$
counts BPS vector multiplets and massless vector multiplets (as they
have the same field content) and $n_{\rm h}$ counts BPS or massless
hypermultiplets (as they also have the same field content). We use
$\cdot$ to denote either massless or BPS multiplets. Inserting the
spectrum  \eqref{masslessN4}--\eqref{massiveN2} one finds the
consistency conditions 
\begin{equation}
\label{s3:countdofgravitini}
n_{\rm v} = n-3-n'_{\rm v}\ ,\qquad n_{\rm h} = n-2-n'_{\rm v}\ .
\end{equation}
Thus in this case there have to be at least three $\cN=4$ vector
multiplets in the spectrum, i.e.\ $n\ge 3$. In this minimal case with 
also $n'_{\rm v}=0$ there are, apart from the $\cN=2$ gravitational
multiplet and the two heavy gravitino multiplets, one massive or
massless hypermultiplet after the symmetry breaking. 

In case that the heavy $\cN=2$ gravitini are contained in a BPS
multiplet one has  
\begin{equation}
M_{4,2,0} + n M_{4,1,0} \to
M_{2,2,0}+ M_{2,3/2,BPS}+ n'_{\rm v} M_{2,1,m\neq 0}+ n_{\rm v}
M_{2,1,\cdot} 
+ n_{\rm h}M_{2,1/2,\cdot}\ ,
\end{equation}
with the consistency conditions
\begin{equation}
\label{s3:countdofBPSgravitini}
n_{\rm v} = n+1-n'_{\rm v}\ ,\qquad n_{\rm h} = n-1-n'_{\rm v}\ ,
\end{equation}
and thus there has to be at least one $\cN=4$ vector multiplet in
the spectrum, i.e.\ $n\ge 1$. In this minimal case with $n'_{\rm
  v}=0$, one finds after the symmetry breaking the $\cN=2$
gravitational multiplet, the BPS gravitino multiplet, and two
massless/BPS vector multiplets. Note that according to equations
\eqref{s3:countdofgravitini} and \eqref{s3:countdofBPSgravitini} 
the case with two long massive gravitino multiplets $M_{2,3/2,m\neq
  0}$, relative to the BPS case, yields one fewer hypermultiplet and
four fewer vector multiplets in the spectrum.

\subsubsection{The linear conditions}
\label{s3.0:solvinglinconstraints}
In this section we first solve the linear $\cN=2$ conditions 
\eqref{s2:N=2A1} and then embark on solving the quadratic constraints
\eqref{s2:quadconstr}. While the linear equations can easily be
solved, it is hard to find the general solution for the quadratic
constraints. 

Let us first focus on the zero entries in $A_1 (=A_2)$. Using the
explicit form given in Appendix \ref{a:amatrices} one easily finds
that only four of the $\fp_{mnp}$ can be non-zero and they depend on only 
two parameters which we denote by $c$ and $d$. More precisely one
finds 
\begin{equation}\label{cddef}
  \fp_{234}  =\fp_{456} =: c \ ,\qquad
  \fp_{126} =\fp_{135} =: d\ ,
\end{equation}
while all other $\fp_{mnp}$ vanish. Moreover, $A_1^{33}$ and
$A_1^{44}$ which are related to the gravitini mass parameters $\mu_1$
and $\mu_2$ introduced in \eqref{s2:N=2A1} also depend on $c$ and
$d$ via 
\begin{equation}
  \label{s3:A1diagmassgravi}
  A_1^{33}  =  -\ft 3 2 \,\cV_- (c+d) = \mu_1 > 0 \ ,\qquad
  A_1^{44}  =  -\ft 3 2 \,\cV_- (c-d) = \mu_2  \ge \mu_1 \ ,
\end{equation}
where as pointed out before $\mu_2 \ge \mu_1$ is chosen without loss
of generality. Let us now turn to the last set of equations in
\eqref{s2:N=2A1} and solve the system of linear equations for the
shift matrices $({A_{2ai}}^j)$. Using \eqref{s2:A2a} and
\eqref{s2:A2atrans} the potentially non-trivial components of
$\fp_{amn}$ turn out to be
\begin{equation}\label{efgdef}
  \fp_{a25}  =  -\fp_{a36} =: e_a \ ,\qquad
  \fp_{a23} =  \fp_{a56} =: f_a \ ,\qquad 
  \fp_{a26}  =  \fp_{a35} =: g_a \ ,
\end{equation}
while $\fp_{a1n}=\fp_{a4n}=0$ for all $a$ and $n$.  Thus, for any $a$,
the matrix ${A_{2ai}}^j$ is non-trivial only in its lower right block and given by
\begin{equation}
\label{s3:A2aij}
  ({A_{2ai}}^j)=
\begin{pmatrix}
  0 & 0  \\
  0 & Z_a 
\end{pmatrix}, \qquad
Z_a = f_a
\begin{pmatrix}
  0 & 1  \\
  -1 & 0 
\end{pmatrix}+i
\begin{pmatrix}
  -e_a & g_a  \\
  g_a & e_a 
\end{pmatrix}\ .
\end{equation}
This concludes our analysis of the linear equations arising from
the Killing spinor equations \eqref{s2:N=2A1}. Let us now turn to the quadratic constraints.

\subsubsection{Partial solution of the quadratic conditions} 
\label{s3:solvingquadconstraints}

In order to ensure that a given choice of gauging is consistent with
supersymmetry and gauge invariance of the Lagrangian, we need  to
impose the quadratic constraints \eqref{s2:quadconstr}
$\fp_{R[MN}\, {\fp_{PQ]}}^R=0$
\cite{deWit:2005ub,Schon:2006kz,Weidner:2006rp}. However, in practice
it is difficult to solve these constraints in general and we will have
to make much use of their symmetry properties. For instance,
\eqref{s2:quadconstr} are $SO(6,n)$ tensor equations and it will prove
crucial to exploit all the symmetries. 

Let us first look at the component $(M,N,P,Q)=(m,n,p,q)=(1,2,4,5)$ of the
quadratic constraints \eqref{s2:quadconstr} and insert \eqref{cddef} to arrive at
\begin{equation}\label{quadfirst}
c\cdot  d=0\ . 
\end{equation}
Since $c=0$ is inconsistent with the gauge choice of
\eqref{s3:A1diagmassgravi}, we need to have $d=0$ and $c<0$.
This in turn implies a first result, namely that the two heavy
gravitini have to be degenerate in mass 
\begin{equation}
  \label{s3:mgravitino}
  m_{3/2} := \ft{2}{3}\, A_1^{33} = \ft{2}{3}\, A_1^{44} = -c \,\cV_-
  \ , 
\end{equation}
as one expects when some fraction of supersymmetry is preserved in a
Minkowski background. Let us also note that  \eqref{quadfirst}
immediately implies that in electrically gauged theories one can never
break $\cN=4$ to $\cN=3$ since $A_1^{33}=0, A_1^{44}\neq 0$ requires
$c=-d\neq 0$, as was first shown in
\cite{Wagemans:1987zy}.

In order to proceed, it is necessary to make some simplifying
assumptions. By inspection, one finds that for $g_a=0$ the equations
simplify considerably and therefore some of them can be solved. On the
other hand, the $g_a\neq 0$ case is much more involved and solutions
--- should they exist --- would have to be more sophisticated, as we
point out in Appendix \ref{a:ganeq0}. In what follows we will
therefore assume that $g_a=0$, which also implies $e_a=0$ due to the
quadratic constraint for $(M,N,P,Q)=(b,n,p,q)=(b,2,4,6)$. This choice
corresponds to turning-off certain components of the $A$-matrices and
minimizes the coupling between gravitini and gaugini in the Lagrangian
\eqref{32couplings}. Indeed, we shall see later that with this choice
it is only the ``first'' $\cN=4$ vector multiplet that
contributes to the gravity/Goldstini sector. The fact that it is the components $g_a =
\fp_{a26}  =  \fp_{a35} =0$ and $e_a = \fp_{a25}  =  -\fp_{a36}=0$
that allow for this simplification is due to our particular  $SU(4)$
gauge choice for which gravitini remain massless \eqref{s2:N=2A1},
suitably translated into $SO(6)$ indices using the 't Hooft matrices
(see \eqref{a:scalars}).

Let us now consider the quadratic constraint
$(M,N,P,Q)=(m,n,p,q)=(2,3,5,6)$. Inserting \eqref{cddef} and
\eqref{efgdef} we find
\begin{equation}
  \label{s3:qc_2356}
  \sum_a f_a^2 = c^2 > 0\ ,
\end{equation}
i.e.\ at least one $f_a$ must be different from zero.
This implies (via \eqref{s3:A2aij}) that $({A_{2ai}}^j)$ has non-zero
entries and from \eqref{32couplings} and \eqref{s3:massspin12}  
we see that additional fermionic couplings have to be non-zero and
related to the gravitino mass. As we will see in Section
\ref{s4:masses}, \eqref{s3:qc_2356} is necessary for the super-Higgs
mechanism and the appropriate couplings of the Goldstone fermions to
the gravitinos.  
In  order to simplify the analysis we use an $SO(n)$ transformation
that leaves the origin invariant and choose $f_a=c\,\delta_{a7}$ which
obviously solves \eqref{s3:qc_2356}. The quadratic constraints
$(M,N,P,Q)=(b,n,p,q)$ then imply  
\begin{equation}
  \fp_{7bm}=0\ ,\quad \forall\, b,m\ .
\end{equation}

In Appendix \ref{a:detailsquadconstr} we list the remaining
non-trivial quadratic constraints. A subset of them,
\eqref{s3:qc_bc12'} - \eqref{s3:qc_bc76'}, can be written in terms
of the antisymmetric real $(n-1)\times(n-1)$ matrices 
\begin{equation}
  G_m = (\fp_{\tb\tc m}) \, \qquad\textrm{and}\qquad  G_7 =
  (\fp_{\tb\tc 7}) \qquad\textrm{with } \tb,\tc=8,\ldots,6+n\ .
\end{equation}
which satisfy
\begin{equation}\ba
  \label{s3:liealg}
  [G_2,H_+] &=-2c\, G_3\ , \qquad
  [G_3,H_+] =+2c\, G_2\ ,\qquad
  [G_2,G_3] = cH_-\ ,\\
  [G_5,H_+] &=-2c\, G_6\ , \qquad
  [G_6,H_+] =+2c\, G_5\ , \qquad
  [G_5,G_6] = c\, H_-\ ,
\ea\end{equation}
where $H_{\pm}=G_4\pm G_7$ and with the remaining commutators all
vanishing. 
\eqref{s3:liealg} defines a Lie bracket on the
7-dimensional real vector space spanned by abstract elements
$\{G_1,G_2, G_3, G_5, G_6, H_+, H_-\}$ and it can be  checked
that the Jacobi identities are satisfied.

Note that  $G_1$ commutes with all other elements and thus we have a
real 7-dimensional Lie algebra $\mathfrak{g}$ which decomposes into a
sum of two ideals,  
\begin{equation}
  \mathfrak{g} \cong \mathbb{R} \oplus \mathfrak{s},
\end{equation}
spanned by $G_1$ and $\{G_2,G_3,G_5,G_6,H_+,H_-\}$, respectively. It
can be further checked that $\mathfrak{s}$ is a solvable Lie algebra
of dimension 6.\footnote{Recall that a Lie algebra $\mathfrak{g}$ is solvable if
  and only if the (upper) derived series of Lie algebras
  $(\mathfrak{g},[\mathfrak{g},\mathfrak{g}],
  [[\mathfrak{g},\mathfrak{g}],[\mathfrak{g},\mathfrak{g}]]\ldots)$
  terminates after finitely many steps.}  
The problem of finding solutions to the quadratic constraints
\eqref{s3:qc_bc12'} - \eqref{s3:qc_bc76'} is now equivalent to finding
antisymmetric finite-dimensional representations of
$\mathfrak{g}$. One obvious class of solutions is given by   
\begin{equation}
  \label{s3:onlysol}
  G_2=G_3=G_5=G_6=H_-=0
\end{equation}
and an arbitrary, antisymmetric $H_+$ that commutes with $G_1$. In
this case one has $G_4=G_7$. In Appendix \ref{a:proof} we will prove
that no other solution exists. Our proof is based on Lie's theorem
concerning complex representations of complex solvable Lie algebras. 


The remaining equations \eqref{s3:qc_bc12'} to \eqref{s3:qc_bcde'}
to be solved now simplify to
\begin{subequations}
\begin{align}
\label{s3:qc_bc14''}
  \fp_{\ta\tb 4}\, \fp_{\ta\tc 1} - \fp_{\ta\tb 1}\,
  \fp_{\ta\tc 4} & = 0 \\
\label{s3:qc_bcd1''}
  \fp_{\ta\tb\tc}\, \fp_{\ta\td 1} + \fp_{\ta\tb 1}\,
  \fp_{\ta\tc\td} - \fp_{\ta\tb\td}\, \fp_{\ta\tc 1} & = 0\\
\label{s3:qc_bcd4''}
  \fp_{\ta\tb\tc}\, \fp_{\ta\td 4} + \fp_{\ta\tb 4}\,
  \fp_{\ta\tc\td} - \fp_{\ta\tb\td}\, \fp_{\ta\tc 4} & = 0\\
\label{s3:qc_bcde''}
  \fp_{\ta\tb\tc}\,\fp_{\td\te \ta} + \fp_{\ta\tb\te}\,
  \fp_{\tc\td \ta} - \fp_{\ta\tb\td}\, \fp_{\ta\tc\te} &= \fp_{1\tb\tc}\,\fp_{\td\te
    1} + \fp_{1\tb\te}\, \fp_{\tc\td 1} - \fp_{1\tb\td}\, \fp_{1\tc\te}\,.
\end{align}
\end{subequations}
Note that the gravitino mass parameter $c$ has disappeared from the
equations. Unfortunately, it is still hard to solve these equations in
generality for any given integer $n$.

Let us first consider $G_1=G_4=0$. In this case the only remaining
non-trivial equation is  
\begin{equation}
  \label{s3:jacobiid}
  \fp_{\ta\tb\tc}\,\fp_{\td\te \ta} + \fp_{\ta\tb\te}\,
  \fp_{\tc\td \ta} - \fp_{\ta\tb\td}\, \fp_{\ta\tc\te} = 0\,,
\end{equation}
which tallies with the Jacobi identity in the adjoint representation
of the compact form of a reductive Lie algebra of rank $(n-1)$ when
expressed in an appropriate basis. Based on the classification of
simple Lie algebras, solutions to \eqref{s3:jacobiid} are
well-understood. As we will see in Section
\ref{s4:unbrokengaugegroup}, when exponentiated this gives 
rise to a compact reductive Lie group that leaves invariant the vacuum
of the theory and, hence, corresponds to the unbroken gauge group.

Now we turn to the case of non-trivial $G_1$ and $G_4$. In Appendix
\ref{a:G1G4zero} we will solve \eqref{s3:qc_bc14''}, which 
in matrix notation reads
\begin{equation}
  [G_1, G_4]=0\, .
\end{equation}
Here we will only explain the result. The solution of this $SO(n-1)$
tensor equation could be given in terms of $SO(n-1)$ representatives
of an orbit of solutions. However, as it is also an $O(n-1)$ tensor
equation, it is more convenient to give its solution in terms of
$O(n-1)$ representatives, up to an additional simple reflection, so as
to obtain this gauge by a $SO(n-1)$ rotation. Regardless of this
subtlety our gauge choice proves useful in the following analysis. One
finds that the most general solution consists of simultaneously
block-diagonal $G_1$ and $G_4$ with blocks that square to a matrix
proportional to the identity of the block. The explicit form of $G_1$
and $G_4$ in our gauge is given as follows: First of all, we have
\begin{equation}
\label{s3:gauge_g1'}
  G_1 = (D\otimes\varepsilon)\oplus 0 =
\begin{pmatrix}D\otimes \varepsilon & 0\\ 0&0\end{pmatrix} \,,
\end{equation}
where $D=diag(x_1,\ldots,x_1,x_2,\ldots,x_2,\ldots)$ is a diagonal
matrix with ordered positive eigenvalues $x_1 > x_2 > \ldots > 0$ and
$\varepsilon$ is the antisymmetric $2\times 2$ matrix with
$\varepsilon_{12} = 1$; the zeros in \eqref{s3:gauge_g1'} denote zero
matrices of appropriate dimensions. Then, we have
\begin{equation}
  \label{s3:commG1G4_G4'}
  G_4 = \begin{pmatrix} A & 0 \\ 0 & (D'\otimes \varepsilon) \oplus 
    0\end{pmatrix}  \,,
\end{equation}
where $A$ is an antisymmetric matrix (of the same matrix dimensions
as $D\otimes \varepsilon$) satisfying
\begin{equation}
  [D\otimes \varepsilon,A]=0 \,,
\end{equation}
and $D'$ is another invertible diagonal matrix. Furthermore, we show
in Appendix \ref{a:G1G4zero} that both $D\otimes \varepsilon$ and $A$
are block-diagonal. Furthermore, as a result, we list the four
different types of blocks that can appear in Table \ref{table_G1G4}.  

\begin{table}[htb]
\begin{center}
{\small\begin{tabular}{|l|l|}
\hline
  $G_1$ block & $G_4$ block \\
\hline
  $x_i\, \mathbb{1}\otimes \varepsilon$ & $0\cdot \mathbb{1}\otimes
  \mathbb{1}_2$\\ 
\hline
  $x_i
\begin{pmatrix}
  \mathbb{1} & 0\\
  0 & \mathbb{1}'  
\end{pmatrix} \otimes \varepsilon$ &
  $|y_{ij}| 
\begin{pmatrix}
  \mathbb{1} & 0\\
  0 & -\mathbb{1}'
\end{pmatrix} \otimes\varepsilon$ \\
\hline
  $x_i\, \mathbb{1}\otimes (\mathbb{1}_2\otimes \varepsilon)$ &
  $|y_{ij}|\, \mathbb{1}\otimes \left (\cos\phi_{ijk}  
\begin{pmatrix}
  1 & 0\\
  0 & -1
\end{pmatrix} \otimes\varepsilon + \sin\phi_{ijk}
\begin{pmatrix}
  0 & 1\\
  -1 & 0
\end{pmatrix} \otimes\mathbb{1}_2
 \right)$ \\
\hline
  $x_i\, \mathbb{1}\otimes (\mathbb{1}_2 \otimes 
  \varepsilon)$ & $D^{(ij0)}\otimes(\varepsilon \otimes
  \mathbb{1}_2)$\\  
\hline
\end{tabular}}
\caption{The four different types of blocks appearing in the solution
  of $[D\otimes \varepsilon, A]=0$. The label $i$ refers to blocks in
  $G_1$ with eigenvalues $-x_i^2\neq 0$ of $G_1^2$. Similarly, the
  label $j$ is associated to subblocks in $G_4$ with eigenvalues
  $-y_{ij}^2\neq 0$ of $G_4^2$. Moreover, $D^{(ij0)}$ is a diagonal
  matrix with eigenvalues $\pm y_{ij}$ and $\phi_{ijk}\in
  (0,\pi/2)$. Finally, $k$ labels different possible angles
  $\phi_{ijk}$.}   
\label{table_G1G4}
\end{center}
\end{table}

We will now solve the tensor equation given in
\eqref{s3:qc_bcd1''}. For a given $G_1$, these equations are linear in
$f_{\ta\tb\tc}$ and can easily be solved for the latter in the gauge
\eqref{s3:gauge_g1}. Before we state the result, we introduce some
index notation in that we distinguish $SO(n-1)$ indices
$\ta,\tb,\ldots$ depending on whether or not they correspond to
non-zero or zero blocks in $G_1$: Components of non-zero $2\times 2$
blocks shall have subindices, e.g.\ $\ta_1=1,2$, indicating the block
they belong to. On the other hand, components associated to the zero
block in $G_1$ shall be denoted by $\ta_0$. Furthermore, we introduce
matrices
\begin{equation}
  G_{\ta_0}^{(x_1)} = ({G_{\ta_0}}_{\tb_1\tc_2}) = f_{\ta_0\tb_1\tc_2}\,,
\end{equation}
where $\tb_1, \tc_2$ run over all indices associated to blocks with
$x_1$ in $G_1$.  The solution of \eqref{s3:qc_bcd1''} is given in
terms of three classes of potentially non-trivial components
$f_{\ta\tb\tc}$. First, 
\begin{equation}
\label{s3:fa0b0c0inreals}
  f_{\ta_0\tb_0\tc_0} \in\mathbb{R}\,,
\end{equation}
can be arbitrary; then one finds
\begin{equation}
\label{s3:Gatilde0}
  G_{\ta_0}^{(x_1)} = S^{(x_1)}\otimes \varepsilon + A^{(x_1)}\otimes
  \mathbb{1}_2 \,,
\end{equation}
for a symmetric matrix $S^{(x_1)}$ and an antisymmetric $A^{(x_1)}$;
finally components $f_{\ta_1 \tb_2 \tb_3}$ are given in terms of two
real numbers,
\begin{eqnarray}
\label{s3:fa1b2c3components}
  f_{2_1 1_2 2_3} & = & f_{1_1 1_2 1_3}\,, \nonumber\\
  f_{1_1 2_2 2_3} & = & -f_{1_1 1_2 1_3}\,, \nonumber\\
  f_{2_1 2_2 1_3} & = & f_{1_1 1_2 1_3}\,, \nonumber\\
  f_{2_1 1_2 1_3} & = & -f_{2_1 2_2 2_3}\,, \nonumber\\
  f_{1_1 2_2 1_3} & = & f_{2_1 2_2 2_3}\,, \nonumber\\
  f_{1_1 1_2 21_3} & = & f_{2_1 2_2 2_3}
\end{eqnarray}
for $x_1 = x_2 + x_3$ ($x_1 \ge x_2 \ge x_3)$. 

Unfortunately we are unable to solve equations \eqref{s3:qc_bcd4''}
and \eqref{s3:qc_bcde''} in full generality. We will therefore proceed
by discussing certain special solutions of them (still in the case
$g_a=0$). 

\subsection{Special solutions}
\label{s3:specsols}
We will discuss two special classes of solutions to the equations
given in \eqref{s3:qc_bc14''} to \eqref{s3:qc_bcde''}. First we will
give all solutions in the case of $n\leq 6$, and secondly we construct
special but physically non-trivial solutions that work for any
$n\in\mathbb{N}$.  

\subsubsection{Solutions for $n\leq 6$}
\label{s3:sols_n6}
In Appendix \ref{a:ganeq0} we show that for $n\leq 6$ consistency
requires $g_a=0$. As a consequence, the equations to be solved are
precisely the ones in \eqref{s3:qc_bc14''} to \eqref{s3:qc_bcde''}. As
in \eqref{s3:gauge_g1'}, we will bring $G_1$ to the following gauge  
\begin{equation}
  \label{s3:sols_n6_g1}
  G_1 = \left [\begin{pmatrix}m_1&0\\0&m_2\\\end{pmatrix}\otimes
    \varepsilon \right]
  \oplus 0\in\text{Mat}_{5,5}
\end{equation}
for $n=6$ with $m_1,m_2\in\mathbb{R}$, or to obvious truncations of
\eqref{s3:sols_n6_g1} to matrices in $\text{Mat}_{n-1,n-1}$ for $n\leq
5$. As discussed in Appendix \ref{a:G1G4zero}, we distinguish between
the following two cases: Given that matrices $(G_1)^2$ and $(G_4)^2$
have four nonzero degenerate eigenvalues each (which can only happen
for $n\ge 5$), $G_4$ can be written as  
\begin{equation}
\label{s3:g4phi}
  G_4 = \pm n_1 \left
    [\cos\phi \begin{pmatrix}1&0\\0&-1\\\end{pmatrix}\otimes
    \varepsilon     +
    \sin\phi \begin{pmatrix}0&1\\-1&0\\\end{pmatrix}\otimes\mathbb{1}_2
  \right]\oplus 0
\end{equation}
for $n=6$ or its obvious truncation in the case of $n=5$, while
otherwise we can write 
\begin{equation}
  G_4 = \left [\begin{pmatrix}n_1&0\\0&n_2\\\end{pmatrix}\otimes
    \varepsilon \right]
  \oplus 0
\end{equation}
for $n=6$ or truncations thereof for $n\le 5$. Here, we introduced
$n_1,n_2\in\mathbb{R}$ and $\phi\in [0,\pi/2]$. Note that the
dimension of the matrices $G_1$ and $G_4$ being smaller than 6 does
not allow for non-trivial deformation components of the kind given in
\eqref{s3:fa1b2c3components}. However, in general we will find
components as in \eqref{s3:fa0b0c0inreals} that, as we will see,
correspond to the structure constants of the unbroken gauge Lie
algebra, as well as components as in \eqref{s3:Gatilde0} that in some
cases for $n=6$ are required to be non-trivial. 

We state the result for $n\le 5$ in terms of representatives of
$SO(n-1)$ orbits in Table~\ref{table_sols_n12345}. In anticipation of
phenomenological aspects to be discussed in Section \ref{s4:pheno}, we 
also list some physical properties for the consistent solutions. Note
that for $n\leq 4$ consistency is trivially given. Furthermore, in the
case of $n=5$ one cannot have $m_1,m_2\neq 0$ which excludes
solutions of the type \eqref{s3:g4phi}.   

\begin{table}[!htb]
\begin{center}
{\small\begin{tabular}{|c|l|l|l|}
\hline
  $n$ & non-trivial components & $\cN=2$ multiplets & unbroken gauge group \\
\hline
  1 & no $G_1,G_4$ & $2\times M_{2,1,0}$ & $U(1)^3$\\
\hline
  2 & $G_1=G_4=0$ & $3\times M_{2,1,0}$, & $U(1)^{3+1}$\\
    &                   & $1\times M_{2,1/2,BPS}$ of mass $|c|$ & \\
\hline
  3 & $G_1=G_4=0$ & $4\times M_{2,1,0}$, & $U(1)^{3+2}$\\
    &             & $2\times M_{2,1/2,BPS}$ of mass $|c|$ & \\
    & $m_1\neq 0 \vee n_1\neq 0$ & $2\times M_{2,1,0}$, & $U(1)^3$\\
    &                           & $2\times M_{2,1,BPS}$ of mass$^2$
    $(m_1^2+n_1^2)$,&\\
    &             & $1\times M_{2,1/2,.}$ of mass$^2$
    $m_1^2+(|c|-n_1)^2,$ & \\
    &             & $1\times M_{2,1/2,BPS}$ of mass$^2$
    $m_1^2+(|c|+n_1)^2$ & \\
\hline
 4 & $G_1=G_4=0, g_{\y 1\y 2\y 3}\neq 0$ & \ldots & $U(1)^3\times
 SU(2)$\\
   & $G_1=G_4=0, g_{\y 1\y 2\y 3}= 0$ & \ldots & $U(1)^{3+3}$\\
   & $m_1\in\mathbb{R}, n_1\neq 0, g_{\y 1\y 2\y 3}\in\mathbb{R}$
   &\ldots & $U(1)^{3+1}$\\
\hline
 5 & $G_1=G_4=0, g_{\y 1\y 2\y 3}\neq 0$ & \ldots & $U(1)^{3+1}\times
 SU(2)$\\
   & $G_1=G_4=0, g_{\y 1\y 2\y 3}=0$ & \ldots & $U(1)^{3+4}$\\
   & $G_1=0, n_1,n_2\neq 0$ & \ldots & $U(1)^3$\\
   & $m_1\neq 0 \vee n_1\neq 0, m_2,n_2=0$ & \ldots & $U(1)^{3+2}$\\
   & \hspace{0.5cm} and $g_{\y 1\y 2\y 3}\in\mathbb{R}$ & & \\
   & $m_1\neq 0, m_2=0, n_2\neq 0$ & \ldots & $U(1)^3$ \\
\hline
\end{tabular}}
\caption{Consistent electric gaugings with $\cN=2$ vacuum for
  $n\leq 5$. Explanations are given in Section \ref{s3:sols_n6}. We
  also always have the $\cN=2$ gravity multiplet $M_{2,2,0}$ and the
  $\cN=2$ BPS gravitino multiplet $M_{2,3/2,BPS}$ of mass $|c|$. For
  brevity for $n\ge 4$ we do not list the $\cN=2$ spectrum (the
  $\ldots$). Note that here for convenience we set $\cV_-=1$.}  
\label{table_sols_n12345}
\end{center}
\end{table}

The result for $n=6$ is given in terms of $SO(5)$ gauge
representatives in Table \ref{table_sols_n6}.\footnote{There exist
  also solutions that are obtained from the ones given in Table
  \ref{table_sols_n6} by a reflection\\ 
  $T=\begin{pmatrix}0&1\\1&0\\\end{pmatrix}\oplus\mathbb{1}_3\in 
  O(5)$.}   
We observe that consistent solutions may still have non-trivial
deformation spaces.  

\begin{table}[!htb]
\begin{center}
{\small\begin{tabular}{|c|c|c|l|}
\hline
  $G_1$ & $G_4$ & solutions: non-trivial $f_{\ta\tb\tc}\,$, etc. &
  unbr.\ g. group\\   
  & & & \vspace{-0.3cm}\\
  \hline

  $m_1,m_2=0$ & $n_1,n_2=0$ & $f_{\tilde{1}\tilde{2}\tilde{3}}=0$ &
  $U(1)^{3+5}$\\ 
  & & $f_{\tilde{1}\tilde{2}\tilde{3}}\neq 0$ & $U(1)^{3+2} \times
  SU(2)$\\
  \hline

  $m_1,m_2=0$ & $n_1\neq 0,\, n_2=0$ & 
  $f_{\tilde{1}\tilde{2}\tilde{3}}\in\mathbb{R}$ & 
  $U(1)^{3+3}$\\ 
  & & $f_{\tilde{3}\tilde{4}\tilde{5}}\neq 0$ & $U(1)^3\times SU(2)$\\
  \hline

  $m_1,m_2=0$ & $0\neq n_1^2\neq n_2^2 \neq 0$ & 
  $f_{\y 1\y 2\y 5}\in\mathbb{R}$
  & 
  $U(1)^{3+1}$\\ 
  \hline

  $m_1,m_2=0$ & $0\neq n_1^2= n_2^2 \neq 0$ & 
  $G_{\y 5}=\begin{pmatrix}
\frac{f_{\y 2\y 4\y 5}^2}{f_{\y 3\y 4\y5}} & 0\\ 
   0 & f_{\y 3\y 4\y 5}\\
   \end{pmatrix}\otimes \varepsilon + 
   \begin{pmatrix}
   0 & f_{\y 2\y 4\y 5}\\
   -f_{\y 2\y 4\y 5}&0\\
   \end{pmatrix}\otimes \mathbb{1}_2$ & 
  $U(1)^{3+1}$\\ 
  & & $f_{\y 1\y 2\y 5}\in\mathbb{R}$ & $U(1)^{3+1}$\\
  \hline

  $m_1\neq 0, m_2=0$ & $n_1\in\mathbb{R}, n_2=0$ & 
  $f_{\y 1\y 2\y 3}\in\mathbb{R}$ & $U(1)^{3+3}$\\
  & & $f_{\y 3\y 4\y 5}\neq 0$ & $U(1)^3\times SU(2)$\\
  \hline 

  $m_1\neq 0, m_2=0$ & $n_1\in\mathbb{R}, n_2\neq 0$ & 
  $f_{\y 1\y 2\y 5}\in\mathbb{R}$ & $U(1)^{3+1}$\\
  \hline

  $\scriptsize 0\neq m_1^2\neq m_2^2\neq 0$ & $n_1,n_2\in\mathbb{R}$ & 
  $G_{\y 5}=\begin{pmatrix}
   \frac{m_1m_2}{f_{\y 3\y 4\y5}} & 0\\ 
   0 & f_{\y 3\y 4\y 5}\\
   \end{pmatrix}\otimes \varepsilon$ & 
  $U(1)^{3+1}$\\ 
  \hline

  $m_1= m_2\neq 0$ & $n_1, n_2= 0$ & 
  $G_{\y 5}=\begin{pmatrix}
   f_{\y 1\y 2\y 5} & 0\\ 
   0 & f_{\y 3\y 4\y 5}\\
   \end{pmatrix}\otimes \varepsilon  + 
   \begin{pmatrix}
   0 & f_{\y 2\y 4\y 5}\\
   -f_{\y 2\y 4\y 5}&0\\
   \end{pmatrix}\otimes \mathbb{1}_2$ & 
  $U(1)^{3+1}$\\
  & & with $m_1^2=f_{\y1 \y2 \y5}\,f_{\y3 \y4 \y5} -f_{\y2
    \y4 \y5}^2$ &\\ 
  \hline

  $m_1= m_2\neq 0$ & $n_1\neq 0, n_2= 0$ & 
  $G_{\y 5}=\begin{pmatrix}
   f_{\y 1\y 2\y 5} & 0\\ 
   0 & f_{\y 3\y 4\y 5}\\
   \end{pmatrix}\otimes \varepsilon$ & 
  $U(1)^{3+1}$\\
  & & with $m_1^2=f_{\y1 \y2 \y5}\,f_{\y3 \y4 \y5}$ &\\ 
  \hline

  $m_1= m_2\neq 0$ & $0\neq n_1^2\neq n_2^2\neq 0$ & 
  $G_{\y 5}=\begin{pmatrix}
   f_{\y 1\y 2\y 5} & 0\\ 
   0 & f_{\y 3\y 4\y 5}\\
   \end{pmatrix}\otimes \varepsilon$ & 
  $U(1)^{3+1}$\\
  & & with $m_1^2=f_{\y1 \y2 \y5}\,f_{\y3 \y4 \y5}$ &\\ 
  \hline

  $m_1= m_2\neq 0$ & $0\neq n_1^2= n_2^2\neq 0$ & 
  $G_{\y 5}=\begin{pmatrix}
   f_{\y 1\y 2\y 5} & 0\\ 
   0 & f_{\y 3\y 4\y 5}\\
   \end{pmatrix}\otimes \varepsilon$ & 
  $U(1)^{3+1}$\\
  & with $\sin\phi=0$ & with $m_1^2=f_{\y1 \y2 \y5}\,f_{\y3 \y4 \y5}$ &\\ 
  \hline

  $m_1= m_2\neq 0$ & $0\neq n_1^2= n_2^2\neq 0$ & 
  $G_{\y 5}=\begin{pmatrix}
   f_{\y 3\y 4\y 5} & 0\\ 
   0 & f_{\y 3\y 4\y 5}\\
   \end{pmatrix}\otimes \varepsilon  + 
   \begin{pmatrix}
   0 & f_{\y 2\y 4\y 5}\\
   -f_{\y 2\y 4\y 5}&0\\
   \end{pmatrix}\otimes \mathbb{1}_2$ & 
  $U(1)^{3+1}$\\
  & with $\cos\phi=0$ & with $m_1^2=f_{\y3 \y4 \y5}^2 -f_{\y2
    \y4 \y5}^2$ &\\ 
  \hline
  $m_1= m_2\neq 0$ & $0\neq n_1^2= n_2^2\neq 0$ & 
  $\scriptsize G_{\y 5}=$ & 
  $U(1)^{3+1}$\\
  & $\sin\phi,\cos\phi\neq 0$ & $\scriptsize\begin{pmatrix}
   2\cot\phi f_{\y 2\y 4\y 5} + f_{\y3 \y4 \y5} & 0\\ 
   0 & f_{\y 3\y 4\y 5}\\
   \end{pmatrix}\otimes \varepsilon  + 
   \begin{pmatrix}
   0 & f_{\y 2\y 4\y 5}\\
   -f_{\y 2\y 4\y 5}&0\\
   \end{pmatrix}\otimes \mathbb{1}_2$ &\\ 
   & & with $m_1^2=f_{\y3 \y4 \y5}^2 -f_{\y2
    \y4 \y5}^2+2\cot\phi f_{\y2 \y 4\y5}\,f_{\y3 \y 4\y5}$ &\\
  \hline
\end{tabular}}
\caption{Consistent electric gaugings with $\cN=2$ vacuum for
  $n=6$. Explanations are given in Section \ref{s3:sols_n6}.} 
\label{table_sols_n6}
\end{center}
\end{table}

\subsubsection{Special solutions with $g_a=0$ and $G_1=0$ for
  arbitrary $n\in\mathbb{N}$} 
\label{s3:specsolforanyn}
A class of special solutions with $g_a=0$ for arbitrary $n$ is
obtained by setting $G_1=0$ which drastically simplifies the equations 
\eqref{s3:qc_bc14''} to \eqref{s3:qc_bcde''}. Similarly to the
discussion for general $G_1$ in Section
\ref{s3:solvingquadconstraints}, we can write $G_4$ as  
\begin{equation}
  G_4 = (D\otimes\varepsilon)\oplus 0 =
\begin{pmatrix}D\otimes \varepsilon & 0\\ 0&0\end{pmatrix} \,,
\end{equation}
where $D=diag(y_1,\ldots,y_1,y_2,\ldots,y_2,\ldots)$ is a diagonal
matrix with ordered positive eigenvalues $y_1>y_2>\ldots>0$. In doing
so, the full solution to equation \eqref{s3:qc_bcd4''} is analogous to
the one given in terms of the (a priori) non-trivial components in
\eqref{s3:fa0b0c0inreals}, \eqref{s3:Gatilde0}, and
\eqref{s3:fa1b2c3components}. For general such components, it is still
hard to solve the last equations \eqref{s3:qc_bcde''}. However, an
interesting class of solutions is obtained after setting all but
$f_{\ta_0  \tb_0 \tc_0}$ to zero since then \eqref{s3:qc_bcde''} is
just the Jacobi identity \eqref{s3:jacobiid} for the gauge Lie algebra
with structure constants $f_{\ta_0  \tb_0 \tc_0}\in\mathbb{R}$. As
stated above many non-trivial solutions to these equations are known,
each of which corresponds to a compact reductive group $G$. As we will
see in Section \ref{s4:unbrokengaugegroup} in those cases the unbroken
gauge group that leaves the vacuum invariant is
\begin{equation}
  U(1)^3\times G_{\cN=2}\, .
\end{equation}
Finally, anticipating the discussion of mass terms, we list the
$\cN=2$ spectrum for such solutions in Table \ref{table_sols_anyn}. 
\begin{table}[!htb]
\begin{center}
{\small\begin{tabular}{|c|c|l|}
\hline
  block in $G_4$ & mass & $\cN=2$ multiplets\\
  \hline
  $\mathbb{0}_{k}$ & 0 & $k\times M_{2,1,0}$\\
  & $|c|$ & $k \times M_{2,1/2,BPS}$\\
  \hline
  $y_i\otimes \varepsilon$ & $y_i$ & $2\times M_{2,1,BPS}$\\
  & $\big ||c|-y_i\big|$ & $1\times M_{2,1/2,.}$\\
  & $|c|+y_i$ & $1\times M_{2,1/2,BPS}$\\
  \hline
\end{tabular}}
\caption{$\cN=2$ multiplets in the matter sector for the solutions in
  Section \ref{s3:specsolforanyn}. In the gravity sector one has the
  $\cN=2$ gravity multiplet $M_{2,2,0}$, the $\cN=2$ BPS gravitino
  multiplet $M_{2,3/2,BPS}$ of mass $|c|$, and two more $\cN=2$ vector
  multiplets $M_{2,1,0}$. The consistency condition given in
  \eqref{s3:countdofBPSgravitini} is fulfilled with $n'_{\rm v}=0$,
  i.e.\ no non-BPS massive vector multiplets. Furthermore, note that
  for blocks with $y_i=|c|$ one obtains massless hypermultiplets. This
  is of interest because together with massless vector multiplets
  these give rise to a non-trivial geometry of the scalar manifold in
  the effective $\cN=2$ theory.}  
\label{table_sols_anyn}
\end{center}
\end{table}



\section{Aspects of the $\cN=2$ low-energy effective theory}
\label{s4:pheno}
In an $\cN=2$ vacuum of $\cN=4$ supergravity the low-energy effective
theory should be consistent with $\cN=2$ supersymmetry. In particular,
we will show that the various fields can be consistently embedded into
complete $\cN=2$ multiplets. We will then discuss the unbroken gauge
group and, finally, we will comment on the effective Lagrangian below
the scale of partial supersymmetry breaking. Bearing in mind that we
have not yet fully solved the quadratic constraint equations, we will
start generally but then restrict ourselves to the solutions with
$g_a=0$. 

\subsection{Mass terms in the gauged theory}
\label{s4:masses}
The fermionic mass terms of the theory emerge from the fermion
bilinears given in equations \eqref{32couplings} and
\eqref{s3:massspin12} after evaluating the $A$-matrices at the critical
point \eqref{s3:origin}. By construction, the gravitini mass matrix is
diagonal and its two non-zero eigenvalues are given by 
\eqref{s3:mgravitino}. 
Masses for vector bosons arise from the gauge-covariant derivative
acting on the scalar fields. At the same time, the mixed couplings of
vector bosons and scalar fields single out the pseudo-Goldstone fields
that provide the longitudinal polarization of massive vector
bosons. In the case of electric gaugings the
scalars in the gravity multiplet are neutral ($D_{\mu}M_{\alpha\beta}
= \partial_{\mu}M_{\alpha\beta}$) and thus the pseudo-Goldstone fields
can only arise from the scalars of the vector multiplets. Using
\eqref{s2:DMMN} together with all the information about the $f_{MNP}$ 
obtained in the previous section, the gauged kinetic term of those
scalars yields  
\begin{eqnarray}
  \label{s3:vector_masses}
  \frac{1}{16} (D_{\mu}M_{MN}) \,(D^{\mu}M^{MN}) & =
  & \frac{1}{16} (\partial_{\mu}M_{MN})\, (\partial^{\mu}M^{MN})
  \nonumber\\
  & & -\frac{g^2}{2} \sum_{a=1}^n (e_a^2 + f_a^2 + g_a^2)
  \sum_{m\in\{2,3,5,6\}}A_{\mu}^{ m} A^{\mu m}\nonumber\\
  & & -\frac{g^2}{2} \sum_{b,c=1}^n O_{bc}\, A_{\mu}^{b} A^{\mu c} +
  \ldots \,,
\end{eqnarray}
where we introduced a symmetric and positive semi-definite matrix
$(O_{ab}) \in\text{Mat}_{n,n}$ with components
\begin{equation}
  \label{s3:Obc}
  O_{bc}\equiv \sum_{a=1}^{n}\sum_{m=1}^6 f_{abm}\, f_{acm}.
\end{equation}
The $\ldots$ in \eqref{s3:vector_masses} denote couplings of vectors
and Goldstone bosons. Note that in \eqref{s3:vector_masses} the terms
mixing $A^{\mu m}$ and $A^{\mu b}$ are absent due to the quadratic
constraints $(b,m,n,p)$ for $m,n,p\in \{2,3,5,6\}$. 

Before reading off the masses of the vector
bosons one has to canonically normalize their kinetic terms in
\eqref{Ldef}. To this end, we redefine $A'^{\mu M} = \sqrt{\Im\tau} 
A^{\mu M}$, for a given background value $\tau$, which amounts to
scaling all mass terms in \eqref{s3:vector_masses} by a factor of
$1/\Im\tau$ as required by supersymmetry, cf.\ Section
\ref{s3:section_origin}. It is then apparent that only four 
gauge bosons ($A^{\mu 2}, A^{\mu 3}, A^{\mu 5}, A^{\mu 6}$) of the
gravity multiplet become heavy and, due to \eqref{s3:qc_2356},
\eqref{a:fa0_2356}, their masses are degenerate and equal to the
gravitino mass 
\eqref{s3:mgravitino}:    
\begin{equation}
  m^2_{A^2,A^3,A^5,A^6} = \cV_-^2\,\sum_{a=1}^n (e_a^2 + f_a^2 + g_a^2)
  = c^2\,\cV_-^2 =(m_{3/2})^2~.
\end{equation}
Thus, an $\cN=2$ vacuum with two non-BPS gravitino multiplets
would require at least four vector multiplets (i.e.\ $n\ge 4$), as in 
this case eight massive vector bosons are contained in the two
gravitino multiplets \eqref{massiveN2}. Eventually, the
symmetric mass matrix $(O_{ab})$ will be diagonalized by means of an
$SO(n)$ transformation and being positive semi-definite it will give
rise to well-defined mass terms. Note that for the solutions discussed
in Section \ref{s3:specsols} we always have $g_a = e_a=0$ and
$G_2 = G_3 = G_5 = G_6=0$ and the above expressions are much simpler.   

In order to analyze the potential \eqref{s2:scalarpot} in a
neighborhood of the origin of the scalar manifold, we employ the
following chart   
\begin{eqnarray}
\label{s3:scalarchart}
  \mathbb{R}^{6n} \supset U & \quad \rightarrow \quad & W \subset
  \nicefrac{SO(6,n)}{SO(6)\times SO(n)}, \nonumber\\
  \phi^{ma} & \mapsto &
  \exp{\left (\sum_{m,a}\phi^{ma}[t_{ma}] \right )}\equiv{\mathcal 
    V}(\phi^{ma})\equiv{\mathcal V} ~,
\end{eqnarray}
where ${[t_{ma}]_M}^N = {\delta_{[m}}^N \eta_{a]M}$ are the
non-compact generators of the coset space associated to the vector 
multiplets. We can then express the scalar kinetic term as 
\begin{equation}
  \frac{1}{16} \left (\partial_{\mu}M_{MN} \right)
  \left (\partial^{\mu}M^{MN} \right) =-\frac{1}{2} \left
  (\partial_{\mu}\phi^{ma} \right) \left(\partial^{\mu}\phi^{ma}
\right) + \mathcal{O}((\partial \phi)^2\phi^2)~.
\end{equation}
As this kinetic term is canonically normalized, we can identify the
coordinates $\phi^{ma}$ with the scalar degrees of
freedom. Geometrically, these can be interpreted as fluctuations in
$SO(6,n)/[SO(6)\times SO(n)]$ around the critical point
\eqref{s3:origin}. It turns out that in the case of electric gaugings
the two scalars of the gravity 
multiplet remain massless. Therefore, in an infinitesimal neighborhood
of the origin where higher-order interactions are negligible, the
scalar manifold of the gravity multiplet remains unaffected and thus
can be ignored in what follows. Up to cubic terms, one finds: 
\begin{align}
  \label{s4:potentialquadratic}
  \mathcal{L}_{pot} =
  -\ft {\cV_-^2} 2 \Big [& \sum_c (e_c^2 + f_c^2 + g_c^2) \sum_a \sum_{m\in
    \{2,3,5,6\}} (\phi^{ma})^2 + \sum_{b,c} O_{bc}\sum_{m=1}^6 \phi^{mb}\phi^{mc}
  \nonumber\\  
  & +\sum_{a,b}\sum_{l,k=1}^6 \left (\sum_c f_{abc}f_{lkc} +
    \sum_{m=1}^6 f_{abm} f_{lkm} + \sum_c f_{akc} f_{lbc} +
    \sum_{m=1}^6 f_{akm} f_{lbm} \right) \phi^{la}\phi^{kb}
  \Big]\nonumber\\ 
   &\hspace{-1.1cm} + \mathcal{O}(\phi^3)~.
\end{align}
Note that the absence of linear terms in \eqref{s4:potentialquadratic}
is a necessary condition for metastability. Furthermore, the fact that
the cosmological constant vanishes is due to the quadratic constraint
\eqref{a:fa0_2356}, as we have seen earlier.

Now that we know all mass terms we can check the super-Higgs mechanism
that is required by partial supersymmetry breaking. First, we will
consider the gravity/Goldstini sector, and secondly, we will discuss
the matter sector. As a result, we will also show that the vacuum
solutions are metastable, as required by the preserved $\cN=2$
supersymmetry. We will restrict ourselves to the case $g_a=0$, which
as we have seen in Section  \ref{s3:solvingquadconstraints} implies
$e_a=0$ and $G_2=G_3=G_5=G_6=0$.  For such solutions the potential
simplifies to  
\begin{align}
  \label{s4:potentialquadratic'}
  \mathcal{L}_{pot} =
  -\ft {\cV_-^2} 2 \Bigg [& c^2\hspace{-0.3cm} \sum_{m\in
    \{2,3,5,6\}} \phi^{m \ta}\phi^{m \ta} + \sum_{m\in
    \{2,3,5,6\}} O_{\ta\tb} \phi^{m\ta}\phi^{m\tb} + 4c\, f_{\ta\tb 4}
  \,(\phi^{2\ta}\phi^{3\tb}+\phi^{5\ta}\phi^{6\tb})
  \nonumber\\  
  & + f_{\tc\ta 4}\, f_{\tc\tb 4}\, \phi^{1\ta}\phi^{1\tb} + f_{\tc\ta 1}\,
  f_{\tc\tb 1}\, \phi^{4\ta}\phi^{4\tb} -2f_{1\ta\tc}\, f_{4\tb\tc}\,
  \phi^{4\ta}\phi^{1\tb} \Bigg]\nonumber\\ 
   &\hspace{-1.1cm} + \mathcal{O}(\phi^3),
\end{align}
where as before we denote the potentially non-trivial embedding tensor
components by $f_{\ta\tb m}$ for $SO(n-1)$ indices $\ta,\tb,$ etc.

\subsubsection{Gravity/Goldstini sector}
In the gauge where $f_a=c\, \delta_{a7}$ it is only the ``first'' $\cN=4$ 
vector multiplet that contributes to the gravity/Goldstini
sector. After canonically diagonalizing the kinetic terms of the
fermions by means of the field redefinition $  \chi'^i = \ft 1
{\sqrt{2}} \chi^i$ we find that the fermionic mass terms in this
sector read\footnote{From now on we will drop the overall scaling
  factor of $\cV_-^2$.} 
\begin{align}
\label{s4:gravity_goldstini_sector}
  c \big [&\psi_{\mu}^3 \,\epsilon\, \sigma^{\mu\nu} \psi_{\nu}^3 +
    \ft 1 2 \sqrt{2}\, \bar{\eta}^{(3)} \,\sigma^\mu\, \psi_\mu^3 \nonumber\\
    &+ \psi_{\mu}^4 \,\epsilon\, \sigma^{\mu\nu} \psi_{\nu}^4 +
    \ft 1 2 \sqrt{2}\, \bar{\eta}^{(4)} \,\sigma^\mu\, \psi_\mu^4
    \nonumber\\
    &-\sqrt{2}\, \chi'^3 (\lambda^{74})^* - \ft 1 2 (\lambda^{74})^*
    \epsilon \, (\lambda^{74})^*  \nonumber\\
    &+\sqrt{2}\, \chi'^4 (\lambda^{73})^* - \ft 1 2 (\lambda^{74})^*
    \epsilon \, (\lambda^{74})^*
    \big ] + \text{h.c.}~,
\end{align}
where the would-be Goldstino combinations eaten by the massive
gravitini are 
\begin{align}
  \bar{\eta}^{(3)} &= \bar{\eta}^{(3)\dot{A}} = \epsilon^{\dot{A}\dot{B}} 
  \chi'^3_{\dot{B}} + \sqrt{2}
  (\lambda^{74A})^* ,\quad \dot{A}, \dot{B}=1,2\nonumber\\
  \bar{\eta}^{(4)} &= \bar{\eta}^{(4)\dot{A}} = \epsilon^{\dot{A}\dot{B}} 
  \chi'^4_{\dot{B}} - \sqrt{2} (\lambda^{73A})^*\, .
\end{align}
The mass terms for the spin-1/2 fermions 
$\chi'_1,\chi'_2,\lambda_1^7, \lambda_2^7$ are absent in
\eqref{s4:gravity_goldstini_sector} and thus these fermions are
massless. As in \cite{Wess:1992}, mixed terms involving both a
gravitino and a spin-1/2 fermion can be removed by means of the
following gravitino shifts 
\begin{align}
\label{s4:gravitinoshifts}
  \tilde{\psi}^3_\mu  &= \tilde{\psi}^{3A}_\mu
   = \psi^{3A}_\mu + \ft {\sqrt{2}} 6
  \bar{\sigma}_\mu^{A\dot{B}} \bar{\eta}^{(3)}_{\dot{B}} +
  \mathcal{O}(\partial\eta^{(3)}), \nonumber\\
  \tilde{\psi}^4_\mu  &= \tilde{\psi}^{4A}_\mu
   = \psi^{4A}_\mu + \ft {\sqrt{2}} 6 
  \bar{\sigma}_\mu^{A\dot{B}} \bar{\eta}^{(4)}_{\dot{B}} +
  \mathcal{O}(\partial\eta ^{(4)})~,
\end{align}
yielding additional contributions to the mass matrix
of the spin-1/2 fermions. As a result, their mass terms 
read 
\begin{align}
\label{s4:gravgold_fermionmasses}
  \ft c 2 \Bigg [
  &\left ((\lambda^{74A})^*, \epsilon^{\dot{A}\dot{B}}
    \chi'^3_{\dot{B}} \right) \epsilon_{\dot{A}\dot{C}}\,
  M^{(-)} \begin{pmatrix} (\lambda^{74C})^* \\
    \epsilon^{\dot{C}\dot{D}} 
      \chi'^3_{\dot{D}} \end{pmatrix} \nonumber\\
  &+\left ((\lambda^{73A})^*, \epsilon^{\dot{A}\dot{B}}
    \chi'^4_{\dot{B}} \right) \epsilon_{\dot{A}\dot{C}}\,
  M^{(+)} \begin{pmatrix} (\lambda^{73C})^* \\
    \epsilon^{\dot{C}\dot{D}} 
      \chi'^4_{\dot{D}} \end{pmatrix}
  \Bigg ] + \text{h.c.}~,
\end{align}
where the mass matrices $M^{(\pm)}$ are given
by
\begin{equation}
  M^{(\pm)} = \ft 1 3 \begin{pmatrix}1 & \pm\sqrt{2}\\ \pm\sqrt{2} &  
    2 \end{pmatrix}\,,
\end{equation}
and both have eigenvalues $0$ and $1$. In fact, the two zero
eigenvalues give rise to two massless helicity-1/2 fermions to be
identified as the would-be Goldstini associated to the broken
supersymmetry. On the other hand, one finds two spin-1/2 fermions of
mass $|c|$ that together with the two massive gravitini fit into the
$\cN=2$ BPS gravitino multiplet.  

As to the bosons in this sector, \eqref{s3:vector_masses} shows that
the only massive vectors are $A_{\mu}^{2}$, $A_{\mu}^{3}$,
$A_{\mu}^{5}$,$A_{\mu}^{6}$ while the massless ones are $A_{\mu}^{1}$,
$A_{\mu}^{4}$, $A_{\mu}^{7}$. The four massive vectors belong to
the $\cN=2$ BPS gravitino multiplet as we shall show in
\ref{s4:bps}. Finally, all eight scalars of this sector are massless,
as can be seen from \eqref{s4:potentialquadratic'}, four of which are
to be interpreted as the would-be Goldstone bosons. In an
infinitesimal neighborhood around the critical point these
fluctuations are described by $\phi^{27}, \phi^{37}, \phi^{57},
\phi^{67}$.  

To conclude, we have shown that the fields in the massive BPS
gravitino multiplet all have the same mass, consistent with $\cN=2$
supersymmetry. Furthermore, in the gravity/Goldstini sector the $\cN=2$
gravity multiplet and the massive $\cN=2$ BPS gravitino multiplet are
accompanied by two massless $\cN=2$ vector multiplets, which are the
remnants of the minimal $\cN=4$ multiplets required for 
spontaneous partial supersymmetry breaking to $\cN=2$. 

\begin{table}[htb]
\begin{center}
{\small\begin{tabular}{|l|l|}
\hline
  $\cN=2$ multiplets & mass squared \\
\hline
  $M_{2,2,0}$ gravity & 0\\
  $M_{2,3/2,BPS}$ BPS gravitino & $c^2$\\
  $2\times M_{2,1,0}$ vector & 0 \\
\hline
\end{tabular}}
\caption{Gravity/Goldstini sector of the $\cN=2$ spectrum.} 
\label{table_N2spectrum_grav}
\end{center}
\end{table}

\subsubsection{Matter sector}
\label{s4:mattersector}
The mass squared matrix for vector bosons $A^{\mu\ta}$ defined in
\eqref{s3:Obc} now reads 
\begin{equation}
  \label{s4:Omatrix}
  O = - G_1^2 - G_4^2~,
\end{equation}
which according to the discussion in Section
\ref{s3:solvingquadconstraints} is already diagonal. For each block in
$G_1$ and $G_4$ with degenerate eigenvalues
\begin{equation}
  \label{s4:G1G4degenerateblock}
  (G_1^{(ij)})^2 = -x^2\, \mathbb{1}_l, \qquad (G_4^{(ij)})^2 = -y^2\, \mathbb{1}_l\,,
\end{equation}
where $x,y\in\mathbb{R}$, one finds $l$ vectors of mass squared
$x^2+y^2$.

Using the explicit expression given for the A-matrices in \eqref{s3:Aabij} the mass terms \eqref{s3:massspin12} for the fermions $\lambda^{1\ta},
\lambda^{2\ta}$ are given by
\begin{equation}
  \ft 1 2 \left ( (\lambda^{\ta 1})^*, (\lambda^{\ta 2})^* \right)
  \epsilon \,  U
  \begin{pmatrix} (\lambda^{\tb 1})^* \\ \lambda^{\tb
      2})^* \end{pmatrix} + \text{h.c.}\,,
\end{equation}
with
\begin{equation}
  U = \begin{pmatrix} 0 & i G_1+G_4 \\ -iG_1-G_4 & 0 \end{pmatrix}\, .
\end{equation}
Thus, their mass squared matrix
\begin{equation}
  UU^\dag = \begin{pmatrix} O & 0 \\ 0 & O \end{pmatrix}
  = \begin{pmatrix}  - G_1^2 - G_4^2 & 0 \\ 0 &  - G_1^2 -
    G_4^2 \end{pmatrix} \,,
\end{equation}
is also diagonal by virtue of the quadratic constraints
\eqref{s3:commG1G4}. Similarly, the mass terms for $\lambda^{3\ta},
\lambda^{4\ta}$ in \eqref{s3:massspin12} are given by
\begin{equation}
  \ft 1 2 \left ( (\lambda^{\ta 3})^*, (\lambda^{\ta 4})^* \right)
  \epsilon \, V
  \begin{pmatrix} (\lambda^{\tb 3})^* \\ \lambda^{\tb
      4})^* \end{pmatrix} + \text{h.c.}\,,
\end{equation}
where
\begin{equation}
  V=\begin{pmatrix} -c & -i G_1+G_4 \\ iG_1-G_4 & -c \end{pmatrix}\, .
\end{equation}
The corresponding mass squared matrix reads
\begin{equation}
  VV^\dag = \begin{pmatrix} c^2 - G_1^2 - G_4^2 & -2c\, G_4 \\ 2c\,
    G_4 & c^2 - G_1^2 - G_4^2 \end{pmatrix}.
\end{equation}
As in \eqref{s4:G1G4degenerateblock}, it can be shown that for each
block in $G_1$ and $G_4$ the eigenvalues are
\begin{equation}
\label{s4:xcymass}
  x^2 + (|c| \pm |y|)^2\,,
\end{equation}
with degeneracy $l$ each.

We can read off the mass terms for the scalar fields
$\phi^{1\ta},\phi^{4\ta}$ directly from
\eqref{s4:potentialquadratic'}, 
\begin{equation}
\label{s4:scalars_phi14}
  -\ft 1 2 \left (\phi^{1\ta}, \phi^{4\ta} \right)
  Z
  \begin{pmatrix} \phi^{1\tb} \\ \phi^{4\tb} \end{pmatrix}\,,
\end{equation}
where
\begin{equation}
\label{s4:Z}
  Z = \begin{pmatrix} -G_4^2 & G_4 G_1 \\ G_1 G_4 & -
    G_1^2 \end{pmatrix} \, .
\end{equation}
Obviously, for the trivial block in $G_1$ and $G_4$ with $x,y=0$ one
obtains $(2l)$ massless scalars. On the other hand, for each block
with $x\neq 0$ or $y\neq 0$, $l=2l'$ has to be even and the
eigenvalues of $Z$ turn out to have ($2l'$)-fold degenerate
eigenvalues 
\begin{equation}
  0~, \qquad (x^2+y^2).
\end{equation} 
The zero eigenvalue set precisely corresponds to the would-be Goldstone
modes eaten by the ($2l'$) vector bosons that become
massive. Finally, the mass terms for the remaining scalars
$\phi^{2\ta}, \phi^{3\ta}, \phi^{5\ta}, \phi^{6\ta}$ turn out to be
\begin{align}
\label{s4:phi2356}
  &-\ft 1 2 \left (\phi^{3\ta}, \phi^{2\ta} \right)
  \begin{pmatrix} c^2 - G_1^2 -G_4^2 & -2c\, G_4 \\ 2c\,
      G_4 & c^2 - G_1^2 - G_4^2 \end{pmatrix}
  \begin{pmatrix} \phi^{3\tb} \\ \phi^{2\tb} \end{pmatrix}\nonumber\,,\\
  &-\ft 1 2 \left (\phi^{6\ta}, \phi^{5\ta} \right)
  \begin{pmatrix} c^2 - G_1^2 -G_4^2 & -2c\, G_4 \\ 2c\,
      G_4 & c^2 - G_1^2 - G_4^2 \end{pmatrix}
  \begin{pmatrix} \phi^{6\tb} \\ \phi^{5\tb} \end{pmatrix}~,
\end{align}
where the mass squared matrices are precisely $VV^\dag$. As a result,
one has $(2l)$ scalars for each mass in
\eqref{s4:xcymass}. It is then clear that all masses-squared are positive and therefore
metastability is guaranteed, as required for a supersymmetric theory
with Minkowski background. Furthermore, one finds that all degrees of
freedom in the matter sector fit into complete $\cN=2$
supermultiplets. The resulting $\cN=2$ spectrum is summarized in Table 
\ref{table_N2spectrum}. Note that blocks in $G_1$ and $G_4$ with $x=0$
and $|y|=|c|$ give rise to massless $\cN=2$ hypermultiplets.

\begin{table}[htb]
\begin{center}
{\small\begin{tabular}{|l|l|l|}
\hline
  block & $\cN=2$ multiplets & mass squared \\
\hline
  $G_1^{(ij)} = G_4^{(ij)} = 0 \cdot \mathbb{1}_l$ & $(l)\times
  M_{2,1,0}$ massless vector & 0\\
  & $(l)\times
  M_{2,1/2,BPS}$ BPS hyper & $c^2$\\
\hline
  $(G_1^{(ij)})^2=-x^2\,\mathbb{1}_{2l'},$&
  $(2l')\times M_{2,1,BPS}$ BPS vector &
  $(x^2+y^2)$ \\
  $(G_4^{(ij)})^2=-y^2\,\mathbb{1}_{2l'}$ & $(l') \times
  M_{2,1/2,BPS}$ BPS 
  hyper & $x^2+(|c|+|y|)^2$ \\
  with $x\neq 0$ or $y\neq 0$ & $(l') \times M_{2,1/2,.}$ (BPS)
  hyper & $x^2+(|c|-|y|)^2$ \\
\hline
\end{tabular}}
\caption{Matter sector of the $\cN=2$ spectrum. The matrices $G_1,
  G_4\in\text{Mat}_{n-1,n-1}$ are simultaneously block-diagonal with
  non-trivial blocks of the type given in Table \eqref{table_G1G4} or
  zero blocks.}  
\label{table_N2spectrum}
\end{center}
\end{table}

\subsubsection{BPS multiplets}
\label{s4:bps}
So far in the discussion of mass terms we have only shown that all
fields fit into complete $\cN=2$ multiplets. In particular, according
to our assignments in Tables \ref{table_N2spectrum_grav} and
\ref{table_N2spectrum} all massive fields lie in BPS
representations. In the generic case where the masses of the various
$\cN=2$ superfields are all different, the above assignments are
obviously correct. However, in the case of mass degeneracies between
various short $\cN=2$ superfields one should exclude the case where
short multiplets combine in order to form long multiplets. In fact, in
what follows we will show that in the case of $g_a=0$ all massive
fields have to be in BPS representations and that no long $\cN=2$
multiplet can occur in this super-Higgs mechanism. To this end we will
study the crucial parts of the supersymmetry transformations of the
bosonic fields that we take from
\cite{Bergshoeff:1985ms}.\footnote{While our proof is somewhat
  indirect, it does not require the supersymmetry transformations of
  the fermions which are not fully given in \cite{Bergshoeff:1985ms}.}
It suffices to analyze the supersymmetry transformations of the
massive bosons.

We first consider the massive vectors $A_\mu^2,A_\mu^3,
A_\mu^5,A_\mu^6$ in the gravity/Goldstini sector. Evaluating their
supersymmetry transformations at the origin \eqref{s3:origin} of
$SO(6,n)$ one finds
\begin{equation}
  \delta_\epsilon A_\mu^m \sim [G_m]_{ij} (\epsilon^i \epsilon\,
  \psi^j_\mu + \epsilon^i\epsilon\bar{\sigma}_\mu \chi^j) +
  \text{h.c.} 
\end{equation}
for $m=2,3,5,6$. Moreover, as in \eqref{sec2:var_ferm}, $\epsilon^i =
q^i \eta$ contains the $SU(4)$ vector $q^i$ and $[G_m]_{ij}$ denote
the 't Hooft matrices given in \eqref{thooft}. In our gauge, cf.\
\eqref{s2:N=2A1}, the unbroken supersymmetry directions are given by 
linear combinations of $q^1$ and $q^2$ (or $\epsilon^1$ and
$\epsilon^2$). As a result, for $m=2,3,5,6$ the massive vectors
$A_\mu^m$ transform into the fermions $\psi_\mu^3, \psi_\mu^4,
\chi^3,\chi^4$. While massive scalars are not present in the
gravity/Goldstini sector, we will now inspect the transformations of
the four Goldstone bosons that provide the longitudinal polarization
of the massive vector bosons. In an infinitesimal neighborhood of the
origin these fluctuations are described by the scalars $\phi^{27},
\phi^{37}, \phi^{57}, \phi^{67}$. Using the explicit chart
\eqref{s3:scalarchart} of $SO(6,n)$ one finds    
\begin{equation}
  \delta_\epsilon {\cV_m}^{a} = \delta_\epsilon\phi^{ma} +
  \mathcal{O}(\phi\, \delta\phi)\,,
\end{equation}
which when evaluated at the origin can again be expressed in terms of
the 't Hooft matrices as 
\begin{equation}
\label{s4:susyphi}
  \delta_\epsilon {\phi}^{ma} \sim [G_m]_{ij}\, \epsilon^i
  \epsilon\, \lambda^{aj} + \text{h.c.}\, .
\end{equation}
In particular, we find that the Goldstone bosons $\phi^{27},
\phi^{37}, \phi^{57}, \phi^{67}$ transform under $\cN=2$ into fermions
$\lambda^{73}, \lambda^{74}$. As a result, the massive bosons of the
gravity/Goldstini sector transform into the massive fermions of the
same sector. Note that the gravitino shifts in
\eqref{s4:gravitinoshifts} also only involves the aforementioned
fermions. 

Next, we will analyze the supersymmetry transformations of the bosonic
fields in the matter sector. The supersymmetry transformations of the
massive vectors $A_\mu^{\hat{a}}$ evaluated at the origin are given
by\footnote{As in Section \ref{s4:unbrokengaugegroup} indices
  $\hat{a},\hat{b},\ldots$ denote $SO(n-1)$ indices $\ta,\tb,\ldots$
  associated to massive vector bosons, i.e.\ to non-trivial blocks in 
  either $G_1$ or $G_4$.}
\begin{equation}
  \delta_\epsilon A_\mu^{\hat{a}}  \sim \epsilon^i \epsilon
  \bar{\sigma}_\mu \epsilon\, (\lambda^{\hat{a}i})^* + \text{h.c.}\,
  . 
\end{equation} 
As a consequence, restricting the transformations to $\cN=2$ one finds
that each massive vector boson $A_\mu^{\hat{a}}$ rotates into the
gaugini $\lambda^{\hat{a}1}$ and $\lambda^{\hat{a}2}$ but not into 
$\lambda^{\hat{a}3}$ and $\lambda^{\hat{a}4}$. Furthermore, as we
discussed below \eqref{s4:scalars_phi14}, the associated Goldstone
bosons are accompanied by massive scalars. Infinitesimally, all of
them are described by linear combinations of the scalar fields 
$\phi^{1\hat{a}}$ and $\phi^{4\hat{a}}$. Their transformations can be
read off from \eqref{s4:susyphi}. Owing to the fact that $[G_m]_{ij}$
for $m=1$ or $m=4$ is block-diagonal, one finds that under $\cN=2$
supersymmetry transformations the scalars $\phi^{1\hat{a}}$ and
$\phi^{4\hat{a}}$ only rotate into fermions $\lambda^{\hat{a}1}$ and  
$\lambda^{\hat{a}2}$. In particular, this also shows that neither the
would-be Goldstone combinations nor the massive scalars in
\eqref{s4:scalars_phi14} transform into $\lambda^{\hat{a}3}$ and
$\lambda^{\hat{a}4}$. Furthermore, it is worth mentioning that neither
$A_\mu^{\hat{a}}$ nor the massive scalars in \eqref{s4:scalars_phi14}
transform into the spin-1/2 fermions in the gravity/Goldstini sector
given in \eqref{s4:gravgold_fermionmasses}, let alone into the massive
gravitini. Finally, the only remaining potentially massive bosons are
the scalars $\phi^{2\hat{a}}, \phi^{3\hat{a}}, \phi^{5\hat{a}},
\phi^{6\hat{a}}$ in \eqref{s4:phi2356}. As can again be seen from
\eqref{s4:susyphi}, they only transform  into fermions
$\lambda^{\hat{a}3}, \lambda^{\hat{a}4}$ and never into
$\lambda^{\hat{a}1}, \lambda^{\hat{a}2}$, let alone into fermions of
the gravity/Goldstini sector. 

We can now conclude that all massive $\cN=2$ supermultiplets have to
be BPS multiplets. The argument goes as follows: We found that the
massive fields in the gravity/Goldstini sector and the massive fields
in the matter sector are not related by supersymmetry transformations
acting on the bosonic fields. This implies that the massive fields in
the gravity/Goldstini sector have to lie in a BPS gravitino multiplet
as massive long gravitino multiplets can never be decomposed into two
non-trivial sets of bosons and fermions such that within each set the
bosons only mix into the fermions, respectively. This follows from the
construction of supermultiplets as representations of the Clifford
algebra. Furthermore, by the same token, the remaining massive vector
bosons have to be in $\cN=2$ BPS vector multiplets.

\subsection{Unbroken gauge group}
\label{s4:unbrokengaugegroup}
We shall now investigate the unbroken gauge group at the $\cN=2$
critical point, i.e.\ the group that leaves the scalar vacuum 
configuration for consistent electric gaugings with $g_a=0$
invariant. First, we note that the critical point in $SL(2)/SO(2)$ is
not affected by gauge transformations. However, on the scalar matter
fields a generic gauge transformations parametrized by a gauge
parameter $\theta^{P}$ acts as   
\begin{equation}
  M_{MN} \rightarrow M_{MN} + 2\, \theta^{P} {f_{P(M}}^Q M_{N)Q}\,,
\end{equation} 
and, in particular, the coset representative of the origin of $SO(6,n)/[SO(6)\times
SO(n)]$ transforms as
\begin{equation}
  \label{s3:gaugetrafoorigin}
  \mathbb{1}_{MN} \rightarrow \mathbb{1}_{MN} + 2 \, \theta^{P}
  ({f_{PM}}^N + {f_{PN}}^M)\, .
\end{equation} 
In demanding invariance of the origin under
\eqref{s3:gaugetrafoorigin}, the gauge parameters are restricted to
the ones with $\theta^m=0$ for $m=2,3,5,6$, and $\theta^{\ta}=0$
for each massive vector boson $A_{\mu}^{\ta}$, the latter of which
requires a non-zero block in $G_1$ or $G_4$. Gauge transformations of
vector fields read \cite{deWit:2005ub,Weidner:2006rp}
\begin{equation}
  \label{s3:gaugetrafovectors}
  \delta A_{\mu}^{M} = \partial_{\mu} \theta^{M} +
  {X_{PQ}}^{M} A_{\mu}^{P} \,\theta^{Q}\,,
\end{equation}
where one has
\begin{equation}
  {X_{MN}}^P = -{f_{MN}}^P.
\end{equation}
Using our knowledge of certain embedding tensor components in the case
of $g_a=0$ one can compute the gauge transformation for the massless
vector bosons, which in this section we will denote as
$A^{\mu\bar{a}}$ so as to distinguish them from massive vectors
$A^{\mu\hat{a}}$.  While we dropped the $\,\tilde{}\,$ above indices,
$\bar{a}$ and $\hat{a}$ are still understood as $SO(n-1)$ indices. One
finds 
\begin{align}
  \label{s4:gaugetrafovectorsdetail}
  \delta A_{\mu}^{1} & = \partial_{\mu} \theta^{1}\,, \nonumber\\
  \delta A_{\mu}^{4} & = \partial_{\mu} \theta^{4}\,,
  \nonumber\\
  \delta A_{\mu}^{7} & = \partial_{\mu} \theta^{7}\,,
  \nonumber\\
  \delta A_{\mu}^{\bar{a}} & = \partial_{\mu} \theta^{\bar{a}}
  -\, f_{\bar{a}\bar{b}\bar{c}} \, A_{\mu}^{\bar{b}}
  \,\theta^{\bar{c}} \, .
\end{align}
Note that in the last line of \eqref{s4:gaugetrafovectorsdetail} we
made use of  $f_{\bar{a}\bar{b}\hat{c}}=0$, which we learned from
the quadratic constraints $(\tb,\bar{c},\bar{d},1)$ and
$(\tb,\bar{c},\bar{d},4)$. The transformations \eqref{s4:gaugetrafovectorsdetail} imply that we can interpret the 
three fields $A_{\mu}^{1}, A_{\mu}^{4}, A_{\mu}^{7}$ as the
vector bosons of a gauge group $U(1)^3$. On the other hand, the
embedding tensor components  $f_{\bar{a}\bar{b}\bar{c}}$ amount
to the structure constants of the gauge Lie algebra associated to
the massless vector bosons $A^{\mu \bar{a}}$. In fact, as already
pointed out in the simple case of \eqref{s3:jacobiid}, the quadratic
constraints for $(\bar{b},\bar{c},\bar{d},\bar{e})$ are simply
the Jacobi identity 
\begin{equation}
  \label{s4:jacobiid'}
  f_{\bar{a}\bar{b}\bar{c}}\,f_{\bar{d}\bar{e}\bar{a}} +
  f_{\bar{a}\bar{b}\bar{e}}\, f_{\bar{c}\bar{d}\bar{a}} -
  f_{\bar{a}\bar{b}\bar{d}}\, f_{\bar{a}\bar{c}\bar{e}} =
  0 ~,
\end{equation}
that gives rise to a gauge Lie group $G_{\cN=2}$. Its dimension equals the
number of  massless vector bosons $(\le n-1)$. If $n$ is sufficiently
large, any compact reductive Lie group can be chosen in order to
satisfy \eqref{s4:jacobiid'}. As a result, the
full unbroken gauge symmetry is    
\begin{equation}
  U(1)^3\times G_{\cN=2}.
\end{equation}
On the other hand, it is important to note that there is an additional
set of constraints on the components $f_{\bar{a}\bar{b}\bar{c}}$
coming from the quadratic equations for $(\bar{b},
\bar{c},\hat{d},\hat{e})$:
\begin{equation}
  \label{s4:addcon}
  f_{\bar{a} \bar{b} \bar{c}} f_{\bar{a}\hat{d}\hat{e}}
  + f_{\hat{a} \bar{b} \hat{e}}
  f_{\bar{c}\hat{d}\hat{a}}
  - f_{\hat{a} \bar{b} \hat{d}}
  f_{\hat{a}\bar{c}\hat{e}} = 0~.
\end{equation}
As we have seen in Section \ref{s3:specsols}, it is not always possible
to set all $f_{\bar{a}\hat{b}\hat{c}}$ (i.e.\ the components given in
\eqref{s3:Gatilde0}) to zero such that \eqref{s4:addcon} is
trivially satisfied. However, we have already shown in Section
\ref{s3:specsolforanyn} that consistent examples exist for any given
compact reductive Lie group $G_{\cN=2}$. 

\subsection{Scalar manifold in the effective theory}
\label{seceffact}
Below the scale  of supersymmetry breaking $m_{3/2}$ we may integrate
out heavy particles and, in doing so, arrive at an $\cN=2$
supersymmetric effective action. We are particularly interested in the
geometry of the scalar manifold of this effective action.
As before, we will consider the case of electric gaugings with
$g_a=0$. In the limit where momenta $p\ll m_{3/2}$ can be neglected,
the equations of motion for the massive vectors are purely
algebraic and can be solved for the massive vector bosons since their 
mass terms are automatically diagonal, as we discussed in Section
\ref{s4:mattersector}. One finds 
\begin{align}
  \label{s4:vectoreom}
  A_{\mu}^{n} &= -\ft 1 {2c^2} \sum_{m\in \{2,3,5,6\}} \left
    (\partial_{\mu} M_{m7} \right) f_{7nm}\, , \nonumber\\
  A_{\mu}^{\hat{b}} &= -\ft 1 {2{m_{\left (\hat{b} \right)}^2}}
\sum_{m\in \{1,4\}} \left (\partial_\mu M_{m\ta} \right)
f_{\ta  \hat{b}m}
\end{align}
for each $n\in \{2,3,5,6\}$ and massive vectors with index
$\hat{b}$.  When inserted back into the Lagrangian and using our
knowledge about certain embedding tensor components, the scalar 
kinetic term yields\footnote{Repeated indices are summed over their
  full index range unless otherwise specified by explicit summation
  symbols.}   
\begin{align}
  \label{s4:leff}
  \cal{L}_{\text{eff}} = &\ft {1}{16} \Bigg
  [2\sum_{m\in\{2,3,5,6\}}\left 
    (\partial_{\mu}M_{m\ta} \right)\left (\partial^{\mu}M^{m\ta}
  \right) + 2\sum_{m\in \{1,4\}} \left (\partial_{\mu}M_{m7}
  \right)\left (\partial^{\mu}M^{m7} \right)\nonumber\\
&\quad +  2\sum_{m\in \{1,4\}}\left (\partial_{\mu}M_{m\ta}
  \right)\left (\partial^{\mu}M^{m\ta} \right)+ \sum_{\hat{b}}
  ({m_{\left (\hat{b} \right)}})^{-2} \sum_{m,n\in \{1,4\}} (\partial_\mu 
M_{m\hat{a}}) (\partial^\mu M_{n\hat{c}}) \, f_{\hat{a}\hat{b} m}
f_{\hat{c}\hat{b} n}  \nonumber\\
&\quad+\left (\partial_{\mu}M_{mn} \right)\left
    (\partial^{\mu}M^{mn} \right) + \left (\partial_{\mu}M_{ab}
  \right)\left (\partial^{\mu}M^{ab} \right)\Bigg]. 
\end{align}
Using the chart \eqref{s3:scalarchart} one finds
\begin{align}
\label{s4:scalarkineticseffth}
& -\ft {1} {2} \sum_{m\in\{2,3,5,6\}} \left(\partial_{\mu}
    \phi^{m\ta} \right) \left(\partial^{\mu} \phi^{m\ta} \right)
   -\ft {1} {2} \sum_{m\in\{1,4\}} \left(\partial_{\mu}
    \phi^{m7} \right) \left(\partial^{\mu} \phi^{m7} \right)
  \nonumber\\
& -\ft {1} {2} \sum_{m\in\{1,4\}} \left(\partial_{\mu}
    \phi^{m\bar{a}} \right) \left(\partial^{\mu} \phi^{m\bar{a}}
  \right) 
  \nonumber\\
&
  -\ft {1}{2} \left (\partial_\mu \phi^{1\hat{a}},\, \partial_\mu
    \phi^{4\hat{a}} \right)
  \left ((O^{\text{(massive)}})^{-2}Z^{(\text{massive})} \right)_{\hat{a}\hat{b}}
    \begin{pmatrix}\partial^\mu \phi^{1\hat{b}} \\ \partial^\mu
      \phi^{4\hat{b}}\end{pmatrix}+
    \mathcal{O}\left((\partial\phi)^2\phi^2\right) \,,
\end{align}
where $O^{\text{(massive)}}$ is the truncation of \eqref{s4:Omatrix} to
an invertible matrix obtained after deleting all its zero rows and
columns, and similarly, $Z^{(\text{massive})}$ is the analogous
truncation of the mass matrix $Z$ defined in \eqref{s4:Z}. Note that
kinetic terms for the Goldstone modes $\phi^{m7}$ for $m=2,3,5,6$ are
absent in \eqref{s4:scalarkineticseffth} as these scalars have been
eaten by the massive vector bosons $A^{\mu m}$ for
$m=2,3,5,6$. Moreover, the same diagonalization scheme of Section 
\ref{s4:mattersector} also diagonalizes the kinetic terms of the
scalars $\phi^{1\hat{a}}$ and $\phi^{4\hat{a}}$ associated to
massive vectors with indices $\hat{a}$. As before, the zero
eigenvalues of $Z^{(\text{massive})}$ ensure that the kinetic terms of
the Goldstone modes in the matter sector vanish (again the Goldstone
modes are eaten by the vector bosons $A^{\mu\hat{a}}$ that acquire
mass). On the other hand, its nonzero eigenvalues are such that the
remaining kinetic terms are canonically normalized, which justifies the
mass assignment in Section \ref{s4:mattersector}.

Let us now summarize the dynamical degrees of freedom in an
infinitesimal neighborhood of the origin. The scalars $\phi^{m\ta}$
for $m=2,3,5,6$ lie in light (with respect to $m_{3/2}$) $\cN=2$ (BPS)
hypermultiplets, while $\phi^{17}$ and $\phi^{47}$ and the two scalars
of $SL(2)/SO(2)$ lie in the two massless $\cN=2$ multiplets that
descend from the gravity/Goldstini sector. The scalars
$\phi^{1\bar{a}}, \phi^{4\bar{a}}$ form $\cN=2$ massless vector
multiplets, while the non-Goldstone modes of the $\phi^{1\hat{a}},
\phi^{4\hat{a}}$ belong to $\cN=2$ BPS vector multiplets. Note,
however, that in the effective theory below the scale of partial
supersymmetry breaking $m_{3/2}$, all scalars (and their supersymmetry
partners) with masses larger than $m_{3/2}$ should also be integrated
out.

As the scalars of $SL(2)/SO(2)$, described by $\tau$, are moduli that
lie in a massless $\cN=2$ vector multiplet, the
$SL(2)/SO(2)$ factor of the $\cN=4$ scalar manifold descends without
change to the scalar field space of the massless
$\cN=2$ vector multiplets in the low-energy theory. If the
number of these vector multiplets is $(k+1)$, we conjecture
that the vector multiplet field space of the $\cN=2$ low-energy
theory is the following product of coset spaces,  
\begin{equation}
\label{s4:mfsp}
  \nicefrac{SL(2)}{SO(2)} \times \nicefrac{SO(2,k)}{SO(2)\times
    SO(k)\, ,}
\end{equation}
which is known to be the only series of special K\"ahler product
manifolds including a factor of $SL(2)/SO(2)$
\cite{Ferrara:1989py}. Moreover, since we only analyze the potential
to quadratic order, we can only infer that the moduli space is a
submanifold of \eqref{s4:mfsp}. To see this explicitly, one should
reconstruct the metric of the scalar manifold order by order (due to
the power expansion of the exponential map in
\eqref{s3:scalarchart}). 
As we saw in Section~\ref{s3:specsolforanyn}, it is also 
possible to have light or massless hypermultiplets,
in which case  $\cN=2$ supersymmetry requires the field space to be 
quaternionic K\"ahler.  However, a complete analysis  of the scalar geometry 
is beyond the scope of this paper.


\section{Conclusion}

We have studied $\cN=2$ vacua of gauged 
$\cN=4$ supergravity theories  focussing on the class of theories
with only electric gaugings i.e.\ vanishing de Roo-Wagemans angles. 
We
reviewed the early result that in such an electrically gauged  $\cN=4$
theory, vacua which preserve $\cN =1,2$ or $4$ are necessarily
Minkowski and that $\cN=3$ vacua do not exist. Following the
observation in \cite{Borghese:2010ei}, we discussed in detail how the
homogeneity of the scalar manifold and the symmetry of the Lagrangian
allows one to carry out the
analysis of the gravitino mass matrices and supersymmetry conditions
at the origin, 
which leads to significant
simplifications when studying supersymmetry breaking.  

In order to construct explicit solutions with spontaneous partial
supersymmetry breaking, we then focussed on $\cN=2$ vacua. We
discussed the possible branching rules for $\cN=4$ supermultiplets,
showing that it was possible to have an $\cN=2$ spectrum with one
short massive BPS gravitino multiplet or two long massive gravitino
multiplets. 
We then constructed the solutions to the linear conditions that follow
from the Killing spinor equations for an $\cN=2$ vacuum, given in
terms of a set of embedding tensor components (charges). 
Consistency of the corresponding gaugings
with supersymmetry and gauge invariance required that this 
set of embedding tensor components satisfy the quadratic constraints \eqref{s2:quadconstr}.

We believe that it is difficult to solve the quadratic
constraints in general (as argued to some extent in Appendix
\ref{a:ganeq0}) and so we focussed on the case where a subset of the
embedding tensor components vanish ($g_a=0$), which holds
automatically when the number of $\cN=4$
vector multiplets $n$ is less or equal than six. In the appendix we
showed that if a solution with $g_a \neq 0$ were to exist, then it
would necessarily require the number of vector multiplet $n$ to be
greater than $6$. Setting $g_a=0$ corresponds to minimizing the
couplings between the gaugini and the gravitini in the $\cN=4$
Lagrangian, and therefore heuristically should make it easier to
guarantee supersymmetry and gauge invariance. We showed that when
$g_a=0$ one can arrange for only one $\cN = 4$ vector multiplet to
contribute to the Goldstini. For the class of gaugings with $g_a=0$
and $n\leq 6$ we gave the solutions of the quadratic constraints and
the unbroken gauge groups. We also found solutions for $n > 6$ with an
additional set of gaugings (and couplings) turned off.  

We then analyzed the mass terms and showed that 
all fields assembled in $\cN=2$
multiplets with appropriate mass degeneracies. 
In particular, all massive $\cN=2$ multiplets (including
the gravitino multiplet) have to be BPS.
We further showed that vacua exist with unbroken gauge group 
\begin{equation}
  U(1)^3\times G_{\cN=2}\,,
\end{equation}
where $G_{\cN=2}$ can be any compact reductive Lie group if the number $n$ of
$\cN=4$ vector multiplets is sufficiently large. 

Finally, we computed the effective $\cN=2$ action which is valid below
the scale of supersymmetry breaking. We found that the complex scalar
$\tau$ of the $\cN=4$ gravity multiplet cannot contribute to the
super-Higgs mechanism i.e.\ it is not charged under the $\cN=4$ gauge
group. This implies that the $SL(2)/SO(2)$ factor parametrized by
$\tau$ in the $\cN=4$ moduli space descends directly to an
$SL(2)/SO(2)$ factor in the $\cN=2$ moduli space. For vacua with
additional $(k+1)$  massless $\cN=2$ vector multiplets we therefore
conjectured that the vector multiplet moduli space 
has to be 
 \begin{equation}
  \nicefrac{SL(2)}{SO(2)} \times \nicefrac{SO(2,k)}{SO(2)\times SO(k)}
\end{equation}
as this series is the only possible special K\"ahler
manifolds that are product manifolds \cite{Ferrara:1989py}. 
We also found that it is possible to have massless hypermultiplets.
In this case  $\cN=2$ requires a field space which is quaternionic
K\"ahler. We leave a complete analysis of the scalar geometry for
future work.

\section{Acknowledgements}
We would like to thank Vicente Cort\'{e}s, Diederik Roest, Henning
Samtleben, Claudio Scrucca, Wolfgang Soergel, Hagen Triendl and Owen
Vaughan for useful discussions. This work was partly supported by the
German Science Foundation (DFG) under the Collaborative Research
Center (SFB) 676 ``Particles, Strings and the Early Universe''. The
research of P.S. is supported by the Swiss National Science Foundation
under the grant PP00P2-135164.

\newpage
\appendix
\noindent
{\bf\Large Appendix}
\section{Conventions}\label{a:indices} 

The spacetime metric $g_{\mu\nu}$ used in this paper has signature
$(-,+,+,+)$ and the totally antisymmetric tensor
$\epsilon^{\mu\nu\rho\lambda}$ is defined with
$\epsilon^{0123}=e^{-1}, \epsilon_{0123}=-e= -\sqrt{|{\rm det}\,
  g|}$.

We use the following indices:
\begin{center}
  \begin{tabular}{l|l}
    indices & group \\
    \hline
    $\alpha,\beta,\gamma,\ldots \in \{-,+\}$ & $SL(2)$ \\
    $M,N,P,\ldots \in\{1,\ldots,6+n \}$ & $SO(6,n)$ \\
    $m,n,p,\ldots \in\{1,\ldots,6 \}$ & $SO(6)$ \\
    $i,j,k,\ldots \in\{1,\ldots,4 \}$ & $SU(4)$ \\
    $a,b,c,\ldots \in\{1,\ldots,n\}$ & $SO(n)$ \\
    $\ta,\tb,\tc,\ldots \in\{1,\ldots, n-1\}$ & $SO(n-1)\subset SO(n)$ 
  \end{tabular}
\end{center}
All indices other than the ones of $SU(4)$ transform under the
fundamental representation of the given groups. In the case of $SU(4)$
upper/lower indices transform under the $\bf 4$ ($\bf \bar{4}$),
respectively. Upon complex conjugation such upper and lower indices
are interchanged, e.g.\ $({X_i}^{j})^* = {X^i}_j$.

In addition, for the $g_a=0$ solutions discussed in Sections
\ref{s4:unbrokengaugegroup} and \ref{seceffact} we use $SO(n-1)\subset
SO(n)$ indices, $\bar{a}\ldots$ and $\hat{a}\ldots$ which are
associated to massless vectors $A^{\mu   \bar{a}}$ and massive vectors
$A^{\mu \hat{a}}$, respectively. 

\subsection{Coset space representatives}
\label{a:scalars}
The coset space $\nicefrac{SO(6,n)}{SO(6)\times SO(n)}$ is represented by a
matrix $\cV = ({\cV_M}^N) \in SO(6,n)$. Raising/lowering $SO(6,n)$
indices is defined via the $SO(6,n)$ invariant metric 
$$\eta =
(\eta_{MN}) = (\eta^{MN}) = \text{diag}(\underbrace{-\ldots -}_{
  \text{6 times}}, \underbrace{+\ldots +}_{n \text{ times}})\,,$$
and ${\cV}^{-1T}=({{\cV}^M}_N)$. $\cV$ transforms as
\begin{equation}
  \cV \rightarrow g \, \cV \, h(x)\,,
\end{equation}
which in terms of indices reads
\begin{eqnarray}
  {\cV_M}^N & \rightarrow & {g_M}^P \, {\cV_P}^Q \, {h(x)_Q}^N\,, \nonumber\\
  {\cV^M}_N & \rightarrow & {g^M}_P \, {\cV^P}_Q \, {h(x)^Q}_N\,,
\end{eqnarray}
where $g=({g_M}^P) \in SO(6,n)$ and a spacetime dependent
$h(x)=({h(x)_Q}^N) \in SO(6)\times SO(n)$ and ${g^M}_P$ and ${h(x)^Q}_N$ are
obtained from the former via lowering/raising indices. It is apparent
that global $SO(6,n)$ acts only on the first index of ${\cV_M}^N$ and
${\cV^M}_N$ while local $SO(6)\times SO(n)$ acts only on the
second. The bosonic part of the Lagrangian can be conveniently
expressed in terms of a symmetric positive definite matrix  
\begin{equation}
  M = (M_{MN}) := \cV \cV^T\,,
\end{equation}
which transforms as a tensor of $SO(6,n)$, i.e.\
\begin{equation}
  M_{MN} \rightarrow {g_M}^Q {g_N}^R M_{QR}\,,
\end{equation}
and is manifestly invariant under local $SO(6)\times SO(n)$
transformations. One also has $M^{-1}=(M^{MN})$ transforming as
\begin{equation}
  M^{MN} \rightarrow {g^M}_Q {g^N}_R M^{QR}.
\end{equation}

In describing the supergravity theory index calculus seems to be
indispensable because $SO(6,n)$ indices associated to $SO(6)\times
SO(n)$ need to be decomposed into those of $SO(6)$ and $SO(n)$, of which
the $SO(6)$ indices are to be transferred to indices of the universal
cover $SU(4)$ in order to describe the coupling of scalar
representatives to fermions. The relation between these indices is due
to the fact that in terms of representations of their common complex
Lie algebra one has $({\bf 4}\otimes {\bf 4})_{\text{antisymmetric}}
\cong {\bf 6}$. As in the Appendix of \cite{Borghese:2010ei}, we
therefore associate to every vector index $m$ of $SO(6)$ a pair of
anti-symmetric $SU(4)$ indices $[ij]$ in the following way
\begin{align}
  \
  \phi_{ij} = \tfrac{1}{2} \sum_{m=1}^{6} \phi_{m} \, [G_{m}]_{ij} \, , \qquad
\phi^{ij} & = - \tfrac{1}{2} \sum_{m=1}^{6} \phi_{m} \, [G_{m}]^{ij} \, ,
\end{align}
where $\phi_m$ shall be a generic $SO(6)$ vector and the $G$'s are the
't Hooft matrices 
\begin{eqnarray}
\label{thooft}
[G_{1}]_{ij} = \begin{matr}{cccc} 0 & i & 0 & 0 \\ -i & 0 & 0 & 0 \\ 0 & 0 & 0 & -i \\ 0 & 0 & i & 0 \end{matr} & , & [G_{2}]_{ij} = \begin{matr}{cccc} 0 & 0 & i & 0 \\ 0 & 0 & 0 & i \\ -i & 0 & 0 & 0 \\ 0 & -i & 0 & 0 \end{matr} \, ,\nonumber\\
\left[ G_{3} \right]_{ij} = \begin{matr}{cccc} 0 & 0 & 0 & i \\ 0 & 0 & -i & 0 \\ 0 & i & 0 & 0 \\ -i & 0 & 0 & 0 \end{matr} & , & [G_{4}]_{ij} = \begin{matr}{cccc} 0 & -1 & 0 & 0 \\ 1 & 0 & 0 & 0 \\ 0 & 0 & 0 & -1 \\ 0 & 0 & 1 & 0 \end{matr} \, ,\nonumber\\
\left[ G_{5} \right]_{ij} = \begin{matr}{cccc} 0 & 0 & -1 & 0 \\ 0 & 0 & 0 & 1 \\ 1 & 0 & 0 & 0 \\ 0 & -1 & 0 & 0 \end{matr} & , & [G_{6}]_{ij} = \begin{matr}{cccc} 0 & 0 & 0 & -1 \\ 0 & 0 & -1 & 0 \\ 0 & 1 & 0 & 0 \\ 1 & 0 & 0 & 0 \end{matr} \, .
\end{eqnarray}
Furthermore, for every $m = 1, \, \ldots, \, 6$ one defines
\begin{align}
  [G_{m}]^{ij} = - \tfrac{1}{2} \epsilon^{ijkl} \, [G_{m}]_{kl} = -
  ([G_{m}]_{ij})^{*}\,,
\end{align} 
so as to obtain $(\phi_{ij})^* = \phi^{ij}$.

At the origin
of $SO(6,n)$, cf.\ \eqref{s3:origin}, one finds $\cV =
{\cV}^{-1T}=\mathbb{1}$ which in components reads  
\begin{eqnarray}
  {\cV_m}^n = {\cV^n}_m = \delta_m^n, & \quad & {\cV_m}^a = {\cV^m}_a
  = 0, \nonumber\\ 
  {\cV_a}^b = {\cV^b}_a = \delta_a^b, & \quad  & {\cV_a}^m = {\cV^m}_a
  = 0.
\end{eqnarray}
In terms of $SU(4)$ indices one now has
\begin{eqnarray}
  {\cV_M}^{ij} = \left\{\begin{matrix}
\frac{1}{2}[G_m]^{ij}, & \mbox{if } M=m   \\
0, & \mbox{if } M=a   \end{matrix}\right.\;, \qquad 
  {\cV^M}_{ij} = \left\{\begin{matrix}
\frac{1}{2}[G_m]_{ij}, & \mbox{if } M=m   \\
0, & \mbox{if } M=a   \end{matrix}\right. .
\end{eqnarray}

As to $\nicefrac{SL(2)}{SO(2)}$, a generic representative would be
$\cV=({\cV_\alpha}^{\beta})\in SL(2)$. Raising/lowering indices is
defined via the antisymmetric matrix
$\epsilon=(\epsilon_{\alpha\beta}) = (\epsilon^{\alpha\beta})$ 
with $\epsilon^{12}=1$ in such a way that
\begin{equation}
  ({\cV^\alpha}_\beta) = (\epsilon^{\alpha\gamma} {\cV_\gamma}^\delta
  \epsilon_{\delta\beta}) = \epsilon \cV \epsilon = -{\cV}^{-1T}.
\end{equation}
As before, transformations in terms of indices are
\begin{eqnarray}
  \cV=({\cV_\alpha}^\beta) & \rightarrow & g\cV h(x) = ({g_\alpha}^\gamma
  {{\cV}_\gamma}^\delta {{h(x)}_\delta}^\beta) \,,\nonumber\\
  -{\cV}^{-1T}=({\cV^\alpha}_\beta) & \rightarrow & ({g^\alpha}_\gamma
  {{\cV}^\gamma}_\delta {{h(x)}^\delta}_\beta)\,,
\end{eqnarray}
and the bosonic Lagrangian can be written in terms of the symmetric
positive definite matrix
\begin{equation}
  M:=\cV {\cV}^T = (M_{\alpha\beta})\,,
\end{equation}
that can be expressed in terms of $\tau\in\mathbb{C}$ with $\Im \tau
>0$ as
\begin{align}
\label{s2:Malphabeta}
   M_{\aa\ab} &= \frac 1 {\Im(\tau)} \begin{pmatrix} |\tau|^2 & \Re(\tau) \\ \Re(\tau) & 1  \end{pmatrix}.
\end{align}
Its inverse is $M^{-1}=(M^{\alpha\beta})$ and transforms
accordingly. The fermionic sector of the supergravity theory requires
a different representation of cosets, namely, in terms of 
$(\cV_\alpha)\in\mathbb{C}^2$ such that
\begin{equation}
  M_{\alpha\beta} = \Re (\cV_\alpha (\cV_\beta)^*)\,.
\end{equation}
For \eqref{s2:Malphabeta} one can always find appropriate
$\cV_\alpha$. Letting them transform as vectors under
global $SL(2)=SL(2,\mathbb{R})$ gives the right transformation for
$M_{\alpha\beta}$. For a given $\tau$ as above, $\cV_\alpha$ is unique
up to local $U(1)$ transformations
\begin{equation}
  \cV_\alpha \rightarrow e^{i\phi(x)} \cV_\alpha
\end{equation}
for arbitrary $\phi(x)\in\mathbb{R}$ (and up to a sign
ambiguity\footnote{Fixing the gauge such that $\mathbb{R}\ni\cV_1>0$,
  one finds a sign ambiguity in the imaginary part of $\cV_2$ as is
  apparent from $M_{\aa\ab}=(\operatorname{Re}
  \cV_\aa)(\operatorname{Re}\cV_\ab)+(\operatorname{Im} \cV_\aa)(
  \operatorname{Im} \cV_\ab)$.}). As fermions also transform under 
this $U(1)$, they couple to coset representatives $\cV_\alpha$. At the 
origin $\cV=\mathbb{1}$ and thus in an appropriate gauge one finds  
$(\cV_\alpha) = (1, i)^T$.


\section{Weyl \& Dirac spinor conventions}
\label{a:weyl}
While we find it more convenient to work with Weyl spinors, the
fermionic terms in the literature
\cite{Weidner:2006rp,Bergshoeff:1985ms} are given in terms of Dirac
spinors. Based on the conventions given in \cite{Weidner:2006rp} we
express Dirac spinors in terms of Weyl spinors. In what follows we
will first summarize their conventions and then express fermionic
terms using Weyl spinors.

The metric $(\eta_{\mu\nu})$ has signature $(-,+,+,+)$. The
$\gamma$-matrices $\Gamma_{\mu}$ satisfying     
\begin{equation}
\{\Gamma_{\mu}, \Gamma_{\nu} \} = 2\eta_{\mu\nu}
\end{equation}
are (chirally) represented by
\begin{equation}
\Gamma_{\mu} = \begin{pmatrix}0 & \sigma^{\mu}\\ \sigma_{\mu} &
  0\end{pmatrix} = \begin{pmatrix}0 & \bar{\sigma}_{\mu}\\
  \sigma_{\mu} & 0\end{pmatrix} .
\end{equation}
where
\begin{equation}
\sigma_{\mu} = (\mathbb{1},\vec{\sigma}) = \bar{\sigma}^{\mu},
\qquad \sigma^{\mu} = \eta^{\mu\nu} \sigma_{\nu} =
(-\mathbb{1},\vec{\sigma}) = \bar{\sigma}_{\mu} \,,\nonumber\\
\end{equation}
and $\vec{\sigma} = (\sigma_1,\sigma_2,\sigma_3)$ is built from the
usual $\sigma$-matrices. One then has
\begin{equation}
\Gamma_5 =
i\Gamma_0\Gamma_1\Gamma_2\Gamma_3 = \begin{pmatrix}\mathbb{1} & 0\\
  0 & -\mathbb{1}\end{pmatrix}
\end{equation}
and
\begin{eqnarray}
(\Gamma_{\mu})^\dag = \eta^{\mu\nu}\Gamma_{\nu} = \Gamma_0
\Gamma_{\mu} \Gamma_0,\qquad (\Gamma^{\mu})^\dag =
(\eta^{\mu\nu}\Gamma_{\nu})^\dag = \Gamma_0 \Gamma^{\mu} \Gamma_0,
\nonumber\\
\Gamma_0^\dag = -\Gamma_0,\qquad (\Gamma^{\mu\nu})^\dag = \ft 1 2
[\Gamma^\mu,\Gamma^\nu]^\dag = -\Gamma_0 \Gamma^{\nu\mu}
\Gamma_0.
\end{eqnarray}
In particular,
\begin{equation}
\Gamma^{\mu\nu} = 2 \begin{pmatrix}\sigma^{\mu\nu} & 0\\0 &
  \bar{\sigma}^{\mu\nu}\end{pmatrix}\,,
\end{equation}
where
\begin{equation}
\sigma^{\mu\nu} = \ft 1 4 (\bar{\sigma}^\mu \sigma^\nu -
\bar{\sigma}^\nu \sigma^\mu),\qquad
\bar{\sigma}^{\mu\nu} = \ft 1 4 (\sigma^\mu \bar{\sigma}^\nu -
\sigma^\nu \bar{\sigma}^\mu).
\end{equation}
Using the charge conjugation matrix
\begin{equation}
\label{a:epsilon}
B = i\Gamma_5 \Gamma_2 = \begin{pmatrix}0 & \epsilon\\ -\epsilon &
  0\end{pmatrix} \qquad\text{with}\qquad \epsilon=\begin{pmatrix}0 &
  1\\ -1 & 0\end{pmatrix}\,,
\end{equation}
one defines for a generic Dirac spinor $\phi^i$ transforming in
the $\bf 4$ of $SU(4)$
\begin{equation}
\phi_i = B(\phi^i)^* \,, 
\end{equation}
which transforms again as Dirac spinor, but now in the complex
conjugate representation $\bf\bar{4}$ of $SU(4)$. For a chiral spinor
with $\Gamma_5 \phi^i = \pm \phi^i$, one finds $\Gamma_5 \phi_i = \mp
\phi_i$, i.e.\ charge conjugation also flips the chirality of chiral
spinors. Furthermore, one defines
\begin{equation}
\bar{\phi}_i = (\phi^i)^\dag\Gamma_0,\qquad
\bar{\phi}^i = (\bar{\phi}_i)^* B.
\end{equation}

The fermionic spectrum of $\cN=4$ supergravity in $D=4$ with a gravity
multiplet and $n$ vector multiplets consists of Dirac spinors
$\psi^i_{\mu}$, $\lambda^{ai}$, $\chi^i$ that have the following
chirality:
\begin{eqnarray}
\psi^i_{\mu}  = \begin{pmatrix}(\psi^{i}_{\mu})^A \\0 \end{pmatrix}
& \qquad & \Gamma_5 \psi^i_{\mu} = \psi^i_{\mu},\nonumber\\
\lambda^{ai} = \begin{pmatrix}(\lambda^{ai})^A \\0 \end{pmatrix}
& \qquad & \Gamma_5 \lambda^{ai} = \lambda^{ai},\nonumber\\
\chi^i  = \begin{pmatrix}0 \\ (\chi^{i})_{\dot{A}} \end{pmatrix}
& \qquad & \Gamma_5 \chi^i = -\chi^i.
\end{eqnarray}
Note that we have not introduced new symbols for Weyl spinors but the
latter are recognized in the van der Waerden notation by undotted
$(A,\ldots)$ and dotted indices $(\dot{A},\ldots)$ transforming with
respect to the two different $SU(2)$ groups of the Lorentz group. We
can now express all the fermionic mass terms in terms of Weyl spinors  
\begin{eqnarray}
\bar{\psi}_{\mu i} \Gamma^{\mu\nu} \psi_{\nu j} + \text{h.c.} & = &
2 (\psi^{i}_\mu)^* \bar{\sigma}^{\mu\nu} \epsilon\, (\psi_{\nu}^{j})^*
- 2(\psi_\nu^j)\,\epsilon\,\sigma^{\nu\mu} (\psi_\mu^i) ,\nonumber\\
\bar{\psi}_{\mu i} \Gamma^\mu \chi_j + \text{h.c.} & = & - (\psi_\mu^i)^*
\sigma^\mu \epsilon\, (\chi^j)^* + (\chi^j)\,\epsilon\,\sigma^\mu
(\psi^i_\mu) ,\nonumber\\
\bar{\psi}_\mu^i \Gamma^\mu \lambda_j^a  + \text{h.c.} & = &
-(\psi^i_\mu)\, \epsilon\,\bar{\sigma}^\mu\epsilon\, (\lambda^{aj})^* -
(\lambda^{aj})\, \epsilon\, \bar{\sigma}^\mu \epsilon\,
(\psi^i_\mu)^*,\nonumber\\
\bar{\lambda}_i^a \lambda_j^b  + \text{h.c}. & = & (\lambda^{ai})^*
\epsilon\,(\lambda^{bj})^* - (\lambda^{bj})\, \epsilon\,
(\lambda^{ai}),\nonumber\\
\bar{\chi}^i \lambda_j^a  + \text{h.c.} & =& (\chi^i)
(\lambda^{aj})^* + (\lambda^{aj}) (\chi^i)^*,
\end{eqnarray}
where on the right hand side we suppressed all dotted/undotted
spinor indices. Note that bilinear terms made from $\bar{\chi}^i
\chi^j$ are absent in gauged $\cN=4$ supergravity, as no such term
exists that is invariant under $U(1)\subset H$ and linear in the
embedding tensor components. In our conventions all $\epsilon$-tensors
with upper/lower, dotted/undotted indices are numerically
identical and given by the one  in \eqref{a:epsilon}.


\section{$A$-matrices at the origin}\label{a:amatrices}
Here we state the results for the $A$-matrices in
\eqref{s2:fermionshiftmatrices} evaluated at the origin
$(\mathbb{1}_2, \mathbb{1}_{6+n})$.\footnote{The result for a critical
  point \eqref{s3:origin} is obtained by multiplying all components by 
  $\cV_-$.}
For $A_1^{ij}=A_2^{ij}$ the result is: 
\begin{eqnarray}
  \label{s2:A_1}
A_1^{11} & = & \frac{3}{4} \left [(-f_{456} +f_{234}
  -f_{135} +f_{126})  
  +i(f_{123} -f_{156} +f_{246}
  -f_{345}) \right] \,,\nonumber\\ 
A_1^{22} & = & \frac{3}{4} \left [(-f_{456} +f_{234}
  +f_{135} -f_{126})  
  +i(f_{123} -f_{156} -f_{246}
  +f_{345}) \right] \,,\nonumber\\ 
A_1^{33} & = & \frac{3}{4} \left [(-f_{456} -f_{234}
  -f_{135} -f_{126})  
  +i(f_{123} +f_{156} +f_{246}
  +f_{345}) \right] \,,\nonumber\\ 
A_1^{44} & = & \frac{3}{4} \left [(-f_{456} -f_{234}
  +f_{135} +f_{126})  
  +i(f_{123} +f_{156} -f_{246}
  -f_{345}) \right] \,,\nonumber\\ 
A_1^{12} & = & \frac{3}{4} \left [(-f_{125} - f_{136}) 
  +i(-f_{346}-f_{245}) \right ] \,,\nonumber\\
A_1^{34} & = & \frac{3}{4} \left [(f_{125} - f_{136}) 
  +i(f_{346}-f_{245}) \right ] \,,\nonumber\\
A_1^{13} & = & \frac{3}{4} \left [(f_{124} - f_{236}) 
  +i(-f_{356}+f_{145}) \right ] \,,\nonumber\\
A_1^{24} & = & \frac{3}{4} \left [(f_{124} + f_{236}) 
  +i(-f_{356}-f_{145}) \right ] \,,\nonumber\\
A_1^{14} & = & \frac{3}{4} \left [(f_{134} + f_{235}) 
  +i(f_{256}+f_{146}) \right ] \,,\nonumber\\
A_1^{23} & = & \frac{3}{4} \left [(-f_{134} + f_{235}) 
  +i(-f_{256}+f_{146}) \right ]\,.
\end{eqnarray} 
The components of the symmetric matrix $(A_1^{ij})$ depend on 20 real
parameters $f_{mnp}$. It is apparent that any symmetric complex
$4\times 4$ matrix can be written in this form. As to $({A_{2ai}}^j)$
for all $a=1,\ldots,n$, the components of $({A_{2ai}}^j)$ read:
\allowdisplaybreaks
\begin{eqnarray}
  \label{s2:A2a}
{A_{2a1}}^1 & = & -\frac{1}{2} i (f_{a14} + f_{a25} + f_{a36})
  \nonumber\,,\\
{A_{2a2}}^2 & = & -\frac{1}{2} i (f_{a14} - f_{a25} - f_{a36})
  \nonumber\,,\\
{A_{2a3}}^3 & = & -\frac{1}{2} i (-f_{a14} + f_{a25} - f_{a36})
  \nonumber\,,\\
{A_{2a4}}^4 & = & -\frac{1}{2} i (-f_{a14} - f_{a25} + f_{a36})
  \nonumber\,,\\
{A_{2a1}}^2 & = & -\frac{1}{2} [(f_{a23}-f_{a56}) + i(f_{a26}-f_{a35})]
  \nonumber\,,\\
{A_{2a3}}^4 & = & -\frac{1}{2} [(-f_{a23}-f_{a56}) + i(-f_{a26}-f_{a35})]
  \nonumber\,,\\
{A_{2a1}}^3 & = & -\frac{1}{2} [(-f_{a13}+f_{a46}) + i(-f_{a16}+f_{a34})]
  \nonumber\,,\\
{A_{2a2}}^4 & = & -\frac{1}{2} [(-f_{a13}-f_{a46}) + i(f_{a16}+f_{a34})]
  \nonumber\,,\\
{A_{2a1}}^4 & = & -\frac{1}{2} [(f_{a12}-f_{a45}) + i(f_{a15}-f_{a24})]
  \nonumber\,,\\
{A_{2a2}}^3 & = & -\frac{1}{2} [(-f_{a12}-f_{a45}) + i(f_{a15}+f_{a24})]\,.
\end{eqnarray}
Moreover,
\begin{equation}
  \label{s2:A2atrans}
  {A_{2a2}}^1 = -\frac{1}{2} [-(f_{a23}-f_{a56}) + i(f_{a26}-f_{a35})]\,,
\end{equation}
etc\ldots where the real part is always multiplied by an extra minus
sign. We conclude that $A_1=A_2$ depends only on $f_{mnp}$ while 
matrices $A_{2a}$ are built from $f_{amn}$. Note that at the origin
$f_{abm}$ and $f_{abc}$ do not appear in the fermion shift matrices
(and therefore also not in the Killing spinor equations).

Finally, we give an explicit result for the antisymmetric $A$-matrices
$({A_{ab}}^{ij})$ for all $a,b$. At the origin of the scalar manifold
they are entirely given in terms of components $f_{abm}$ 
\begin{equation}
  \label{s3:Aabij}
  ({A_{ab}}^{ij}) = \frac{1}{2}
\begin{pmatrix}
  0 & if_{ab1}+f_{ab4} & if_{ab2}+f_{ab5} & if_{ab3}+f_{ab6}  \\
  -* & 0 &-if_{ab3}+f_{ab6} & if_{ab2}-f_{ab5} \\
  -* & -*& 0 & -if_{ab1} + f_{ab4} \\
  -* & -* & -* & 0
\end{pmatrix}
\end{equation}
for all $a,b$.



\section{Partial solution of the quadratic constraints}
\subsection{Discussing constraint equations for $g_a\neq 0$}
\label{a:ganeq0}
The quadratic constraints for electric gaugings in the case of
$g_a\neq 0$ are hard to solve. In fact, so far we have not found any
example of a consistent solution with $g_a\neq 0$. Here we will
discuss the following two aspects: First, we will show that an
electrically gauged $\cN=4$ theory with $\cN=2$ vacuum requires
$f_a\neq 0$; secondly, we will give some details on a lengthy but
elementary calculation that shows that $g_a\neq 0$ solutions, if at
all, exist only in $n>6$. These two aspects illustrate that $g_a\neq
0$ consistent solutions would have to be rather sophisticated. As in 
Section \ref{s3:solvingquadconstraints} we label the quadratic
constraints given in \eqref{s2:quadconstr} by the quadruple
$(M,N,P,Q)$ of $SO(6,n)$-indices.   

\subsubsection{$\cN=2$ vacua require $f_a\neq 0$}
\label{a:fa0impossible}
We will prove this claim by contradiction; we therefore assume
$f_a=0$. The constraint equations to be used in this proof
are
\begin{align}
  \label{a:fa0_2356z}
  (2,3,5,6) &&  \vec{e}^2 + \vec{g}^2 &= c^2 \neq 0  \,,\\
  \label{a:fa0_b245z}
  (b,2,4,5) &&  F_4\, \vec{e} &= 2c\, \vec{g}\,,\\
  \label{a:fa0_b246z}
  (b,2,4,6) &&  F_4\, \vec{g} &= -2c\, \vec{e}\,,\\
  \label{a:fa0_b235z}
  (b,2,3,5) && F_2\,\vec{g} & = F_3\,\vec{e}\,,\\
  \label{a:fa0_b236z}
  (b,2,3,6) && F_3\,\vec{g} & = -F_2\,\vec{e}\,,\\
  \label{a:fa0_bc23z}
  (b,c,2,3) && ([G_2,G_3])_{bc} &= c (F_4)_{bc}+2(e_c\,g_b-e_b\,g_c)\,,
\end{align}
where for better legibility we use a matrix notation with $SO(n)$
vectors $\vec{e},\vec{g}$ and matrices $(F_m)_{ab}=f_{mab}$. It is
obvious from \eqref{a:fa0_2356z}, \eqref{a:fa0_b245z},
\eqref{a:fa0_b246z} that both $\vec{e}$ and $\vec{g}$ must be nonzero
because an $\cN=2$ vacuum requires $c\neq 0$. Thus, without loss of
generality, using first an $SO(n)$ transformation and subsequently a
transformation of the residual $SO(n-1)$ symmetry\footnote{We assume
  that $n$ is large enough.}, one can write  
\begin{equation}
  \vec{e} = \begin{pmatrix}e\\ 0 \\ \vec{0}\end{pmatrix}, \qquad
  \vec{g} = \begin{pmatrix}g'\\ g \\ \vec{0}\end{pmatrix}\,,
\end{equation}
with $e\neq 0, g,g'\in\mathbb{R}$. Then equations \eqref{a:fa0_b245z},
\eqref{a:fa0_b246z} show that $g'=0, g=\sigma e$ with $\sigma=\pm 1$
and 
\begin{equation}
  F_4 = 
  \begin{pmatrix}
  \begin{matrix} 0 & -2c\sigma\\ 2c\sigma & 0\\ \end{matrix} & 0\\
  0 & \tilde{F}_4\\
  \end{pmatrix}\,,
\end{equation}
where $\tilde{F}_4\in\text{Mat}_{n-2,n-2}$. Furthermore,
\eqref{a:fa0_b235z} and \eqref{a:fa0_b236z} imply
\begin{equation}
  F_2 = 
  \begin{pmatrix}
  0 & 0 & * \\
  0 & 0 & * \\
  \vec{v} & \vec{w} & \tilde{F}_2
  \end{pmatrix},\qquad
  F_3 = 
  \begin{pmatrix}
  0 & 0 & * \\
  0 & 0 & * \\
  \sigma\vec{w} & -\sigma\vec{v} & \tilde{F}_3
  \end{pmatrix}\,,
\end{equation}
with $\vec{v},\vec{w}\in \text{Mat}_{n-2,1}$ and antisymmetric
matrices $\tilde{F}_2, \tilde{F}_3\in\text{Mat}_{n-2,n-2}$. As a
consequence, \eqref{a:fa0_2356z} and \eqref{a:fa0_bc23z} yield
\begin{equation}
  \sigma [F_2,F_3]_{78} = -3 c^2 = \vec{v}^2+\vec{w}^2 \ge 0\,,
\end{equation}
which contradicts $c\neq 0$. Hence, $\vec{f}$ cannot vanish in
consistent solutions with $\cN=2$ vacuum. This ends the proof.  

\subsubsection{$g_a\neq 0$ solutions do not exist in $n\leq 6$}
\label{a:ganeq0}
First we will concentrate on the subset of non-trivial quadratic
constraints in \eqref{s2:quadconstr} that can easily be solved: 
\begin{align}
  \label{a:fa0_2356}
  (2,3,5,6) &&  \vec{e}^2 + \vec{f}^2 + \vec{g}^2 &= c^2 \neq 0  \,,\\
  \label{a:fa0_b123}
  (b,1,2,3) &&  F_1\,\vec{f} &= 0  \,,\\
  \label{a:fa0_b125}
  (b,1,2,5) &&  F_1\,\vec{e} &= 0  \,,\\
  \label{a:fa0_b126}
  (b,1,2,6) &&  F_1\,\vec{g} &= 0  \,,\\
  \label{a:fa0_b234}
  (b,2,3,4) &&  F_4\,\vec{f} &= 0  \,,\\
  \label{a:fa0_b245}
  (b,2,4,5) &&  F_4\, \vec{e} &= 2c\, \vec{g}\,,\\
  \label{a:fa0_b246}
  (b,2,4,6) &&  F_4\, \vec{g} &= -2c\, \vec{e}\,,\\
  \label{a:fa0_b235}
  (b,2,3,5) && F_3\,\vec{e} - F_5\,\vec{f} - F_2\,\vec{g}  & = 0\,,\\
  \label{a:fa0_b236}
  (b,2,3,6) && F_2\,\vec{e} - F_6\,\vec{f} + F_3\,\vec{g} & = 0\,,\\
  \label{a:fa0_b256}
  (b,2,5,6) && F_6\,\vec{e} + F_2\,\vec{f} - F_5\,\vec{g}  & = 0\,,\\
  \label{a:fa0_b356}
  (b,3,5,6) && F_5\,\vec{e} + F_3\,\vec{f} + F_6\,\vec{g}  & = 0\, .
\end{align}
Here we use the same matrix notation as in Section
\ref{a:fa0impossible}. Having shown that $\vec{f}=0$ is impossible,
without loss of generality we write it as
\begin{equation}
  \vec{f} = \begin{pmatrix}f\\ \vec{0}\\\end{pmatrix}\,,
\end{equation}
with $f\neq 0$ and due to \eqref{a:fa0_b123} and \eqref{a:fa0_b234}
find 
\begin{equation}
  F_1=
  \begin{pmatrix}0 & 0\\ 0 & *\end{pmatrix},\qquad
  F_4=
  \begin{pmatrix}0 & 0\\ 0 & *\end{pmatrix}\,,
\end{equation}
with certain matrices $*\in\text{Mat}_{n-1,n-1}$. Unlike in Section 
\ref{s3:solvingquadconstraints} here we consider the case where 
$\vec{g}\neq 0$. Analogously to the discussion in Section
\ref{a:fa0impossible}, one can, without loss of generality and using
\eqref{a:fa0_b245} and \eqref{a:fa0_b246}, write 
\begin{equation}
  \label{a:ge}
  \vec{g} = \begin{pmatrix}0 \\ \sigma e\\0 \\ \vec{0}\end{pmatrix},
  \qquad 
  \vec{e} = \begin{pmatrix}0 \\ 0\\ e \\ \vec{0}\end{pmatrix}\,,
\end{equation}
with $e\neq 0$ and $\sigma=\pm 1$ to find
\begin{equation}
  \label{a:F1F4}
  F_1 = \mathbb{0}_{3,3} \oplus \tilde{F}_1, \qquad
  F_4 = \begin{pmatrix}
          \begin{matrix}0 & 0 & 0\\ 0 & 0 & 2\sigma c\\ 0 & -2\sigma
            c & 0\\\end{matrix} &0\\
          0 & \tilde{F}_4\\
        \end{pmatrix}\,,
\end{equation}
with matrices $\tilde{F}_1,
\tilde{F}_4\in\text{Mat}_{n-3,n-3}$. Furthermore, equations
\eqref{a:fa0_b235} to \eqref{a:fa0_b356} are solved by
\begin{align}
  \label{a:F2F3F5F6}
  F_2 = 
  \begin{pmatrix}
    \mathbb{0}_{3,3} & *\\
    \begin{matrix}\vec{a} & \vec{b} & -\sigma\vec{d}+f/e
      \vec{c}'\end{matrix} & \tilde{F}_2\\
  \end{pmatrix},\qquad
  F_3 = 
  \begin{pmatrix}
    \mathbb{0}_{3,3} & *\\
    \begin{matrix}\vec{c} & \vec{d} & \sigma\vec{b}+f/e
      \vec{a}'\end{matrix} & \tilde{F}_3\\
  \end{pmatrix},\nonumber\\
  F_5 = 
  \begin{pmatrix}
    \mathbb{0}_{3,3} & *\\
    \begin{matrix}\vec{a}' & \vec{b}' & -\sigma\vec{d}'-f/e
      \vec{c}\end{matrix} & \tilde{F}_5\\
  \end{pmatrix},\qquad
  F_6 = 
  \begin{pmatrix}
    \mathbb{0}_{3,3} & *\\
    \begin{matrix}\vec{c}' & \vec{d}' & \sigma\vec{b}'-f/e
      \vec{a}\end{matrix} & \tilde{F}_6\\
  \end{pmatrix}\,,
\end{align}
with $\vec{a}, \vec{b}, \vec{c}, \vec{d}, \vec{a}', \vec{b}', \vec{c}',
\vec{d}'\in\text{Mat}_{1,n-3}$ and antisymmetric
$\tilde{F}_{2},\tilde{F}_{3},
\tilde{F}_{5},\tilde{F}_{6}\in\text{Mat}_{n-3,n-3}$.  

There remain a large number of non-trivial quadratic constraints which
we do not know how to fully solve. Here, we list only those that are
useful in our argument:
\begin{align}
  \label{a:fa0_bc1m}
  (b,c,1,m) &&  [F_1,F_m] &=  0  \,,\\
  \label{a:fa0_bc24}
  (b,c,2,4) &&  [F_2,F_4] &= -c\,F_3\,,\\
  \label{a:fa0_bc34}
  (b,c,3,4) &&  [F_3,F_4] &= c\,F_2\,,\\
  \label{a:fa0_bc45}
  (b,c,4,5) &&  [F_5,F_4] &=- c\,F_6\,,\\
  \label{a:fa0_bc46}
  (b,c,4,6) &&  [F_6,F_4] &= c\,F_5\,,\\
  \label{a:fa0_bc23}
  (b,c,2,3) && ([F_2,F_3])_{bc} & = c(F_4)_{bc}-f(F_7)_{bc} +
  2(e_c\,g_b-e_b\,g_c) \,,\\
  \label{a:fa0_bc56}
  (b,c,5,6) && ([F_5,F_6])_{bc} & = c(F_4)_{bc}-f(F_7)_{bc} +
  2(e_c\,g_b-e_b\,g_c) \,,\\
  \label{a:fa0_bc26}
  (b,c,2,6) && ([F_2,F_6])_{bc} & = -g(F_8)_{bc} +
  2(e_b\,f_c-e_c\,f_b) \,,\\
  \label{a:fa0_bc35}
  (b,c,3,5) && ([F_3,F_5])_{bc} & = -g(F_8)_{bc} +
  2(e_b\,f_c-e_c\,f_b) \,,\\
  \label{a:fa0_bc25}
  (b,c,2,5) && ([F_2,F_5])_{bc} & = -e(F_9)_{bc} -
  2(f_c\,g_b-f_b\,g_c) \,,\\
  \label{a:fa0_bc36}
  (b,c,3,6) && ([F_3,F_6])_{bc} & = e(F_9)_{bc} +
  2(f_c\,g_b-f_b\,g_c)\,.
\end{align} 
Here, $(F_7)_{ab} = f_{7ab}$, etc.\ for the first three $SO(n)$
indices denoted by $7,8,9$. Using \eqref{a:F1F4} and
\eqref{a:F2F3F5F6}, equations \eqref{a:fa0_bc24} to \eqref{a:fa0_bc46}
are equivalent to:  
\begin{align}
  \label{a:F4ac}
  \tilde{F}_4 \,\vec{a} & = c\, \vec{c} \,,\\
  \tilde{F}_4 \,\vec{b} & = 3c\, \vec{d} - 2c\sigma \ft f e\, \vec{c'}
  \,,\\
  \tilde{F}_4 \,\vec{c} & = -c\, \vec{a} \,,\\
  \tilde{F}_4 \,\vec{d} & = -3c\, \vec{b} - 2c\sigma \ft f e\,
  \vec{a'} \,,\\
  \tilde{F}_4\, \vec{a'} & = c\, \vec{c'}\,,\\
  \tilde{F}_4\, \vec{b'} & = 3c\, \vec{d'} + 2c\sigma \ft f e\,
  \vec{c}\,,\\
  \tilde{F}_4\, \vec{c'} & = -c\, \vec{a'} \,,\\
  \label{a:F4ddbba}
  \tilde{F}_4\, \vec{d'} & = -3c\, \vec{b'} + 2c\sigma \ft f e\,
  \vec{a} \,,\\
  [\tilde{F}_2, \tilde{F}_4] & = -c\, \tilde{F}_3 \,,\\
  [\tilde{F}_3, \tilde{F}_4] & = c\, \tilde{F}_2 \,,\\
  [\tilde{F}_5, \tilde{F}_4] & = -c\, \tilde{F}_6 \,,\\
  [\tilde{F}_6, \tilde{F}_4] & = c\, \tilde{F}_5\,.
\end{align}
While tedious, it is possible to find the general solution to equations
\eqref{a:F4ac} to \eqref{a:F4ddbba}. Rather than discussing this in
detail we will content ourselves with showing that consistent solutions
require as a necessary condition that $\vec{a},\vec{b},$ etc.\ be at
least nonzero column vectors of dimension 4. This then immediately allows us to
prove the claim of this section (obviously due to
\eqref{a:F2F3F5F6}). To this end, we solve equations
\eqref{a:fa0_bc23} to \eqref{a:fa0_bc36} for $F_7,F_8,F_9$,
respectively, and invoke the antisymmetry of $f_{abc}$. This gives
rise to another set of quadratic constraints. The ones of interest for
this argument are
\begin{align}
\label{a:addconstr1}
  \vec{a}\cdot \vec{d} & = \vec{b}\cdot\vec{c}\,,\\
  \vec{a'}\cdot\vec{d'} & = \vec{b'}\cdot\vec{c'} \,,\\
  \vec{a}\cdot\vec{d'} & = \vec{b}\cdot\vec{c'} \,,\\
  \vec{c}\cdot\vec{b'} & = \vec{d}\cdot\vec{a'} \,,\\
  \sigma (\vec{a}\cdot\vec{b} + \vec{c}\cdot\vec{d}) & = \ft f e
  (\vec{c}\cdot\vec{c'} - \vec{a}\cdot\vec{a'}) \,,\\
  \sigma (\vec{a'}\cdot\vec{b'} + \vec{c'}\cdot\vec{d'}) & = -\ft f e
  (\vec{c}\cdot\vec{c'} - \vec{a}\cdot\vec{a'}) \,,\\
  \sigma(\vec{b}\cdot\vec{b'} + \vec{d}\cdot\vec{d'}) & = \ft f e
  (\vec{a}\cdot\vec{b} + \vec{c'}\cdot\vec{d'}) \,,\\
  \sigma (\vec{b}\cdot\vec{b'} + \vec{d}\cdot\vec{d'}) & = -\ft f e
  (\vec{a'}\cdot\vec{b'} + \vec{c}\cdot\vec{d}) \,,\\
  \sigma(\vec{d}\cdot\vec{a'} - \vec{a}\cdot\vec{d'}) &= \ft f e
  (\vec{a}\cdot\vec{c} + \vec{a'}\cdot\vec{c'}) \,,\\
  \sigma (\vec{b}\cdot\vec{c'} - \vec{c}\cdot\vec{b'}) & = -\ft f e
  (\vec{a}\cdot\vec{c} + \vec{a'}\cdot\vec{c'}) \,,\\
  \sigma (\vec{d}\cdot\vec{b'} - \vec{b}\cdot\vec{d'}) & = \ft f e
  (\vec{b}\cdot\vec{c} + \vec{b'}\cdot\vec{c'}) \,,\\
  \sigma (\vec{b}\cdot\vec{d'} - \vec{d}\cdot\vec{b'}) & = -\ft f e
  (\vec{a}\cdot\vec{d} + \vec{a'}\cdot\vec{d'}) \,,\\
  \vec{a}\cdot\vec{b'} - \vec{b}\cdot\vec{a'} &= \vec{d}\cdot\vec{c'} -
  \vec{c}\cdot\vec{d'} \,,\\
  \sigma (\vec{b}^2+\vec{d}^2) + \ft f e (\vec{b}\cdot\vec{a'} -
  \vec{d}\cdot\vec{c'}) & = \sigma(\vec{b'}^2 + \vec{d'}^2) - \ft f e
  (\vec{a}\cdot\vec{b'} - \vec{c}\cdot\vec{d'}) \,,\\
  \sigma (\vec{a}\cdot\vec{b'} + \vec{d}\cdot\vec{c'}) - \ft f e
  (\vec{a}^2 + \vec{c'}^2) & = -\sigma (\vec{b}\cdot\vec{a'} +
  \vec{c}\cdot\vec{d'}) - \ft f e (\vec{a'}^2 + \vec{c}^2)\,,\\
  \sigma e (6e^2+\vec{b}^2+\vec{d}^2) & = f \left (-\vec{a}\cdot\vec{b'} -
  \vec{b}\cdot\vec{a'} + \ft f e \sigma(\vec{a}^2+\vec{c'}^2) \right)\,,
\\ 
\label{a:addconstr2}
  \sigma e (6e^2+\vec{b}^2+\vec{d}^2) & = f (\vec{a}\cdot\vec{b'} +
  \vec{d}\cdot\vec{c'} - 2 \vec{b}\cdot\vec{a'})\,,
\end{align}
where for the last two equations we also used \eqref{a:fa0_2356} and
\eqref{a:ge}. Those two equations imply that not all
$\vec{a},\vec{b},\ldots$ can vanish because by assumption $e\neq
0$. Furthermore, one finds that solutions satisfying \eqref{a:F4ac} to
\eqref{a:F4ddbba} subject to the additional constraints
\eqref{a:addconstr1} to \eqref{a:addconstr2} necessarily require
nonzero column vectors $\vec{a},\vec{b},$ etc.\ of dimension at least
4. Since $\vec{a},\vec{b},\ldots\in\text{Mat}_{1,n-3}$ we conclude
that $g_a\neq 0$ solutions do not exist in $n\le 6$.

\subsection{Discussing constraint equations for $g_a=0$}
\label{a:detailsquadconstr}
Here we list the quadratic constraint equations that are not trivially
satisfied, c.f.\ Section \ref{s3:solvingquadconstraints}. In what
follows the quadruple $(M,N,P,Q)$ in the first column refers to the
free indices in \eqref{s2:quadconstr}:
\allowdisplaybreaks
\begin{subequations}
\begin{align}
  \label{s3:qc_bc12'}
  (\tb,\tc,1,2) & & f_{\ta\tb 2}\, f_{\ta\tc 1} - f_{\ta\tb 1}\,
  f_{\ta\tc 2} & = 0\,,\\
  (\tb,\tc,1,3) & & f_{\ta\tb 3}\, f_{\ta\tc 1} - f_{\ta\tb 1}\,
  f_{\ta\tc 3} & = 0\,,\\
  (\tb,\tc,1,4) & & f_{\ta\tb 4}\, f_{\ta\tc 1} - f_{\ta\tb 1}\,
  f_{\ta\tc 4} & = 0\,,\\
  (\tb,\tc,1,5) & & f_{\ta\tb 5}\, f_{\ta\tc 1} - f_{\ta\tb 1}\,
  f_{\ta\tc 5} & = 0\,,\\
  (\tb,\tc,1,6) & & f_{\ta\tb 6}\, f_{\ta\tc 1} - f_{\ta\tb 1}\,
  f_{\ta\tc 6} & = 0\,,\\
  (\tb,\tc,2,5) & & f_{\ta\tb 5}\, f_{\ta\tc 2} - f_{\ta\tb 2}\,
  f_{\ta\tc 5} & = 0\,,\\
  (\tb,\tc,2,6) & & f_{\ta\tb 6}\, f_{\ta\tc 2} - f_{\ta\tb 2}\,
  f_{\ta\tc 6} & = 0\,,\\
  (\tb,\tc,3,5) & & f_{\ta\tb 5}\, f_{\ta\tc 3} - f_{\ta\tb 3}\,
  f_{\ta\tc 5} & = 0\,,\\
  (\tb,\tc,3,6) & & f_{\ta\tb 6}\, f_{\ta\tc 3} - f_{\ta\tb 3}\,
  f_{\ta\tc 6} & = 0\,,\\
  (\tb,\tc,3,4) & & f_{\ta\tb 4}\, f_{\ta\tc 3} - f_{\ta\tb 3}\,
  f_{\ta\tc 4} & = c\, f_{\tb\tc 2}\,,\\
  (\tb,\tc,2,4) & & f_{\ta\tb 4}\, f_{\ta\tc 2} - f_{\ta\tb 2}\,
  f_{\ta\tc 4} & = -c\, f_{\tb\tc 3}\,,\\
  (\tb,\tc,4,5) & & f_{\ta\tb 5}\, f_{\ta\tc 4} - f_{\ta\tb 4}\,
  f_{\ta\tc 5} & = c\, f_{\tb\tc 6}\,,\\
  (\tb,\tc,4,6) & & f_{\ta\tb 6}\, f_{\ta\tc 4} - f_{\ta\tb 4}\,
  f_{\ta\tc 6} & = -c\, f_{\tb\tc 5}\,,\\
  (\tb,\tc,2,3) & & f_{\ta\tb 3}\,f_{\ta\tc 2}- f_{\ta\tb2}
  \,f_{\ta\tc 3} & =  c \,( f_{\tb\tc 4} - f_{7\tb\tc})\,,\\
  (\tb,\tc,5,6) & & f_{\ta\tb 6}\,f_{\ta\tc 5}-
  f_{\ta\tb 5}\,f_{\ta\tc 6} & = c \,( f_{\tb\tc 4} - f_{7\tb\tc})\,,\\
  (\tb,\tc,7,1) & & f_{\ta\tb 1}\, f_{\ta\tc 7} - f_{\ta\tb 7}\,
  f_{\ta\tc 1} & = 0\,,\\
  (\tb,\tc,7,4) & & f_{\ta\tb 4}\, f_{\ta\tc 7} - f_{\ta\tb 7}\,
  f_{\ta\tc 4} & = 0\,,\\
  (\tb,\tc,7,2) & & f_{\ta\tb 2}\, f_{\ta\tc 7} - f_{\ta\tb 7}\,
  f_{\ta\tc 2} & = c\, f_{\tb\tc 3}\,,\\
  (\tb,\tc,7,3) & & f_{\ta\tb 3}\, f_{\ta\tc 7} - f_{\ta\tb 7}\,
  f_{\ta\tc 3} & = -c\, f_{\tb\tc 2}\,,\\
  (\tb,\tc,7,5) & & f_{\ta\tb 5}\, f_{\ta\tc 7} - f_{\ta\tb 7}\,
  f_{\ta\tc 5} & = c\, f_{\tb\tc 6}\,,\\
  \label{s3:qc_bc76'}
  (\tb,\tc,7,6) & & \hspace{4.6cm} f_{\ta\tb 6}\, f_{\ta\tc 7} -
  f_{\ta\tb 7}\, f_{\ta\tc 6} & = -c\, f_{\tb \tc 5}\,,
\end{align}
\end{subequations}
\begin{subequations}
\begin{align}
  (\tb,\tc,\td,m) & & 0&= f_{\ta\tb\tc}\, f_{\ta\td m} + f_{\ta\tb m}\,
  f_{\ta\tc\td} - f_{\ta\tb\td}\, f_{\ta\tc m}\,,\\
  (\tb,\tc,\td,7) & & 0&= f_{\ta \tb\tc}\,f_{\td 7\ta} + f_{\ta \tb 7}\,
  f_{\tc\td\ta} - f_{\ta\tb\td}\, f_{\ta\tc 7}\,,\\
  \label{s3:qc_bcde'}
  (\tb,\tc,\td,\te) & & f_{a\tb\tc}\,f_{\td\te a} + f_{a\tb\te}\,
  f_{\tc\td a} - f_{a\tb\td}\, f_{a\tc\te} &= f_{r\tb\tc}\,f_{\td\te
    r} + f_{r\tb\te}\, f_{\tc\td r} - f_{r\tb\td}\, f_{r\tc\te}\,.
\end{align}
\end{subequations}

\subsubsection{The most general solution to equations \eqref{s3:liealg}}
\label{a:proof}
Here we will prove the claim that the most general solution of
equations \eqref{s3:liealg} is given by \eqref{s3:onlysol} and an
arbitrary, antisymmetric $H_+$ that commutes with $G_1$. In fact, it
suffices to consider the Lie subalgebra
$\mathfrak{s}'\subset\mathfrak{g}$ spanned by $\{G_2,G_3,H_+,H_-\}$
which is also solvable. Its non-vanishing Lie brackets are 
\begin{align}
  \label{a:system}
  [G_2,H_+] &=-2c\, G_3, & [G_2,G_3] &= c\, H_-,\nonumber\\
  [G_3,H_+] &=+2c\, G_2\,. &  &
\end{align}
We shall prove the following theorem:\\
\noindent {\bf Theorem:} The most general solution to system
\eqref{a:system} consists of solutions with
\begin{equation}
  \label{a:solution}
  G_2=G_3=H_- = 0, \qquad\quad H_+=-H_+^T \text{ arbitrary}.
\end{equation}
Our proof requires two elementary lemmata about matrices and a
corollary of Lie's theorem concerning finite-dimensional
representations of complex, solvable Lie algebras.\\

\noindent {\bf Lemma:} An antisymmetric matrix
$A\in\text{Mat}(\mathbb{R},m\times m)$ is nilpotent if and only if
$A=0$.\\
{\it Proof: Being antisymmetric $A$ can be brought to diagonal form $P
  A P^{-1}=\text{diag}(\lambda_1,\ldots,\lambda_m)$ with a
  $P\in\text{GL}(\mathbb{C},m\times m)$ and $\lambda_i\in
  i\mathbb{R}$. As $PA^nP^{-1}=(PAP^{-1})^n$ for all $n\in\mathbb{N}$,
  nilpotency is
  basis-independent. It is then obvious that, $$(PAP^{-1})^n=\text{diag}(
  \lambda_1^n,\ldots,\lambda_m^n)~,$$ is nilpotent iff 
  $\lambda_i=0\, \forall i$ which implies $A=0$. The converse is
  trivial.}\\

\noindent {\bf Lemma:} Given matrices
$A_1,\ldots,A_k\in\text{Mat}(\mathbb{C},m\times m)$ for
$k\in\mathbb{N}$. For simultaneously triangularizable matrices
$A_1,\ldots,A_k$ the commutator $[A_i,A_j]$ is nilpotent for all
$i,j=1,\ldots,k$.\\ 
{\it Proof: The commutator of two upper triangular matrices is
  strictly upper triangular and, hence, nilpotent.}\\  

\noindent {\bf Corollary of Lie's theorem\footnote{Given a complex,
    solvable Lie algebra, then all its finite-dimensional irreducible
    representations are one-dimensional.}:} Let $\mathfrak{g}$ be a  
complex, solvable Lie algebra and $(V,\rho)$ a finite-dimensional
representation of $\mathfrak{g}$. Then there exists a basis of $V$
such that all elements of $\mathfrak{g}$ are represented as upper
triangular matrices.\\
{\it Proof: Lecture script by W.\ Soergel \cite{soergel_lie}.}\\

In order to be able to apply this corollary we need to complexify our
real Lie algebra \eqref{a:system}. \\

\noindent {\bf Lemma:} Given a real Lie algebra $\mathfrak{g}$ and a
finite-dimensional real representation $(V,\rho)$ of
$\mathfrak{g}$. Then one finds a finite-dimensional
representation $(V_{\mathbb{C}},\rho_{\mathbb{C}})$ of the
complexified Lie algebra
$\mathfrak{g}_{\mathbb{C}}:=\mathfrak{g}\otimes_{\mathbb{R}}\mathbb{C}$
(with $\mathbb{C}$-linear extension of the Lie bracket) defined by
$V_{\mathbb{C}}:=V\otimes_{\mathbb{R}}\mathbb{C}$
and $$\rho_{\mathbb{C}}(X+iY):= \rho(X)+i\rho(Y),$$ 
for all $X,Y\in\mathfrak{g}$.\\
{\it Proof: $\mathbb{C}$-linearity of $\rho_{\mathbb{C}}$ is obvious
  and so is the proof of $$\rho_{\mathbb{C}}([X+iY,U+iV]) =
  [\rho_{\mathbb{C}}(X+iY),\rho_{\mathbb{C}}(U+iV)]$$ for all
  $X,Y,U,V\in \mathfrak{g}$. As a result,
  $(V_{\mathbb{C}},\rho_{\mathbb{C}})$ is a finite-dimensional
  representation of the complex Lie algebra
  $\mathfrak{g}_{\mathbb{C}}$.}\\ 

Now we can prove the theorem:\\
\noindent {\it Proof of the theorem: Assume that there exists a
solution of \eqref{a:system} with an antisymmetric $G_2\neq
0\in\text{Mat}(\mathbb{R},m\times m)$. Any such solution would be a
finite-dimensional real representation $(\mathbb{R}^m,\rho)$ of our
real solvable Lie algebra $\mathfrak{s}'$. In this proof such a
solution will be denoted by $\rho(G_2),\rho(G_3),\rho(H_-),\rho(H_+)$
with $\rho(G_2)\neq 0$ by assumption, while $G_2, G_3,
H_-,H_+\in\mathfrak{s}'$ shall refer to the abstract elements of the
Lie algebra. We denote the induced representation of the complexified
Lie algebra $\mathfrak{s}_{\mathbb{C}}'$ as
$(\mathbb{C}^m,\rho_{\mathbb{C}})$. Since also
$\mathfrak{s}_{\mathbb{C}}'$ is solvable, we apply the corollary and
find that
$\rho_{\mathbb{C}}(G_2),\rho_{\mathbb{C}}(G_3),\rho_{\mathbb{C}}(H_-), 
\rho_{\mathbb{C}}(H_+)\in\text{Mat} (\mathbb{C},m\times m)$ are
simultaneously triangularizable. Then, according to the second lemma
we find that, in particular ($p=1$), $$[\rho_{\mathbb{C}}(G_3),
\rho_{\mathbb{C}}(H_+)] = 2c\,\rho_{\mathbb{C}}(G_2)$$ is
nilpotent. As $c\neq 0$ one finds $\rho_{\mathbb{C}}(G_2)=\rho(G_2)$ is 
nilpotent. However, being antisymmetric $\rho(G_2)$ must be zero by
the first lemma which is in contradiction with $\rho(G_2)\neq 0$. We
therefore conclude that $\rho(G_2)=0$ which, by means of the Lie
algebra \eqref{a:system}, immediately implies
$\rho(G_3)=\rho(H_-)=0$. As a result, solutions \eqref{a:solution} are
already the most general solutions to \eqref{a:system}. This ends the
proof.}\\  

\subsubsection{Solving $[G_1,G_4]=0$}
\label{a:G1G4zero}
We will now solve \eqref{s3:qc_bc14''}, which in matrix notation reads
\begin{equation}
\label{s3:commG1G4}
  [G_1,G_4]=0\,.
\end{equation}
It is by means of an $O(n-1)$ transformation that, without loss of
generality, any $G_1$ can be written in block-diagonal form as 
\begin{equation}
\label{s3:gauge_g1}
  G_1 = (D\otimes\varepsilon)\oplus 0 =
\begin{pmatrix}D\otimes \varepsilon & 0\\ 0&0\end{pmatrix} \,,
\end{equation}
where $D=diag(x_1,\ldots,x_1,x_2,\ldots,x_2,\ldots)$ is a diagonal
matrix with ordered positive eigenvalues $x_1 > x_2 > \ldots > 0$ and
$\varepsilon$ is the antisymmetric $2\times 2$ matrix with $\varepsilon_{12}
= 1$; the zeros in \eqref{s3:gauge_g1} denote zero matrices of
appropriate dimensions. Note that, in general, this gauge can only be 
obtained by also using reflections (in addition to rotations). While
strictly speaking we are only allowed to use $SO(n-1)\subset G$
rotations, the quadratic constraints \eqref{s3:qc_bc14''} -
\eqref{s3:qc_bcde''} are also $O(n-1)$ tensor equations. We may
therefore also use reflections to arrive, as an intermediate step, at
the gauge \eqref{s3:gauge_g1} --- which simplifies the subsequent
analysis --- as long as, in the end, we return to only using rotations,
in that we apply another reflection that flips two directions but
preserves the block structure (e.g.\ $x_i \rightarrow -x_i$ for one
$2\times 2$ block). Since $D\otimes \varepsilon$ is invertible,
\eqref{s3:commG1G4} implies (also using another gauge choice for the
lower right block)   
\begin{equation}
  \label{s3:commG1G4_G4}
  G_4 = \begin{pmatrix}A & 0 \\ 0 & (D'\otimes \varepsilon) \oplus
    0\end{pmatrix} \,,
\end{equation}
where $A$ is an antisymmetric matrix (of the same matrix dimensions
as $D\otimes \varepsilon$) satisfying
\begin{equation}
\label{s3:commG1G4_DA}
  [D\otimes\varepsilon,\, A]=0
\end{equation}
and $D'$ is another invertible diagonal matrix. In order to solve
\eqref{s3:commG1G4_DA} we note that any even-dimensional antisymmetric
$A$ can be written as 
\begin{equation}
  A = S \otimes \varepsilon + A_1 \otimes \mathbb{1} + A_2\otimes
  \sigma_1 + A_3\otimes \sigma_3\,,
\end{equation}
where $S$ is symmetric, $A_1, A_2, A_3$ are antisymmetric, and
$\sigma_1,\sigma_3$ are the usual Pauli matrices. Now
\eqref{s3:commG1G4_DA} implies\footnote{$\{.,.\}$ denotes the
  anticommutator.} 
\begin{equation}
  [D,S]=0, \qquad [D,A_1]=0, \qquad \{D,A_2\} = 0, \qquad \{D,A_3\} = 0\,,
\end{equation} 
which in the reflection gauge \eqref{s3:gauge_g1} implies $A_2=A_3=0$
and $S_{ij}=(A_1)_{ij}=0$ for all $i,j$ with $x_i\neq x_j$. As a
result, we obtain
\begin{equation}
\label{s3:commG1G4_A}
  A = S\otimes \varepsilon + A_1 \otimes \mathbb{1}\,,
\end{equation}
where now $S$ and $A_1$ are block-diagonal with blocks associated to 
degenerate $x_i$ in $D$. We will now refine the block-structure in
$G_4$. To this end, we will use the residual symmetry of the blocks in
$G_1$ and $G_4$ to bring each $G_4$ block associated to some $x_i$ to
the form
\begin{equation}
  (\text{ith block in $G_4$}) =
  \left(diag(y_{i1},\ldots,y_{i1},y_{i2},\ldots,y_{i2},\ldots)\otimes
  \varepsilon\right ) \oplus 0\,,
\end{equation}
with $y_{i1}>y_{i2}>\ldots>0$. While this, of course, temporarily
spoils the gauge \eqref{s3:gauge_g1}, it is by means of
\eqref{s3:commG1G4} that we find, using the same argument as before,
that the ith block in $G_1$ has a subblock structure with blocks
associated to degenerate $y_{ij}$ or zero in the ith $G_4$
block. Now we apply symmetries that respect these subblocks to bring
$G_1$ back to our gauge \eqref{s3:gauge_g1} and at the same time
maintain the subblock structure in $G_4$. Then, repeating the argument
that lead to \eqref{s3:commG1G4_A}, we know that the subblock
associated to $x_i$ in $G_1$ and $y_{ij}$ in $G_4$ is given by 
\begin{equation}
\label{s3:tmp}
  \left(\text{$(i,j)$ block in $G_4$}\right) = S^{(ij)}\otimes
  \varepsilon + A_1^{(ij)}\otimes \mathbb{1}\,,
\end{equation}
where
\begin{equation}
\label{s3:commG1G4_ysquared}
  \left (S^{(ij)}\otimes \varepsilon + A_1^{(ij)}\otimes \mathbb{1}
  \right)^2 = -(y_{ij})^2 \, \mathbb{1}\otimes \mathbb{1}.
\end{equation}
The $(i,j)$ block in $G_1$ is $x_i\, \mathbb{1}\otimes \varepsilon$ and is
thus invariant under orthogonal transformations that only act on the
first tensor product factor. Such transformations can be used to bring
$S^{(ij)}$ to diagonal form
\begin{equation}
  D^{(ij)} =
  diag(d_{ij1},\ldots,d_{ij1},-d_{ij1},\ldots,-d_{ij1},\ldots)
  \oplus 0\,,
\end{equation}
where $d_{ijk}>0$ and the dimensions of positive and negative
eigenvalues can in general be different. In doing so,
\eqref{s3:commG1G4_ysquared} gives rise to the following system of 
equations  
\begin{equation}
\label{s3:commG1G4_ysquared'}
  (A_1^{(ij)})^2 + (y_{ij})^2 = (D^{(ij)})^2,\qquad\quad \{D^{(ij)},
  A_1^{(ij)}\}=0.
\end{equation}
The second equation gives
\begin{equation}
   (A_1^{(ij)})_{kl}=0 \qquad \vee\qquad (D^{(ij)})_{kk} = -
   (D^{(ij)})_{ll}
\end{equation}
and, hence, the $D^{(ij)}$ and $A_1^{(ij)}$ have the following
block-diagonal form 
\begin{equation}
  D^{(ij)} =
\begin{pmatrix} 
  d_{ij1}\mathbb{1} & 0  &\cdots & 0 \\
  0 & -d_{ij1}\mathbb{1}'  & \cdots &  0 \\
  \vdots & \vdots & \ddots & \vdots \\
  0 & 0 & \cdots & 0 
\end{pmatrix}
\qquad
  A_1^{(ij)} =
\begin{pmatrix} 
  0 & F^{(ij1)}  &\cdots & 0 \\
  -{F^{(ij1)}}^T & 0  & \cdots &  0 \\
  \vdots & \vdots & \ddots & \vdots \\
  0 & 0 & \cdots & F^{(ij0)} 
\end{pmatrix}\,,
\end{equation}
where $F^{(ijk)}$ are rectangular matrices and $F^{(ij0)}$ are
antisymmetric square matrices subject to the following conditions
(from \eqref{s3:commG1G4_ysquared'}):   
\begin{equation}
\label{s3:commG1G4_DA1blocks}
  d_{ijk}^2 +
\begin{pmatrix}
  F^{(ijk)} {F^{(ijk)}}^T & 0 \\
  0 & {F^{(ijk)}}^T F^{(ijk)}
\end{pmatrix}
= y_{ij}^2,
\qquad\qquad (F^{(ij0)})^2 = -y_{ij}^2.
\end{equation}
Without loss of generality we can use the residual symmetry to bring
each $F^{(ij0)}$ into diagonal form
\begin{equation}
  D^{(ij0)}\otimes \varepsilon\,,
\end{equation}
where the eigenvalues of $D^{(ij0)}$ must be $\pm y_{ij}$ in order to
satisfy \eqref{s3:commG1G4_DA1blocks}. In particular, $F^{(ij0)}$ must
have even dimension. As to the $F^{(ijk)}$,
\eqref{s3:commG1G4_DA1blocks} implies that 
\begin{subequations}
\begin{eqnarray}
  \label{s3:commG1G4_FFt}
  F^{(ijk)} {F^{(ijk)}}^T & = & \xi_{ijk}\, \mathbb{1}\,,\\
  \label{s3:commG1G4_FtF}
  {F^{(ijk)}}^T F^{(ijk)} & = & \xi_{ijk} \mathbb{1}'
\end{eqnarray}
\end{subequations}
for some non-negative number $\xi_{ijk}$. In the case where
$\xi_{ijk}=0$ one finds $F^{(ijk)}=0$, and
\eqref{s3:commG1G4_DA1blocks} implies $d_{ijk}=y_{ij}$. On the other
hand, for $\xi_{ijk}>0$, \eqref{s3:commG1G4_FFt},
\eqref{s3:commG1G4_FtF}, respectively, shows that the rows/columns
of $1/{\sqrt{\xi_{ijk}}} F^{(ijk)}$ are orthonormal which, however, is
only possible if $F^{(ijk)}$ is a square matrix. In this case,
$1/{\sqrt{\xi_{ijk}}} F^{(ijk)}$ is an orthogonal matrix that without
loss of generality can be orthogonally transformed to the unit
element: In fact, the $(i,j,k)$ block in $D^{(ij)}$  
\begin{equation}
  \begin{pmatrix}
    d_{ijk}\mathbb{1} & 0 \\
    0 & -d_{ijk}\mathbb{1}
  \end{pmatrix}
\end{equation}
is invariant under an orthogonal transformation
\begin{equation}
  \begin{pmatrix}
    T & 0 \\
    0 & S
  \end{pmatrix}
\end{equation}
that at the same time acts on the $(i,j,k)$ block in $A^{(ij)}$ as 
\begin{eqnarray}
  \begin{pmatrix}
    0 & F^{(ijk)} \\
    -{F^{(ijk)}}^T & 0
  \end{pmatrix} & \rightarrow &
  \begin{pmatrix}
    T^T & 0 \\
    0 & S^T
  \end{pmatrix}
  \begin{pmatrix}
    0 & F^{(ijk)} \\
    -{F^{(ijk)}}^T & 0
  \end{pmatrix}
  \begin{pmatrix}
    T & 0 \\
    0 & S
  \end{pmatrix}\nonumber\\ & = &
  \begin{pmatrix}
    0 & T^T F^{(ijk)} S \\
    -(T^T F^{(ijk)} S)^T & 0
  \end{pmatrix}.
\end{eqnarray}
Choosing $S=\mathbb{1}, T=1/\sqrt{\xi_{ijk}} F^{(ijk)}$ one obtains 
\begin{equation}
  F^{(ijk)} = \sqrt{\xi_{ijk}} \mathbb{1}\,.
\end{equation}
The condition \eqref{s3:commG1G4_DA1blocks} finally reads
\begin{equation}
  d_{ijk}^2 + \xi_{ijk} = y_{ij}^2
\end{equation}
and, hence,
\begin{equation}
  d_{ijk} = |y_{ij}| \cos\phi_{ijk}, \qquad\quad \sqrt{\xi_{ijk}} =
  |y_{ij}| \sin\phi_{ijk}
\end{equation}
for some angle $\phi_{ijk}\in (0,\pi/2)$.
To conclude, we have the following block types,
\begin{eqnarray}
\label{s3:commG1G4_ijkphi}
  G_1^{(ijk)} & = & x_i
\begin{pmatrix}
  \mathbb{1} & 0\\
  0 & \mathbb{1}  
\end{pmatrix} \otimes \varepsilon,\nonumber\\
  G_4^{(ijk)} & = & |y_{ij}| \left (\cos\phi_{ijk}
\begin{pmatrix}
  \mathbb{1} & 0\\
  0 & -\mathbb{1}
\end{pmatrix} \otimes\varepsilon + \sin\phi_{ijk}
\begin{pmatrix}
  0 & \mathbb{1}\\
  -\mathbb{1} & 0
\end{pmatrix} \otimes\mathbb{1}_2
 \right)
\end{eqnarray}
for $\phi_{ijk}\in (0,\pi/2)$, while blocks with $F^{(ijk)}=0$ read
\begin{eqnarray}
  G_1^{(ijk)} & = & x_i
\begin{pmatrix}
  \mathbb{1} & 0\\
  0 & \mathbb{1}'  
\end{pmatrix} \otimes \varepsilon,\nonumber\\
  G_4^{(ijk)} & = & |y_{ij}| 
\begin{pmatrix}
  \mathbb{1} & 0\\
  0 & -\mathbb{1}'
\end{pmatrix} \otimes\varepsilon.
\end{eqnarray}
Finally, zero blocks in $D^{(ij)}$ give rise to the following blocks:
\begin{eqnarray}
\label{s3:commG1G4_ij0}
  G_1^{(ij0)} & = & x_i (\mathbb{1}\otimes \mathbb{1}_2) \otimes 
  \varepsilon,\nonumber\\
  G_4^{(ij0)} & = & (D^{(ij0)}\otimes\varepsilon) \otimes \mathbb{1}_2 \,.
\end{eqnarray}
Using appropriate orthogonal transformations it is possible to write
\eqref{s3:commG1G4_ijkphi} as
\begin{eqnarray}
\label{s3:commG1G4_ijkphi'}
  G_1^{(ijk)} & = & x_i\, \mathbb{1}\otimes
  (\mathbb{1}_2\otimes \varepsilon),\nonumber\\
  G_4^{(ijk)} & = & |y_{ij}|\, \mathbb{1}\otimes \left (\cos\phi_{ijk} 
\begin{pmatrix}
  1 & 0\\
  0 & -1
\end{pmatrix} \otimes\varepsilon + \sin\phi_{ijk}
\begin{pmatrix}
  0 & 1\\
  -1 & 0
\end{pmatrix} \otimes\mathbb{1}_2
 \right),
\end{eqnarray}
and, similarly, we transform \eqref{s3:commG1G4_ij0} to\footnote{Note
  that \eqref{s3:commG1G4_ij0'} yields \eqref{s3:commG1G4_ijkphi'} for 
  $\phi_{ijk} = \pi/2$ provided that $D^{(ij0)}$ has only positive
  eigenvalues. But the latter need not be the case in general.}  
\begin{eqnarray}
\label{s3:commG1G4_ij0'}
  G_1^{(ij0)} & = & x_i\, \mathbb{1}\otimes (\mathbb{1}_2 \otimes 
  \varepsilon),\nonumber\\
  G_4^{(ij0)} & = & D^{(ij0)}\otimes(\varepsilon \otimes \mathbb{1}_2) .
\end{eqnarray}
Note that both \eqref{s3:commG1G4_ijkphi'} and
\eqref{s3:commG1G4_ij0'} are block-diagonal matrices with non-trivial
$4\times 4$ blocks. From these blocks and using \eqref{s3:commG1G4_G4}
we can construct the full solution of \eqref{s3:commG1G4} for the
gauge choice outlined above. As mentioned already, in the end one may
have to apply another reflection so that this gauge can be
obtained from generic matrices $G_1$ and $G_4$ only by rotations,
rather than reflections.

\newpage
\bibliography{myrefs}
\bibliographystyle{JHEP-2}


\end{document}